\documentclass[fleqn,usenatbib]{mnras}

\usepackage{newtxtext,newtxmath}

\usepackage[T1]{fontenc}

\usepackage{graphicx}	

\newcommand{\pmra}{$\mu_{\alpha}\cos{\delta}$}
\newcommand{\pmdec}{$\mu_{\delta}$}
\newcommand{\masyr}{mas yr$^{-1}$}
\newcommand{\G}{{$\rm G$}}
\newcommand{\bprp}{{$\rm BP-RP$}}
\newcommand{\ks}{{$\rm K_{s}$}}
\newcommand{\jks}{{$\rm J-K_{s}$}}
\newcommand{\V}{{$\rm V$}}

\newcommand{\Vt}{{$\rm V_t$}}
\newcommand{\MV}{$\rm M_{V}$}
\newcommand{\AG}{{$\rm A_G$}}
\newcommand{\AV}{{$\rm A_V$}}
\newcommand{\rc}{{$\rm r_c$}}
\newcommand{\rt}{{$\rm r_t$}}
\newcommand{\rhl}{{$\rm r_{h}$}}

\newcommand{\feh}{{$\rm [Fe/H]$}}
\newcommand{\kms}{{$\rm km\ s^{-1}$}}

\title[Low-luminosity GCs towards the MW bulge]{Hidden in the Haystack: \\ Low-luminosity globular clusters towards the Milky Way bulge
\thanks{Based on observations collected at the European Southern Observatory under ESO programmes 0103.D-0386(A), 105.20MY.001, 179.B-2002, and 198.B-2004.}}

\author[F. Gran, M. Zoccali, I. Saviane et al.]{
F. Gran$^{1,2,3}$,
M. Zoccali$^{1,2}$,
I. Saviane$^{3}$,
E. Valenti$^{4,5}$,
A. Rojas-Arriagada$^{1,2}$,
R. Contreras Ramos$^{1,2}$,
\newauthor{
J. Hartke$^{3}$,
J. A. Carballo-Bello$^{6}$,
C. Navarrete$^{3,2}$,
M. Rejkuba$^{4}$,
J. Olivares Carvajal$^{1,2}$}
\\
$^{1}$Instituto de Astrof\'isica, Av. Vicu\~na Mackenna 4860, Santiago, Chile\\ 
$^{2}$Instituto Milenio de Astrof\'isica, Santiago, Chile\\ 
$^{3}$European Southern Observatory, Alonso de C\'ordova 3107, Casilla 19001 Santiago, Chile\\
$^{4}$European Southern Observatory, Karl Schwarzschild-Strabe 2, 85748 Garching bei M\"unchen, Germany\\
$^{5}$Excellence Cluster ORIGINS, Boltzmann-Stra\ss e 2, D-85748 Garching bei M\"unchen, Germany\\
$^{6}$Instituto de Alta Investigaci\'on, Sede Esmeralda, Universidad de Tarapac\'a, Av. Luis Emilio Recabarren 2477, Iquique, Chile
}

\date{Accepted 2021 August 25. Received 2021 August 25; in original form 2021 May 1}

\pubyear{2020}

\begin{document}
\label{firstpage}
\pagerange{\pageref{firstpage}--\pageref{lastpage}}
\maketitle

\begin{abstract}
Recent wide-area surveys have enabled us to study the Milky Way with unprecedented detail. 
Its inner regions, hidden behind dust and gas, have been partially unveiled with the arrival of near-IR photometric and spectroscopic datasets. 
Among recent discoveries, there is a population of low-mass globular clusters, known to be missing, especially towards the Galactic bulge.
In this work, five new low-luminosity globular clusters located towards the bulge area are presented. 
They were discovered by searching for groups in the multi-dimensional space of coordinates, colours, and proper motions 
from the Gaia EDR3 catalogue and later confirmed with deeper VVV survey near-IR photometry. 
The clusters show well-defined red-giant branches and, in some cases, horizontal branches with their members forming a dynamically coherent structure in proper motion space. 
Four of them were confirmed by spectroscopic follow-up with the MUSE instrument on the ESO VLT.
Photometric parameters were derived, and when available, metallicities, radial velocities and orbits were determined. 
The new clusters {\tt Gran~1} and {\tt 5} are bulge globular clusters, while {\tt Gran~2, 3} and {\tt 4} present halo-like properties. 
Preliminary orbits indicate that {\tt Gran~1} might be related to the Main Progenitor, or the so-called ``low-energy’’ group, 
while {\tt Gran~2, 3} and {\tt 5} appear to follow the Gaia-Enceladus/Sausage structure.
This study demonstrates that the Gaia proper motions, combined with the spectroscopic follow-up and colour-magnitude diagrams,
are required to confirm the nature of cluster candidates towards the inner Galaxy.
High stellar crowding and differential extinction may hide other low-luminosity clusters.
\end{abstract}




\section{Introduction}
\label{sec:intro}

Globular clusters (GCs) represent one of the most valuable stellar tracers that can be observed to understand the Milky Way (MW) evolution. 
They allow us to constrain ages, masses and distances with unique precision, in contrast with most stars located in the Galaxy.
Only recently and for a limited sample of stars, a more precise determination of those parameters became available in the second Data Release (DR) of the Gaia catalogue \citep{brown21}.

Their contribution to the MW assembly has been widely explored in numerical simulations \citep{kruijssen19a, kruijssen19b, carlberg20}, 
presenting the proto-GCs properties and suggesting that most of their stellar content is now lost in the inner Galaxy \citep{baumgardt19}. 
From the observational point of view, it seems established that the properties of GCs observed today in the Galaxy are different from those that 
they had when they formed at high redshift \citep{renzini17, carlberg20}.
A similar hypothesis is needed to explain the observations of multiple stellar populations in massive GCs, whereby the first generation of stars must 
have been much more massive than currently observed, to enrich the second generation \citep[see, e.g.][for recent reviews]{bastian18, gratton19}.

Besides, GCs can be used to trace the different components that have been assembled during the history of the build-up of our Galaxy.
A tentative separation of the Galactic GC population was proposed by \cite{forbes10}, \cite{leaman13} and \cite{myeong18}, 
the latter employing orbital energy criteria to classify the {\it in situ} and accreted GCs. 
The accreted group comprises those GCs formed within dwarf galaxies, then accreted by the MW \citep{mackey04, myeong19}.
One of the most important implications of these results is that bulge GCs have an essential role in the characterisation of the early phases of the Galaxy formation, 
as they can trace the fossil record of its early stages \citep{barbuy18, zoccali19}.

Despite the critical role that the Galactic bulge GCs play in the characterisation of the early Galaxy, 
no consensus has been reached on the absolute number of clusters belonging to this component. 
This is primarily due to significant and differential extinction towards the disk and bulge and more considerable contamination by field stars.
Both occurrences hinder our ability to recognise the cluster colour-magnitude diagram (CMD) sequences against the field stars.
To complement this scenario, it was also reported that even in the most distant regions of the MW, the same uncertainty is observed \citep{webb21}.

According to the analysis of \citet[see their Figure 8]{baumgardt19}, we are only detecting the high-mass end of the bulge GCs.
They also derive that a minimum mass was required for a given GC to survive the dynamical processes going on in the early inner Galaxy. 
Recent observational efforts to complete the census of bulge GCs include systematic searches in the near-infrared (near-IR) 2MASS \citep{2mass} 
and VISTA Variables in the V\'ia L\'actea \citep[VVV,][]{vvv} surveys, and more recently, in the Gaia survey \citep[][]{gaia_mission, lindegren18, brown18, brown20}. 
Despite the latter being an optical survey, its very precise astrometry permits to kinematically distinguish cluster from field stars, 
dramatically improving the cluster detection capabilities. 
Only in the direction of the Galactic bulge, in the last few years, the number of publications reporting the discovery of new GC candidates has risen significantly 
\citep[e.g.,][]{minniti11a, monibidin11, borissova14, minniti17a, minniti17b, minniti17c, gran19, palma19, garro20}.

Recently, a large body of evidence of the contribution of GCs to the star content of the disk \citep{pricejones20} and bulge \citep{hughes20, horta21, kisku21} has been revealed. 
It is clear from that view that GCs will dissolve because of the strong MW gravitational potential, producing extended stellar tails or streams.
Nevertheless, those remnants are extremely difficult to isolate from the field population, 
with only a few exceptions towards the inner Galaxy \citep{ibata01, pricewhelan16, ibata18, pricejones20}.

We will focus our study on the compact cores or remnants of these processes, as we can identify them as overdensities in star counts, variable star content or any other dynamical tracer.
However, by definition, a cluster needs to be dynamically bound. 
Thus the only way to confirm an overdensity detection as a real cluster is by verifying that the stars move coherently in space, using either radial velocities (RVs), proper motions (PMs), or both.
In fact, a major fraction of the new bulge GC candidates that have been found based on star counts in previous studies were discarded by \cite{gran19} using dynamical constraints. 
Specifically, the PM dispersion of the putative cluster members was comparable to that of the field stars in a spatial region near the centre of those cluster candidates.
Similar issues arise from the \cite{cantatgaudin20a} analysis of open clusters candidates.

In the present paper, new GC candidates are presented based on a refined version of the technique introduced in \cite{gran19} to detect coherent groups of stars. 
The initial search was performed on the Gaia DR2 catalogue and later confirmed in the Gaia early-DR3 (EDR3) database.
The detected clusters were also searched in the VVV PSF photometry \citep[][Contreras-Ramos in prep.]{contrerasramos17,surot19}, 
which contains deeper and more precise photometry in the most extincted regions close to the Galactic plane. 
Finally, the possible presence of RR Lyrae was verified using the OGLE IV and Gaia catalogues \citep{soszynski19,gaiadr2_vars} if the candidate GC exhibits a well-defined horizontal branch.

The paper is organised as follows: 
Sec.~\ref{sec:data_method} describes the Gaia DR2/EDR3 catalogue that was used to search for new GCs and the algorithm adopted to detect clusters. 
Sec.~\ref{sec:newGCs} presents the structural and dynamical analysis of the newly discovered GCs.
Sec.~\ref{sec:MUSE} presents the spectroscopic confirmation of four GCs, together with their RVs, mean metallicities and derived orbits.
Finally, Sec.~\ref{sec:summary} summarises the discoveries and state the prospects of this population.
Appendix~\ref{sec:AP_Gaia_CMDs}, \ref{sec:AP_GaiaVVV_CMDs}, and \ref{sec:AP_radial_profiles} contain all the plots relative to each of the new GCs, 
while Appendix~\ref{sec:AP_BH261Djorg1} and \ref{sec:AP_C1} contain the diagnostic plots for other analysed clusters.

\section{Gaia DR2: bulge data and methodology}
\label{sec:data_method}

\subsection{Gaia DR2: data selection}

We follow the same method detailed in \cite{gran19}, focusing our search on the Gaia DR2 catalogue \citep{gaia_mission,brown18,lindegren18} which provides coordinates, 
proper motions and magnitudes for a large region of the sky in the bulge area. 
This area includes the whole Galactic bulge within $|\ell, b| \lesssim 10$ deg with a complete photometric and astrometric solution. 

We query for Gaia DR2 sources in the mentioned region, dividing it into small circles of $0.8$ deg radius.
This choice ensures that each coordinate within the selected area is included at least once and up to four times within the circles.
This was done to optimise the clustering execution and avoid losing clusters that may appear close to the edges.
Only the stars from the main {$\tt GAIA\_SOURCE$} catalogue with complete astrometric and photometric information about their positions ($\alpha$, $\delta$), 
PMs ($\mu_{\alpha}\cos{\delta}$,$\mu_{\delta}$) and magnitudes (${\rm G}$, ${\rm RP}$, and ${\rm BP}$) were kept. 

\subsection{A method to discover new GCs}

For each star, the algorithm counts the number of neighbours within 1 unit in the phase-space defined by 1 arcmin in space, 1 \masyr\ in PM, and 1 mag in (\bprp) colour. 
Then, we select stars with at least nine neighbours (i.e., a group with at least 10 members), 
trying to avoid the detection of false-positive clusters due to stochastic overdensities, common in the bulge area.
This step was performed using the K-Dimensional Tree (KDTree) implementation on {\tt scikit-learn} \citep{scikit-learn}. 
Then, we apply a clustering algorithm to the selected stars with more than 10 neighbours to search for their centroids on the sky if any.
The latter task was performed with the DBSCAN \citep{dbscan} routine, also implemented in {\tt scikit-learn}. 
We choose DBSCAN over other algorithms because it does not need an already known number of groups to be found in each region, 
different to the K-means behaviour, and it offers the possibility to apply physical constraints to the cluster determination, as the maximum separation within points of the same group.
Note that the physical size of the resulting cluster may vary, as several groups of stars can form it within the separation constraints, in which DBSCAN will only recognise one cluster. 
In fact, a group of neighbours defines a ``neighbourhood'' that can be significantly larger than the minimum distance between a star and its neighbours. 
Finally, DBSCAN has been intensively tested with Gaia data, especially to search for open clusters.
For a complete DBSCAN description and comparison with other clustering algorithms, we point to the recent detailed studies by 
\cite{cantatgaudin19,cantatgaudin20b}, \cite{castroginard19,castroginard20}, and \citet[and references therein]{hunt21} for the use of DBSCAN to detect open clusters using Gaia DR2.

Then, an initial median centroid of the distribution of the candidate cluster stars in the sky and PM space was derived. 
In a second step, the resulting candidate cluster stars were compared with the surrounding stars within 5 arcmins of the derived centre, 
by means of the on-sky positions, PM diagram (also called Vector Point Diagram; hereafter VPD) and the (BP-RP, G) CMD.

The algorithm found 2614 cluster candidates, and for each one, a visual inspection was made to select groups whose members show outstanding coherence in position, 
motions and, particularly, the characteristic structures in the CMD of a single-age stellar population. 
In the present paper, we will focus on the old clusters; therefore, we selected candidates with a well-populated red giant branch (RGB), red-clump (RC) or hints of a horizontal branch (HB). 
We ended up with seven bonafide old star clusters and one candidate (C1), later discarded. 
Of those seven clusters, two match known objects (BH~261 and Djorg~1), and five are considered original discoveries, 
which we call {\tt Gran~1, 2, 3, 4} and {\tt 5}. 
A possible identification of Gran~1 with ESO~456-29 is later addressed (see Sec.~\ref{sec:known_clusters}).

Because this work considerably improves their parameters, we decided to retain Djorg~1 and BH~261 in the present discussion. 
We also note that {\tt Gran~1} was already reported and characterised in \cite{gran19}, 
but we will include it in the present paper because we perform a complete analysis of its properties. 
Our algorithm was able to recover all but 5 of the confirmed GCs listed in the \cite{kharchenko16} catalogue as described in \cite{gran19}. 
Not surprisingly, given that Gaia works at visible wavelengths, the 5 missed clusters are all very close to the Galaxy midplane, at $|b|\leq 2.5$ deg, 
in regions heavily affected by crowding and extinction. 

\subsection{Known clusters from the literature}
\label{sec:known_clusters}

A crossmatch with 10 arcmin tolerance was performed between the new cluster centroids and other GC catalogues from the literature 
\citep[][online versions]{harris96, harris10, kharchenko16} in order to identify already known clusters. 
We found three matches within the imposed angular separation with our candidates in the \cite{kharchenko16} catalogue.

Of those three, the properties of {\tt Gran~1} differ significantly from the possible literature match, 
although it is located only 38.72 arcsec away from ESO~456-29 (or ESO~456-SC29), discovered by the ESO/Uppsala photometric plates survey \citep{holmberg74, holmberg78, lauberts82} as a star cluster.
However, its nature was questioned by \cite{dias02, dias14} who changed its classification to ``dubious open cluster''. 
Despite this fact, \cite{kharchenko16} and \cite{bica19} still consider ESO~456-29 as an open cluster.
Moreover, \cite{kharchenko16} derived a distance to the candidate cluster of $\sim 3$ kpc and PMs of $\mu_{\alpha}\cos{\delta} = -3.38 \pm 4.15$ \masyr, $\mu_{\delta} = -6.25 \pm 4.72$ \masyr.
This PM is also different from the value of $\mu_{\alpha}\cos{\delta} = 0.558 \pm 4.15$ \masyr, $\mu_{\delta} = -2.96 \pm 4.72$ \masyr quoted in the online-catalogue by \citet{dias14}.

\begin{figure*}
    \centering
    \includegraphics[scale=0.45]{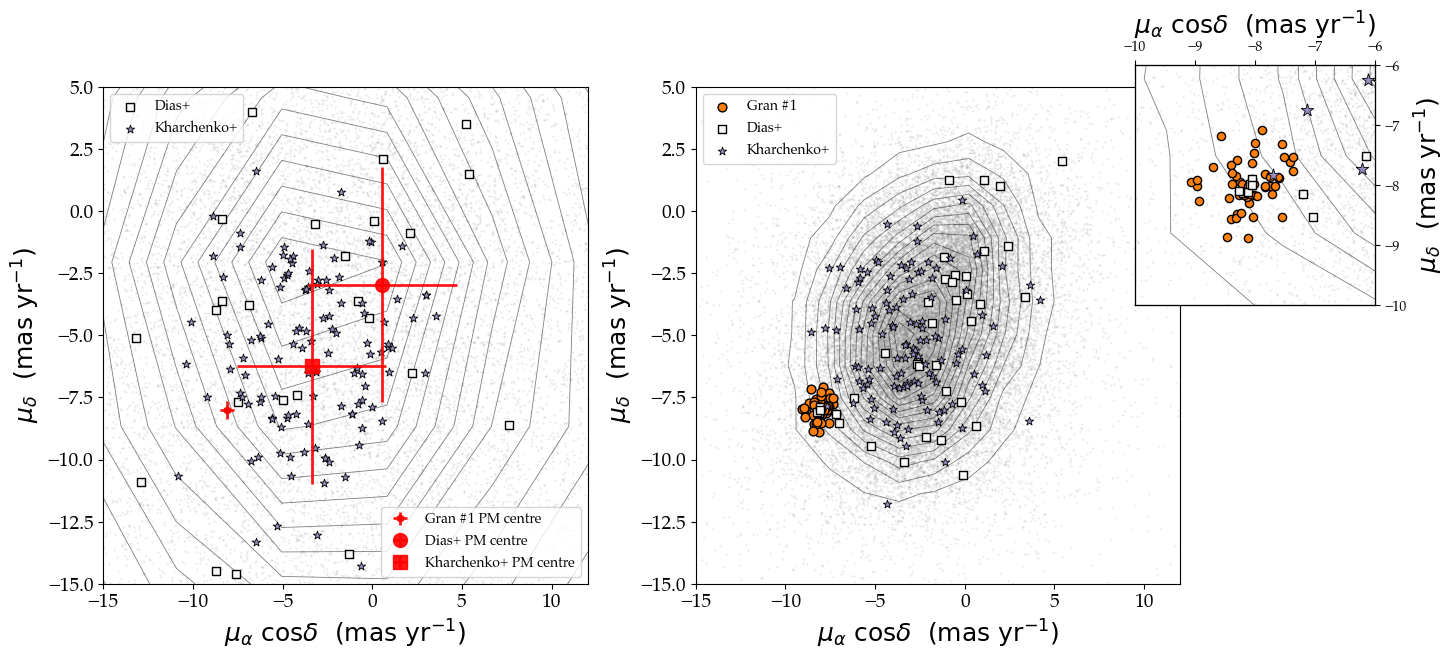}
    \caption{{\bf (Left)}: VPD of the \protect\cite{kharchenko13} stars within 10 arcmin from the centre we derived for {\tt Gran~1}. 
    Background grey points represent all the field stars while the contours mark different density levels.
    Squares and stars are the putative cluster members of ESO~456-29 from Dias+ \protect\citep{dias14}, and Kharchenko+ \citep{kharchenko13}, respectively.
    The red point, circle and square with error bars represent the derived PM centroid of the cluster, for the authors indicated in the labels.
    {\bf (Right)}: Same as the left panel, now with the PM, for the same stars, from the Gaia EDR3 catalogue. 
    Orange circles are the members of {\tt Gran~1} derived in the present work, while squares and stars are the cluster members from the Dias+ \protect\citep{dias14}, and Kharchenko+ \protect\citep{kharchenko13} catalogues, respectively.
    Both panels are in the same scale. 
    Finally, the inset is a zoom-in version of the VPD, that we include because eight stars in \protect\cite{dias14} and one from \protect\cite{kharchenko13} are within our derived PM centre of the cluster.
    As can be seen, only with Gaia data this cluster could be detected. }
    \label{fig:GranESO}
\end{figure*}

In order to verify these measurements, Fig.~\ref{fig:GranESO} shows the VPD of the selected stars from both catalogues \citep{dias14, kharchenko13} as quoted in the original papers (left panel) and as resulting from a match with Gaia (right panel; 1 arcsec tolerance in the match).  
Note that we show all the stars from \cite{dias14} with the ESO~456-29 designation, but only the ones with at least $80\%$ probability of being cluster members according to \cite{kharchenko13}. 

As Fig.~\ref{fig:GranESO} shows, neither \citeauthor{dias14} nor \citeauthor{kharchenko13} recover the cluster in the VPD; 
only with additional data from Gaia EDR3 a total of nine stars (eight from the Dias, and one from the Kharchenko catalogues) are located within the PM centre derived by us. 
With the additional Gaia EDR3 data, most of the putative cluster members identified by \cite{dias14} and \cite{kharchenko13} are now classified as field stars. 
Since the original cluster centre proposed by \cite{lauberts82} agrees with the one derived in this work within the (arguably large) errors, {\tt Gran~1} may indeed be ESO~456-29. 
However, in Sec.~\ref{sec:newGCs} we show that our derived properties differ significantly from those presented in past works, reason why we will maintain the {\tt Gran~1} designation, already introduced in \citet{gran19}.
Nonetheless, as for many GCs, we consider both names equally valid (i.e., {\tt Gran~1} $=$ ESO~456-29).  


One of our derived cluster candidate is 50.58 arcsec away from the known GC BH~261 (also known as ESO 456-78 or AL~3), 
discovered by \cite{andrews67} and analysed by \cite{lauberts82, BH, ortolani06} and \cite{barbuy21}. 
A first CMD was presented by \cite{ortolani06} describing BH~261 by its prominent HB with respect to the field stars, 
deriving a distance of $6.0 \pm 0.5$ kpc and a metallicity of $-1.3 \pm 0.3$ dex.
However, the \cite{ortolani06} observations could not optimally decontaminate the cluster RGB nor turn-off (TO) point.
\cite{bonatto08} and \cite{rossi15} later attempted a characterisation of the cluster properties, however, 
the low number of members and high reddening in the area led to inconclusive results. 
Also, \cite{barbuy21} tried to constrain the cluster metallicity by means of near-IR spectra. 
However, no \feh\ measurement was performed due to the limited wavelength range, adopting the same \cite{ortolani06} metallicity value of $-1.3$ dex.
H. Baumgardt catalogue of MW GCs\footnote{\url{https://people.smp.uq.edu.au/HolgerBaumgardt/globular/}} \citep{baumgardt20, baumgardt21} reports a distance of $6.12 \pm 0.26$  kpc, 
which is an average of the 4 independent literature measurements.  They do not report Gaia EDR3 parallax, nor kinematic distance measurement.
As we have an improved decontamination procedure based on the Gaia EDR3 PMs, we are able to isolate members down to the sub-giant branch (SGB) level, to better constrain BH~261 distance and metallicity.
Sec.~\ref{sec:newGCs} presents our derived properties for this cluster.

Lastly, one of our detected overdensities is located 36.82 arcsec away from the known GC Djorg~1 \citep{djorgovski87}.
Due to the high extinction present towards the cluster, only a few studies have focused on it \citep{ortolani95, davidge00, valenti10}. 
Recently, \cite{vasquez18} determined its radial velocity (RV), \cite{vasilev19} its PM and \cite{ortolani19} presented the first decontaminated CMD showing a prominent HB.
Unfortunately, our analysis could not detect the HB of the cluster, which is below the detection limit of both Gaia or VVV data.
Additionally, our derived CMD shows more scatter in comparison with other discovered clusters.
Nevertheless, the PM overdensity that we detect matches the previous \cite{vasilev19} determination.
We include Fig.~\ref{fig:AP_BH261Djorg1_plots} with both BH~261 and Djorg~1 within the Appendix~\ref{sec:AP_BH261Djorg1} for completeness to show our results.




\subsection{Other cluster candidate}

In addition to the five GCs that we present in Sec.~\ref{sec:newGCs}, we have selected one candidate for which we are not able to decipher whether it is real cluster or only stellar field overdensity. 
The positional properties of the candidate {\tt C1} is presented in Table~\ref{tab:positional}.
In this case, coordinates and PMs are clustered, but the CMD do not display a narrow cluster-like sequence.


For the {\tt C1} overdensity, our conclusion that this is not a real cluster is based on follow-up spectroscopy (see Sec.~\ref{sec:MUSE}), 
which does not display coherent kinematics lacking a clear peak in the RV distribution of its Gaia-selected members, which can be seen in Fig.~\ref{fig:AP_C1_MUSE}.
Appendix~\ref{sec:AP_C1} presents coordinates, PMs and CMD for {\tt C1}.

\section{A new population of GCs towards the MW bulge}
 \label{sec:newGCs}

Following \cite{harris10} and \cite{baumgardt18}, we decided to split our cluster characterisation into three categories: photometric, structural and dynamical.

\subsection{Photometric properties of the discovered clusters}
 \label{sec:photometric}

For each selected cluster, the algorithm returns the median coordinates (RA, Dec) and PMs (\pmra, \pmdec)
calculated from the putative cluster members, i.e., those stars having at least nine neighbours. 
At this point, we expand the selection criteria by including all the stars within 2 arcmin from the cluster centre in the sky and within 1 \masyr\ from the cluster centroid in the VPD. 
It is worth mentioning that while the initial cluster search was performed on the Gaia DR2, during this analysis the new EDR3 became available \citep{brown20}, and therefore we repeated the search on it.
Note that, while the clustering algorithm recovered the same five clusters, more members and tighter PM distribution were found with the EDR3 data. 

As shown in the CMDs (see Figures~\ref{fig:Gaia_CMD}, \ref{fig:GaiaVVV}, \ref{fig:AP_Gaia_CMDs}, and \ref{fig:AP_GaiaVVV_CMDs}), 
by including all the stars within a small circle around the cluster centroid in the sky and in the VPD we do not increase the contamination significantly. 
On the contrary, we need this step in order to include possible members in the cluster outskirts, 
that are ignored by the clustering code because we run it with a strong requirement of spatial concentration. 
The cluster centroids are listed in Table~\ref{tab:positional}.

Note that {\tt Gran~5} was selected due to the presence of a prominent RC, as shown in Fig.~\ref{fig:AP_Gaia_CMDs} and \ref{fig:AP_GaiaVVV_CMDs}.
However, being located very close to the Galactic plane, it is affected by high absolute and differential extinction,
broadening its CMD sequences.
Moreover, the upper RGB (above the RC) seems to be devoid of stars, both in Gaia and VVV data.
Only taking into account spectroscopic observations, we were able to confirm its cluster nature (see Sec.~\ref{sec:MUSE}).

\begin{figure}
    \centering
    \includegraphics[scale=0.45]{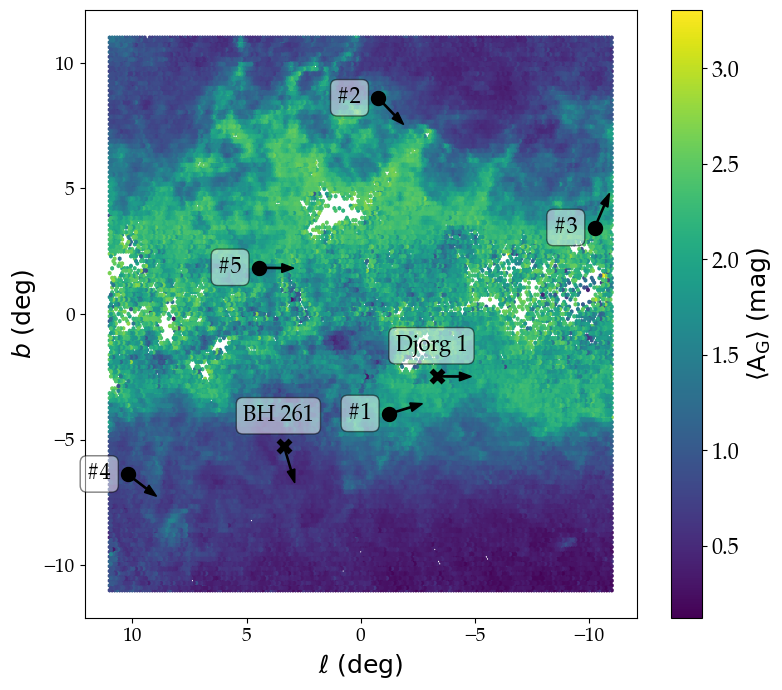}
    \caption{Spatial distribution of the new clusters with a {\tt STARHORSE} mean G-band extinction map as background. 
    The normalised 2D motion of the clusters is also shown as arrows.
    The numbers in the labels are the same as in Table~\ref{tab:positional}.
    The location of BH~261 and Djorg~1 within the bulge area are marked with crosses.}
    \label{fig:map}
\end{figure}

\begin{figure}
    \centering
    \includegraphics[scale=0.41]{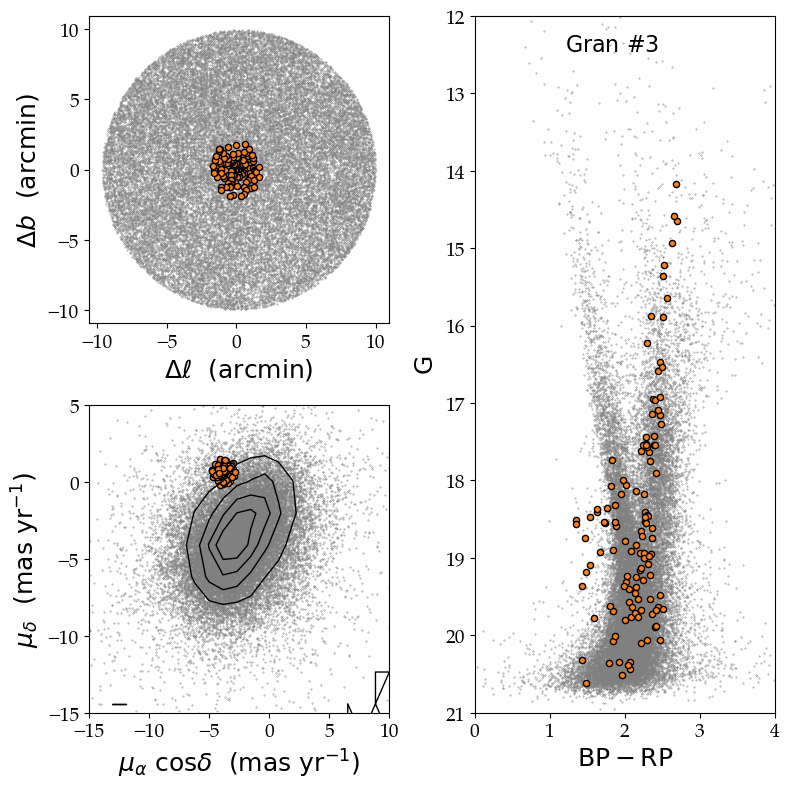}
    \caption{{\bf (Upper left)}: Spatial distribution of {\tt Gran~3} members. 
    Orange coloured circles represent the cluster stars while grey small points account for the foreground/background sources within 10 arcmins from the cluster centre.
    {\bf (Lower left)}: Gaia VPD for the same sources as in other panels. 
    Contours were calculated for the field population (cluster subtracted) to account for the high number of bulge stars that move with similar values and saturate a scatter plot. 
    Note that in this case, the {\tt Gran~3} members are located in the outskirt of the dominant bulge distribution of stars.
    {\bf (Right)}: Gaia CMD of {\tt Gran~3} with the same star sample as described in Sec.~\ref{sec:photometric}. 
    The cluster shows a clear and narrow RGB as well as an HB at ${\rm G}\sim 18.5$ mag, which is confirmed in Sec.~\ref{sec:MUSE}. 
    At magnitudes below ${\rm G=19}$ mag, where bulge and disk sequences merge with those of the cluster, there may be more contamination in the CMD of the cluster members.}
    \label{fig:Gaia_CMD}
\end{figure}

\begin{figure}
    \centering
    \includegraphics[width=0.5\textwidth, height=6cm]{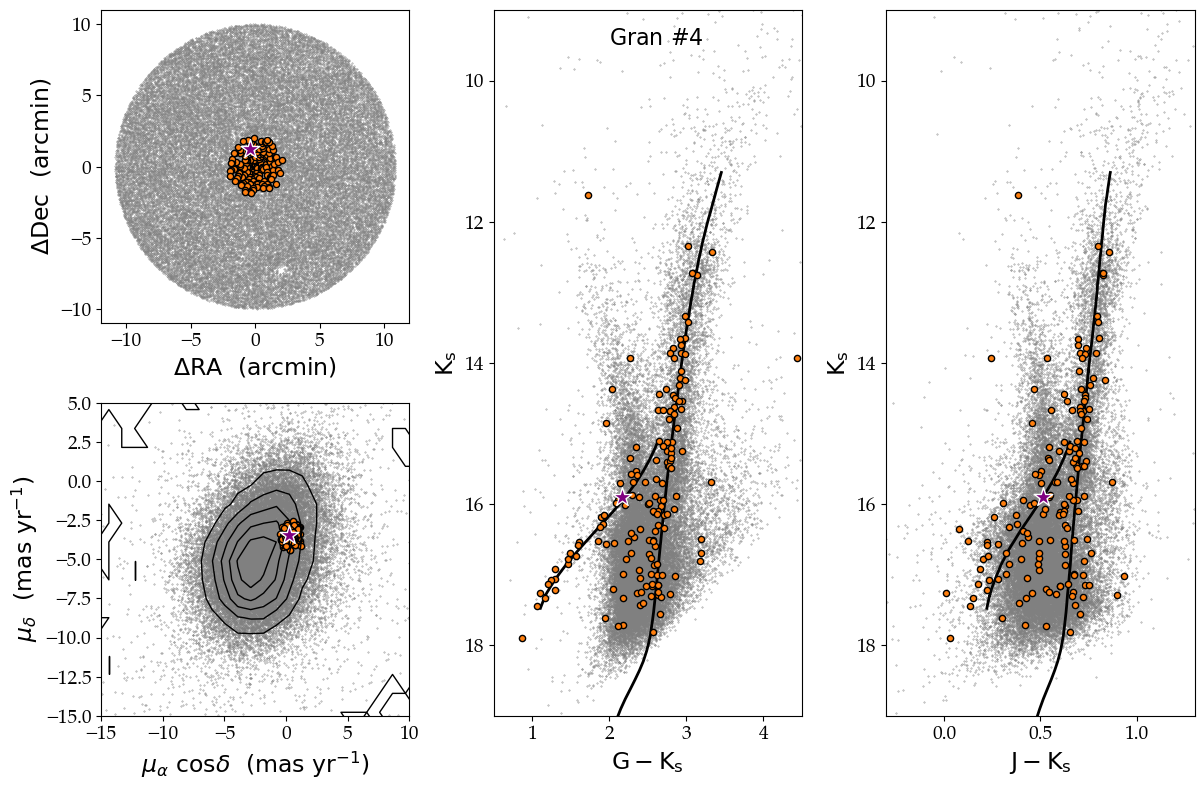}
    \includegraphics[width=0.5\textwidth, height=6cm]{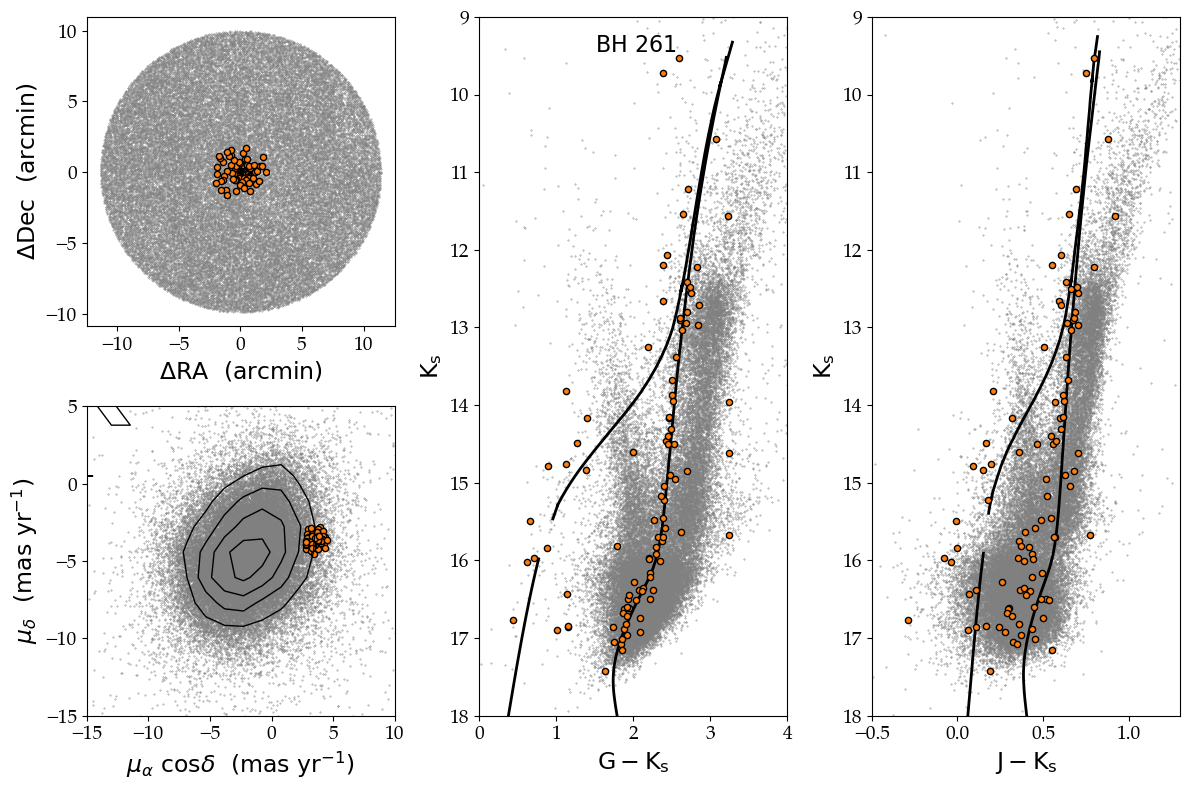}
    \caption{({\bf Upper four panels}): Spatial distribution, VPD, and near-IR-optical (\G$-$\ks\ and \jks) CMD for the {\tt Gran~4} GC with identical panel distribution as in Figure~\ref{fig:Gaia_CMD}.
    A PARSEC isochrone of 12 Gyr and \feh$\sim-2.4$ dex is presented as a reference. 
    The cluster member RR Lyrae star is marked with a star in all the panels.
    ({\bf Lower four panels}): Same as in the upper panels, but for BH~261.
    Note that we draw the AGB phase in the CMD, since the possible association of some stars to that stage, as two parallel sequences can be appreciated in both near-IR-optical CMDs at \ks$\gtrsim 13$ mag.
    The selected isochrone is a composition of a PGPUC (SGB, RGB and HB) and a PARSEC (AGB) models as it is described in Sec.~\ref{sec:photometric}.
    Both isochrones are for a metallicity of $-2.4$ dex and shifted by a distance modulus of $14.80$ mag.} 
    \label{fig:GaiaVVV}
\end{figure}

We also include here BH~261, Djorg~1 and the cluster candidate labelled as ${\rm C1}$ with the parameters derived by the clustering algorithm.
The position of the clusters in the sky is shown in Fig.~\ref{fig:map}, labelled by their IDs, as they appear in Table~\ref{tab:positional}.
The background of Fig.~\ref{fig:map} is a reddening map derived from the publicly available database {\tt STARHORSE}\footnote{Available at \url{https://gaia.aip.de/}} \citep{queiroz18, anders19}
within the inner $2.5$ kpc from the Galactic centre \citep[located at $8.2$ kpc,][]{MWreview16}. 
The query considers stars within our initial search box, with valid {\tt STARHORSE} flags.
A total of $2.5$ million stars were used to derive the mean extinction map in the Gaia G-band (${\rm A_G}$).
Note that all clusters except {\tt Gran~4}, lie in high extinction regions, with a mean ${\rm A_G}\gtrsim2$ mag.

We construct optical (${\rm G}$, ${\rm BP-RP}$) CMDs with the selected members as shown in the right panel of Fig.~\ref{fig:Gaia_CMD} for {\tt Gran~3}.
We include all the other cluster CMDs in Appendix~\ref{sec:AP_Gaia_CMDs}.
The spatial distribution is shown in the top left panel of the same figure, while the VPD, a key tool to identify new clusters, is shown in the bottom left panel. 
A radius of 10 arcmin from the cluster centre was taken to compare the cluster population with the surrounding field (grey points and contours). 
As this figure clearly shows, the new clusters are highly concentrated in space, move coherently in the plane of the sky and define narrow sequences in the CMD. 
In a few cases, the CMD also shows an HB or a RC, helping us to assign a reliable distance to the cluster.
Note that our procedure does not require the cluster mean PM to be different from the mean PM of bulge field stars.
Indeed, half of the new clusters have PMs similar to the bulge (see the {\tt Gran~2} and {\tt 4} PM distribution).

\begin{table*}
\caption{Positional parameters from the newly discovered GCs and discarded candidates. The numbered GC names are maintained across the whole article.
The coordinates are in the Gaia J2016 reference frame as they were taken directly from the available catalogue using median values. 
The reported number of stars considered in $N_{\rm members}$ accounts for the constraints adopted in Sec.~\ref{sec:data_method}.
Parameters for BH~261, Djorg~1, and {\tt C1} are also listed under their respective groups.}
\label{tab:positional}
\begin{tabular}{c|c|c|c|c|c|c|c|c|c}
GC & $\ell$ & $b$ & RA & Dec & $\mu_\alpha \cos{(\delta)}$ & $\mu_\delta$ & $\mu_\ell \cos{(b)}$ & $\mu_{b}$ & $N_{\rm members}$ \\
     &  (deg) & (deg) & (deg) & (deg) & (\masyr) & (\masyr) & (\masyr) & (\masyr) & (number) \\
\hline\hline
Gran  1 &  -1.233 & -3.977 & 269.651 & -32.020 & -8.10 & -8.01 & -10.94 &  3.03 & 57  \\
Gran  2 &  -0.771 &  8.587 & 257.890 & -24.849 &  0.19 & -2.57 &  -1.86 & -1.76 & 102 \\
\hline
Gran  3 & -10.244 &  3.424 & 256.256 & -35.496 & -3.78 &  0.66 &  -1.76 &  3.71 & 118 \\
Gran  4 &  10.198 & -6.388 & 278.113 & -23.114 &  0.46 & -3.49 &  -2.88 & -2.01 & 155 \\
\hline 
Gran  5 &   4.459 &  1.838 & 267.228 & -24.170 & -5.32 & -9.20 & -10.55 & -0.10 &  76 \\
\hline 
\multicolumn{10}{|c|}{Cluster candidates}   \\
\hline
C1      &  -3.589 &  4.174 & 260.151 & -29.673 & -2.90 & -6.11 &  -6.61 & -1.07 & 113 \\
\hline 
\multicolumn{10}{|c|}{Known clusters analysed}   \\
\hline
BH~261  &   3.359 & -5.272 & 273.527 & -28.638 &  3.55 & -3.60 &  -1.55 & -4.80 & 99  \\
Djorg~1 &  -3.324 & -2.485 & 266.871 & -33.066 & -4.67 & -8.41 &  -9.66 & -0.04 & 149 \\
\hline\hline
\end{tabular}
\end{table*}

For all the clusters that are inside the footprint of the VVV survey (namely, {\tt Gran~1, 4} and {\tt 5}, BH~261 and Djorg~1), 
we perform a match with the catalogue presented in \cite{surot19} to obtain deep near-IR colours for the Gaia detected stars.
Figure~\ref{fig:GaiaVVV} contains the Gaia-VVV matched stars for two clusters ({\tt Gran~4} and BH~261).
Note that, the selection is performed exclusively on the Gaia catalogue, given the importance of isolating cluster members from field stars using PMs.
All the Gaia-VVV CMDs are shown in Appendix~\ref{fig:AP_GaiaVVV_CMDs} with the same panel arrangement and symbols as in Figure~\ref{fig:Gaia_CMD}.
Both clusters in Figure~\ref{fig:GaiaVVV} present clearer sequences in the optical-near-IR CMD, showing HBs, and probably hints of an 
asymptotic giant branch (AGB) in BH~261, starting from (\G-\ks, \ks)$\sim(2.0, 13.5)$ mag to $\sim(2.5, 11.)$ mag.
PARSEC isochrones \citep[version 1.2S;][]{PARSEC1,PARSEC2,PARSEC3,PARSEC4,PARSEC5,PARSEC6} were added here just as a reference for the eye.

We searched for RR Lyrae variables near the clusters. 
To this goal, we used the Gaia and OGLE catalogues \citep{soszynski19,gaiadr2_vars}, finding one confirmed dynamical member of {\tt Gran~4} located within $\sim1.3$ arcmin from its centre.
The variable star was found in both catalogues, however we used the period derived by the OGLE team, which is measured from 105 epochs in the I-band, compared to the 15 G-band epochs of Gaia.
The RR Lyrae star ({\tt OGLE-BLG-RRLYR-62550} or {\tt Gaia DR2 4077796986282497664}), with a typical RRab-type saw shaped light curve and a period of $\sim 0.610087$ days, 
allows us to determine a robust distance to the cluster.
Using the PL relations described in \cite{catelan04, alonsogarcia15} for the near-IR bands of VVV, the approximate cluster metallicity derived from the isochrones ($\sim -2.4$ dex), 
and the Cardelli extinction law \citep{cardelli89}, we derived a distance of $20.69 \pm 0.14$ kpc for the RRL in {\tt Gran 4}. 
This value is consistent with the approximate distance of $\sim 22.5$ kpc, obtained by matching the isochrone to the cluster HB. 
Clearly, this is not a bulge GC as it is located on the far side of the Galaxy.

BH~261, on the other hand, shows a rather broad HB, a feature that might be enhanced due to a population of blue straggler (BS) stars. 
This hypothesis was already considered by \cite{ortolani06}, suggesting the presence of a non-negligible BS population within the cluster.
Given the large PM difference of this cluster with respect to bulge stars, field contamination would be very unlikely. 
Based on the (G-K$_{\rm s}$, K$_{\rm s}$) CMD presented in Fig.~\ref{fig:GaiaVVV}, and using
the SGB and the slope of the RGB we were able to estimate a distance of $\sim9.50$ kpc and a metallicity of $\sim -2.4$ dex, 
similar to that of {\tt Gran~4}.

Given the low-metallicity of BH~261, we used the $\alpha$-enhanced isochrones from the PGPUC \citep{PGPUC} database,
finding the best match with [Fe/H]=$-2.4$ dex and ${\rm [\alpha/Fe]}=0.2$ dex. 
We use the \cite{wang19} relations to convert the E(G$-$K$_{\rm s}$) reddening into the A$_{\rm K_s}$ extinction.
Previous estimates reported a distance for this cluster of d=6.12 kpc, versus the 9.12 kpc derived here. Imposing this 
shorter distance is not compatible with the low metallicity we find. A reasonable fit could still be achieved for this
distance and a higher metallicity [Fe/H]=$-1.3$ dex. We show all three fits in Fig.~\ref{fig:AP_BH261_iso}. We favor 
a larger distance and lower metallicity because the fit is better in the lower RGB, and SGB, and also because a shorter
distance would imply unreasonably high dynamical mass (by a factor of 5), mass-to-light ratio and velocity dispersion (both doubled), compared to the typical values in \citep{baumgardt18, baumgardt19}. 
Note that these are estimates and should be confirmed with both high-resolution spectroscopy and deeper photometry.

For {\tt Gran~2} and {\tt Gran~3} we have distance determinations by means of spectra taken with MUSE@VLT, which will be presented in Sec.~\ref{sec:MUSE}.

Finally, we include in this list {\tt Gran~5} and the candidate cluster {\tt C1}.
While the first was confirmed, the latter was eventually ruled out, both based on MUSE spectroscopic follow-up (Sec~\ref{sec:MUSE}).
Nonetheless, we consider important to show how several pieces of evidence favoured its existence as a cluster, in order to illustrate how difficult it is to positively confirm such objects. 
{\tt Gran~5} and {\tt C1} were selected based on the visual inspection of their CMD and its clumped PM using the Gaia DR2. 
Photometry and PMs from the updated Gaia EDR3 catalogue showed that their CMDs are diffuse and probably affected by differential reddening, especially in the case of {\tt Gran~5}. 
However, we determined a distance using the assumed RC located at (\jks, \ks) $\sim$ (1.00, 12.85), as shown in Fig.~\ref{fig:AP_GaiaVVV_CMDs}. 
Using the same approach as in \cite{minniti11b,gran16}, we estimated a distance of $\sim6.82$ kpc, i.e., in the near side of the Galactic bulge. 
The cluster nature is later confirmed by the spectroscopic measurements, while we also refine the distance determination.

For {\tt C1}, in addition to a dubious CMD with rather broad sequences, we also have MUSE spectroscopic observations. 
These allowed us to finally discard this candidate, as the putative cluster members selected from Gaia PMs do not clump in RV. 
In fact, in Fig.~\ref{fig:AP_C1} we could not identify any RV peak when considering all stars detected within the MUSE field of view.
We cannot stress this point enough to warn the reader that even prominent overdensities can be stochastic fluctuations of bulge stars. 
The PMs (or RVs in one of our cases) are a key piece of information to unveil the true nature of these candidates. 
Given the uncertain nature of these candidates, in the following sections we will focus our analysis on the confirmed clusters.

Individual distances and reddening values for optical and near-IR colours are reported in Table~\ref{tab:photometrical}.
Note that we use the \cite{cardelli89} relation to convert the reddening into a ${\rm K_{s}}$-band extinction using ${\rm A_{K_{s}} = 0.689\times E(J-K_{s})}$ to properly define an isochrone distance.
Additionally, we use ${\rm A_{G} = 2.0\times E(BP-RP)}$ derived in \cite{andrae18} to convert optical reddening to extinction.

Another parameter that we can derive from photometry is the integrated luminosity of each cluster. 
In order to compare with previous studies, we convert the Gaia \G\ magnitudes into the Johnson {$\rm V$} system using the publicly available colour transformation in the Calibration 
models of the Gaia documentation webpage.\footnote{\url{https://gea.esac.esa.int/archive/documentation/GEDR3/Data_processing/chap_cu5pho/cu5pho_sec_photSystem/cu5pho_ssec_photRelations.html}}
When deriving the integrated light of the cluster, we need to take into account also the stars that might not be included in the Gaia catalogues due to their incompleteness. 
In order to account for these missing stars, we used the completeness analysis by \cite{Gaiaverse1, Gaiaverse2, Gaiaverse3}, through the {\tt scanninglaw} python package.
We retrieved the average fraction of stars, at a given coordinate and G-magnitude, that were ignored by the Gaia selection function. 
We performed this correction at the position of each cluster, in bins of $\Delta {\rm G} = $ 0.1 mag, across the cluster luminosity function. 
Overall, the completeness of our sources for ${\rm G < 16}$ mag is more than $\sim90\%$, but it drops significantly at fainter magnitudes, reaching less than $\sim 30\%$ at ${\rm G = 21}$ mag.

After this step, we transformed the corrected integrated fluxes into an apparent V magnitude (\Vt, according to the \citealt{harris10} notation).
Distances and extinctions are required to derive the absolute V magnitude of each cluster, for which we used two methods depending on whether we were able to fit an isochrone to the cluster CMD or not.
In the case of {\tt Gran~4} and BH~261, isochrones were fitted to the clusters and are shown in Figure~\ref{fig:GaiaVVV}.
Isochrones were also used for {\tt Gran~1, 2, 3} and {\tt 5} (see Sec.~\ref{sec:MUSE}).

For completeness, Table~\ref{tab:photometrical} lists the integrated and absolute \V\ magnitude in addition to the extinction coefficients in the \G- and \V- bands (\AG\ and \AV) and the 
adopted metallicities for all the clusters.

\begin{table*}
\caption{Photometric properties of the discovered GCs. Distances and optical-near-IR reddening and extinctions are based on isochrone fitting to the CMDs. 
The integrated magnitudes and half-light radius ($r_h$) of the clusters are also presented here.}
\label{tab:photometrical}
\begin{tabular}{c|c|c|c|c|c|c|c|c|c|c|c|c|c|c|c}
GC     &   dm   &  Distance  & ${\rm E(J-K_{s})}$ & ${\rm A_{K_{s}}}$ &    \AG   & \AV   & \Vt   &   \MV   &  $r_{h}$  &   \feh     \\
       &  (mag) &   (kpc)    &       (mag)        &       (mag)       &   (mag)  & (mag) & (mag) &  (mag)  &  (arcmin) &   (dex)    \\
\hline\hline   
Gran 1 &  14.60 &    7.94    &        0.45        &         0.24      &    2.70  &  3.38 & 12.41 &  -5.46  &   0.86    &  -1.19     \\
Gran 2 &  16.10 &   16.60    &         ---        &         ---       &    1.90  &  2.37 & 12.56 &  -5.92  &   1.07    &  -2.12     \\
\hline
Gran 3 &  15.40 &   12.02    &         ---        &         ---       &    2.60  &  3.25 & 12.63 &  -6.02  &   1.05    &  -2.33     \\
Gran 4 &  16.84 &   22.49    &        0.20        &         0.14      &    1.20  &  1.50 & 11.81 &  -6.45  &   1.14    & $\sim$-2.4 \\
\hline 
Gran 5 &  13.25 &    4.47    &        0.63        &         0.43      &    3.24  &  4.05 & 12.11 &	-5.95  &   0.94    &  -1.56     \\
\hline 
\multicolumn{10}{|c|}{Known clusters analysed}   \\
\hline
BH~261 &  14.80 &    9.12    &        0.16        &         0.11      &    1.04  &  1.30 & 10.63 &  -5.56  &   1.09    & $\sim$-2.4 \\
Djorg~1 &  ---  &    ---     &        ---         &         ---       &    ---   &  ---  &  ---  &   ---   &   1.23    &     ---    \\
\hline\hline
\end{tabular}
\end{table*}

\subsection{Structural properties of the discovered clusters}
 \label{sec:structural}
 
\begin{figure}
    \centering
    \includegraphics[scale=0.425]{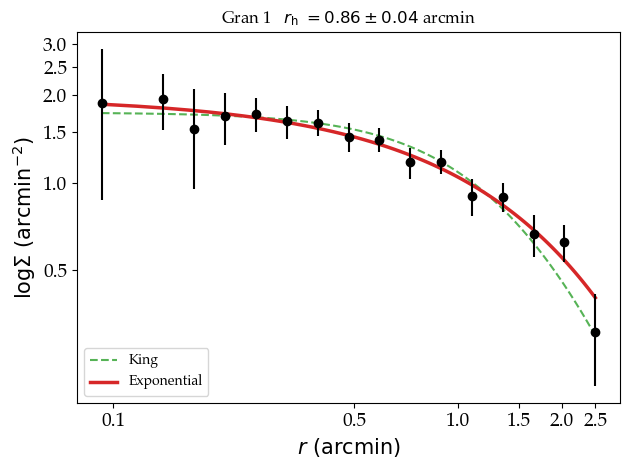}
    \caption{Derived fit for the King and exponential profiles of {\tt Gran 1} as dashed green and continuous red lines, respectively. 
    In this case, the preferred model is the exponential one, which is marked by a more pronounced line, consistent with a \rhl\ of $0.86$ arcmin.}
    \label{fig:gran1_profile}
\end{figure}

Having a clean sample of cluster members, in this section we discuss their radial profile and structural properties. 
We fitted the radial luminosity profile of the clusters with two sets of empirical models, a \cite{king62} model and an exponential one. 
We fitted the model parameters, namely the core (\rc) and the tidal radius (\rt) for the King profile and the scale factor in the exponential one using the SciPy {\tt curve\_fit} routine \citep{scipy}, 
by minimising a classical $\chi^2$ function. 
Figure~\ref{fig:gran1_profile} shows an example of the two fitted profiles for Gran~1, while all the other fits can be seen in Appendix~\ref{sec:AP_radial_profiles}. 
For all the clusters the exponential fit provides a better match to the data. 
In order to derive the luminosity errors $\sigma_\Sigma$, we used the recipe by \cite{gehrels86}, i.e.,  $\sigma_\Sigma = 1 + \sqrt{N + 0.75}$, where $N$ is the number of stars in each bin. 
From the fitted profile, we derived the half-light radius (\rhl) of each cluster as the middle point in the cumulative luminosity histogram of the fit. 
Note that we only report values for the \rhl\ as the exponential profile were preferred for all the clusters.
The values for each cluster can be found in Table~\ref{tab:photometrical}.

\begin{figure}
    \centering
    \includegraphics[scale=0.4]{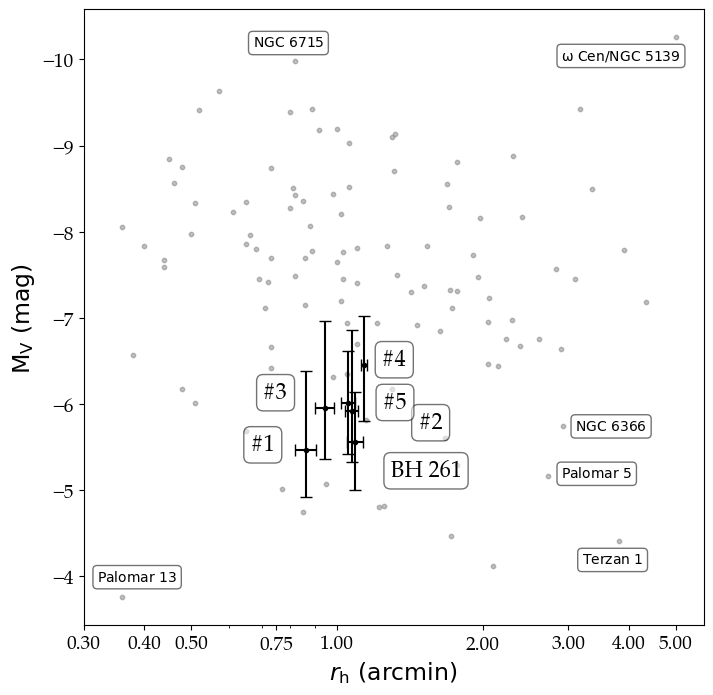}
    \caption{Half-light radius ($r_h$) and total absolute integrated magnitude (\MV) in the V-band of the known MW GCs listed in \protect\cite{baumgardt18}, shown as grey dots, 
    and the new ones presented here as black circles with error bars, labeled with the IDs as in Table~\ref{tab:positional}. 
    A handful of well-known clusters at the edges of this diagram are labelled as comparison in clockwise direction: $\omega$ Cen/NGC~5139, NGC~6365, Palomar~5, Terzan~1, Palomar~13 and 3, and NGC~6715.}
    \label{fig:mvrhl}
\end{figure}

At this point, we can compare directly the newly discovered clusters with the known population of GCs in the MW.
\cite{baumgardt19} re-derived the \rhl\ for almost all the known GCs in the \cite{harris10} catalogue using the Gaia DR2 data to select cluster members. 
Using \cite{harris10} absolute integrated magnitudes, we can compare the integrated luminosity and \rhl\ of our clusters with the other MW GCs directly, as shown in Figure~\ref{fig:mvrhl}. 
The error bars were calculated as follows: the standard deviation of the recovery factor of Gaia has been used to estimate an upper limit for the cluster total integrated magnitude. 
On the other hand, a lower limit has been derived, assuming that up to $\sim10\%$ of the cluster integrated light per magnitude bin might be contamination from field stars.

All our clusters are located at relatively faint absolute magnitudes, as expected for such low mass clusters that were not yet discovered \citep{baumgardt18}. 
For reference, Fig.~\ref{fig:mvrhl} includes the position of some well-known clusters with roughly similar \rhl\ and/or integrated flux as the new clusters. 
All the discovered clusters have similar \rhl, with {\tt Gran~1} and {\tt 4} at the two extremes, with \rhl$\sim$0.86 and $\sim$1.14 arcmin, respectively.

\subsection{Dynamical properties of the discovered clusters}
 \label{sec:dynamical}
 
\begin{figure}
    \centering
    \includegraphics[scale=0.4]{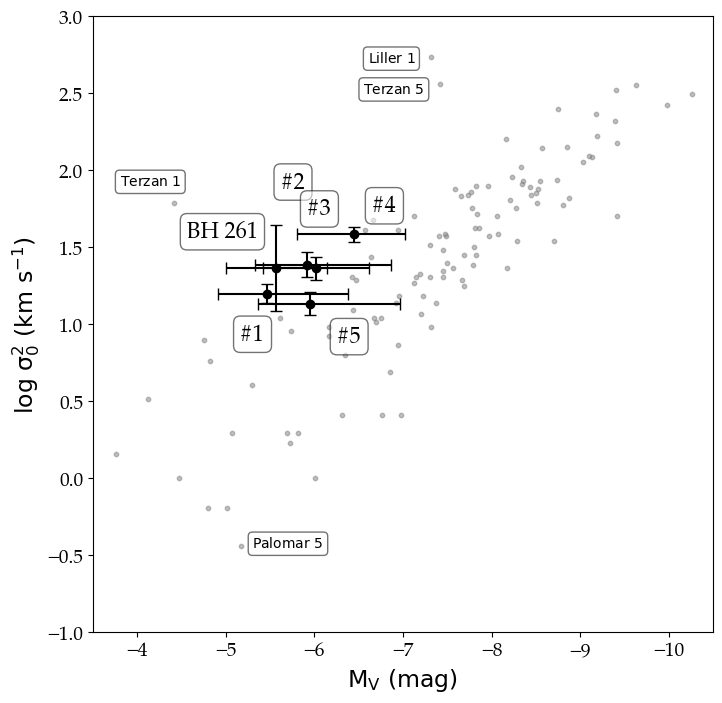}
    \caption{Absolute integrated magnitude (\MV) and velocity dispersion ($\log \sigma_0^2$) for our discovered clusters and the ones analysed by \protect\cite{baumgardt18}. 
    The same symbols and numerations as in Figure~\ref{fig:mvrhl} were used, 
    showing grey points and black circles with error bars for the known MW GCs and the ones described in this study, respectively.}
    \label{fig:logsigmamv}
\end{figure}

In this section we present the cluster velocity dispersions ($\sigma_0$) based on the PMs from Gaia EDR3. 
We included in the calculation only stars with uncertainties less than 1 \masyr\ in both components. 
The intrinsic velocity dispersion of each cluster was derived as $\sigma_0^2 = \sigma_{\rm vel}^2 - \sigma_{\rm errors}^2$,
where $\sigma_{\rm vel}$ is the standard deviation of the velocity distribution of the cluster members and $\sigma_{\rm errors}$ is the mean error of the measurements.
This parameter on its own is critical to establish whether the cluster is real. 
Known clusters have velocity dispersions that increase with their mass, with an upper limit of $\sim 20$ \kms\ \citep{baumgardt18}.
Therefore, a higher $\sigma_0$ would suggest that a cluster is not real.
In order to obtain the velocity dispersion in \kms\, we must first convert the PM into a tangential velocity (${\rm v_t}$) using the relation ${\rm v_t = 4.74 \mu d}$, 
in which $4.74$ is the equivalent in \kms\ of one astronomical unit in one tropical year, ${\rm \mu = \sqrt{(\mu_{\alpha} \cos \delta)^2 + (\mu_\delta)^2}}$ (in arcsec yr$^{-1}$) is the total PM, 
and ${\rm d}$ the distance to the cluster, expressed in parsecs. 
The velocity dispersions derived for the new clusters are listed in Table~\ref{tab:dynamical}.
Figure~\ref{fig:logsigmamv} compares the velocity dispersions of the new clusters with other MW GCs.
The position of the new clusters is compatible with the trend defined by well-studied GCs \citep{harris10,baumgardt18}. 
They stay in the upper part of the main \MV-$\log{\sigma_0^2}$ correlation.
As in Figure~\ref{fig:mvrhl}, we highlight some of the known clusters that stand out of the marked trend, 
such as, e.g., Liller~1 and Terzan~5 \citep{saracino15, ferraro21}.

Finally, we can put constraints to the clusters dynamical mass enclosed within a certain radius through the virial theorem with the estimator described by \cite{errani18}. 
The latter claims that the mass of a given pressure-supported system can be described by the minimum variance estimator 
${\rm M(<1.8r_h) = 3.5 (1.8r_h) G^{-1} \langle\sigma_0^2\rangle}$, with ${\rm M(<1.8r_h)}$ being the enclosed mass within $1.8$ times the half-light radius.
With this relation, we can derive approximate masses for the new clusters, converting ${\rm r_h}$ from arcmin to parsecs by means of the cluster distance.
Masses are listed in Table~\ref{tab:dynamical} and show good agreement with the bulk of MW GCs presented in \cite{baumgardt18}.
As expected from its CMD, {\tt Gran~4} is the most massive cluster in our sample, with a well-populated HB.
Its mass places it among the $\sim20\%$ most massive GCs in the MW, similar to NGC~6656/M~22.
As a final remark, using masses and luminosities computed so far, we have derived estimated mass-to-light (ML) ratios for the new clusters (Table~\ref{tab:dynamical}),
and overall, they behave similarly to other clusters within the same \MV-$\log{\sigma_0^2}$ regime \citep{baumgardt19}.
However, we recognise that ML ratios larger than $\sim4$ may not be valid measurements \citep{bianchini17, baumgardt20}. 
Several additional corrections must be performed in order to derive an accurate ML ratio, from deeper photometric observations, such as a proper completeness analysis, cluster area scaling, mass function decontamination, among others.
All these corrections should account for the missing factor of $\sim2-3$ in the integrated light magnitudes to achieve the expected ML ratios for all the GCs.

\begin{table*}
\caption{Dynamical properties of the clusters derived in Sec.~\ref{sec:dynamical}. 
Cluster name, intrinsic velocity dispersion, dynamical mass within $1.8{\rm r_h}$, and ML ratios are presented.}
\begin{tabular}{c|c|c|c}
 \label{tab:dynamical}
GC     &      $\sigma_0$   & ${M}^{\rm dyn}(<1.8{\rm r_h})$ &         $\Upsilon$        \\
       &      (\kms)       &          ($10^5 M_\odot$)      &  ($M_\odot L_\odot^{-1}$) \\
\hline\hline
Gran 1 &  3.96 $\pm$ 0.29  &          0.45 $\pm$ 0.08       &         3.61 $\pm$ 3.12   \\
Gran 2 &  4.93 $\pm$ 0.47  &          1.84 $\pm$ 0.40       &         9.50 $\pm$ 8.51   \\
\hline
Gran 3 &  4.79 $\pm$ 0.41  &          1.24 $\pm$ 0.25       &         5.84 $\pm$ 3.45   \\
Gran 4 &  6.18 $\pm$ 0.33  &          4.16 $\pm$ 0.61       &        13.15 $\pm$ 7.14   \\
\hline 
Gran 5 &  3.68 $\pm$ 0.32  &          0.37 $\pm$ 0.08       &        1.85 $\pm$ 1.77   \\ 
\hline
\multicolumn{4}{|c|}{Known clusters analysed}   \\
\hline
BH~261 &  3.79 $\pm$ 0.24  &          0.63 $\pm$ 0.11       &         4.53 $\pm$ 2.54   \\
\hline\hline
\end{tabular}
\end{table*}

\section{MUSE reconfirmation of Gran 1, 2, 3 and 5}
\label{sec:MUSE}

Given the importance of the newly discovered clusters, we applied for telescope time to spectroscopically follow-up the most promising candidates and to be able to derive their full orbital parameters. 
MUSE \citep{bacon10} observations were approved and executed during ESO P103 and P105 (June-August 2019 and April 2021; PI: F. Gran) 
for four of our clusters: {\tt Gran 1, 2, 3} and {\tt 5} and one candidate: {\tt C1}.
In Wide-Field Mode (WFM), the integral field unit (IFU) field-of-view (FoV) matches the projected size of our clusters ($\sim$ 1 arcmin), providing spectra for most of the stars in a single pointing.

The observations were carried out with the GALACSI adaptive optics system \citep{stuik06,arsenault08,strobele12,strobele20,hartke20}.
Standard observing blocks of one hour per cube were prepared, and the automatic ESO pipeline was used to pre-reduce the cubes with a typical image quality of $\sim 0.8-0.9$ arcsec per observation.
Finally, all the cubes were convolved with the VRI filter response functions to extract images in each colour and generate CMDs.

In order to extract the spectrum of each star, in these crowded fields, we employed the widely used {\tt PampelMuse} software \citep{kamman13}.
By default, {\tt PampelMuse} needs a stellar input catalogue to process the slices of each MUSE cube.
To make this catalogue, we performed PSF photometry with {\tt photutils} \citep{photutils} over the I-images and select detections above $3\sigma$ over the background.
Major changes to the predefined parameters of {\tt PampelMuse} were the use of {\tt PSFFIT = 30} value, which implies a complete fit of the Moffat PSF model for each star, 
including the FWHM, $\beta$ exponent, ellipticity, and position angle.
Other changes involve the double extraction of the spectra, the first time with a spectral binning of 10 MUSE spaxels width, or {\tt LAYERBIN = 10} within the {\tt PampelMuse} routine.
This binning corresponds to $\Delta \lambda \sim 12.5$ \AA, and it is done to speed up the extraction process of the whole raw spectrum. 
During the first run, the FWHM is derived for each binned image to properly account for its changes across the wavelength, and then this value is used to extract individual frame fluxes without binning.

\begin{figure}
    \centering
    \includegraphics[scale=0.35]{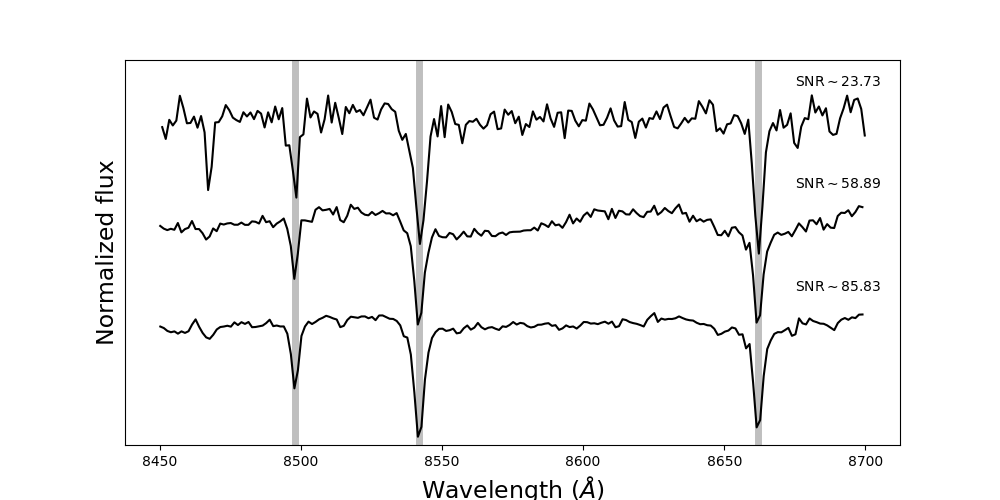}
    \caption{Extracted MUSE spectra of {\tt Gran~2} around the CaT lines for three stars at different SNR regimes. 
    Estimated SNR are given at the upper right of each spectra.
    Theoretical CaT lines are highlighted with grey shaded areas at at 8498, 8542 and 8662 \AA.}
    \label{fig:MUSE_spec}
\end{figure}

After extracting all the spectra, we normalise them using the {\tt fit\_generic\_spectra} routine in the Astropy affiliated package {\tt specutils} \citep{astropy1,astropy2}. 
RVs were derived using a cross-correlation routine implemented in \cite{pyastronomy} between the cube spectra and 6 synthetic templates, 
with the range of temperatures and gravities expected for giant and dwarf stars. 
The main cross-correlation indicator was the Ca triplet (CaT) lines at 8498, 8542 and 8662 \AA, lines for which we also calculate equivalent widths (EWs). 
Figure~\ref{fig:MUSE_spec} shows three {\tt Gran~2} extracted spectra at different signal-to-noise ratios (SNR).
Metallicity was estimated from the spectra applying a well-known relation using the EW of CaT lines \citep[][and references therein]{armandroff91, rutledge97, cenarro01, vasquez15}.
We employ the relation derived for the MUSE instrument by \cite{husser20}, which also extends the applicable region below the HB. 
This relation only requires prior knowledge of the CaT EW and the magnitude of the HB in the Jonhson V filter, 
which are $\rm V = 18.48, 18.59$, $18.65$ and $18.04$ mag for {\tt Gran~1, 2, 3} and {\tt 5}, respectively, based on the CMD.
In order to derive the reduced EW ($W^\prime$) of each star, we use 

\begin{equation*}
    W^\prime = \Sigma {\rm EW_{8542+8662} + 0.0442(V-V_{HB}) -0.058(V-V_{HB})^2},
\end{equation*}\
where $\Sigma {\rm EW_{8542+8662}}$ is the sum of the EWs of the two more prominent Calcium lines at 8542 and 8662 \AA, respectively.
Additionally, to derive the calibrated metallicity from $W^\prime$ we have to apply the following relation, also derived in \cite{husser20}
\begin{equation*}
    \rm [Fe/H] = -2.52 - 0.04W^\prime + 0.07W^{\prime 2}.
\end{equation*}

\begin{figure*}
    \centering
    \includegraphics[scale=0.35]{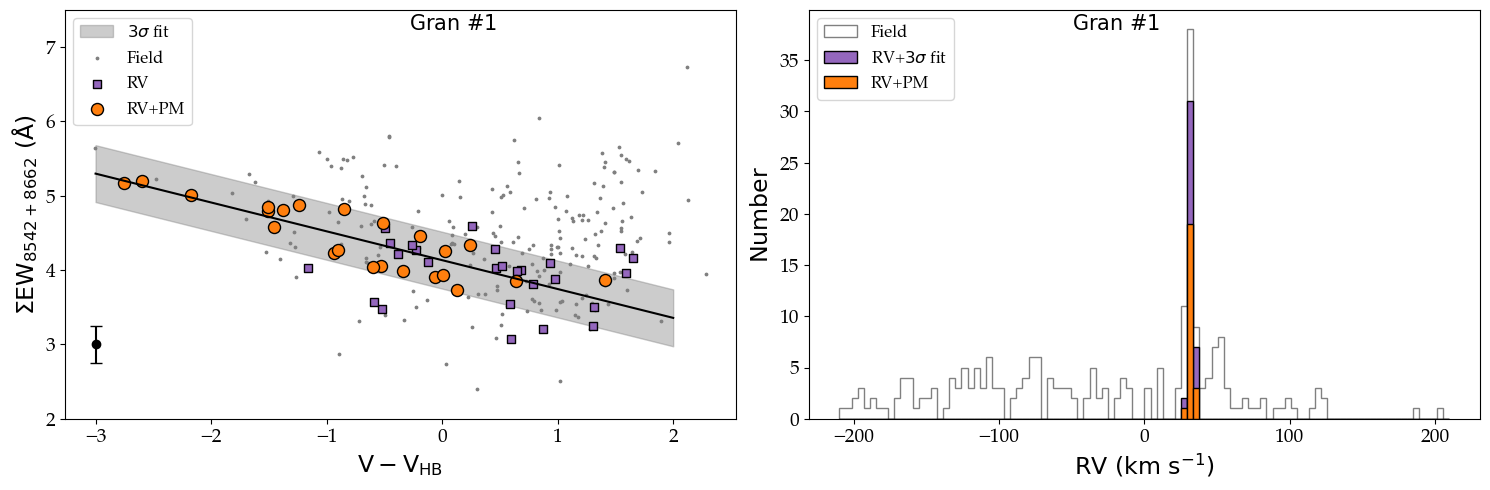}
    \includegraphics[scale=0.35]{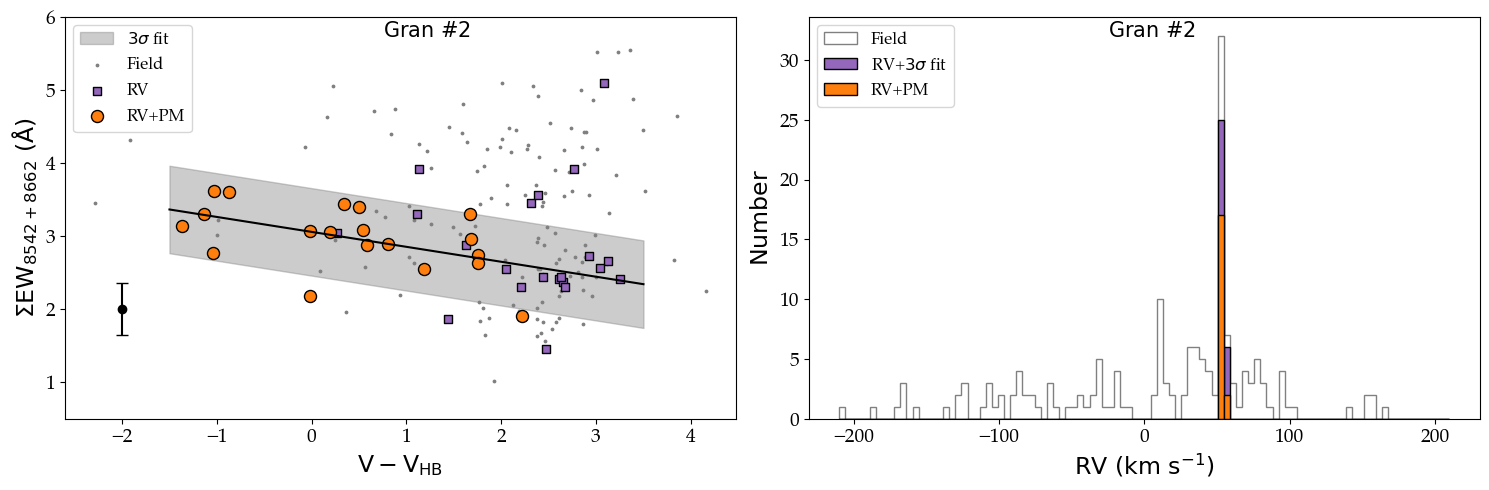}\\
    \includegraphics[scale=0.35]{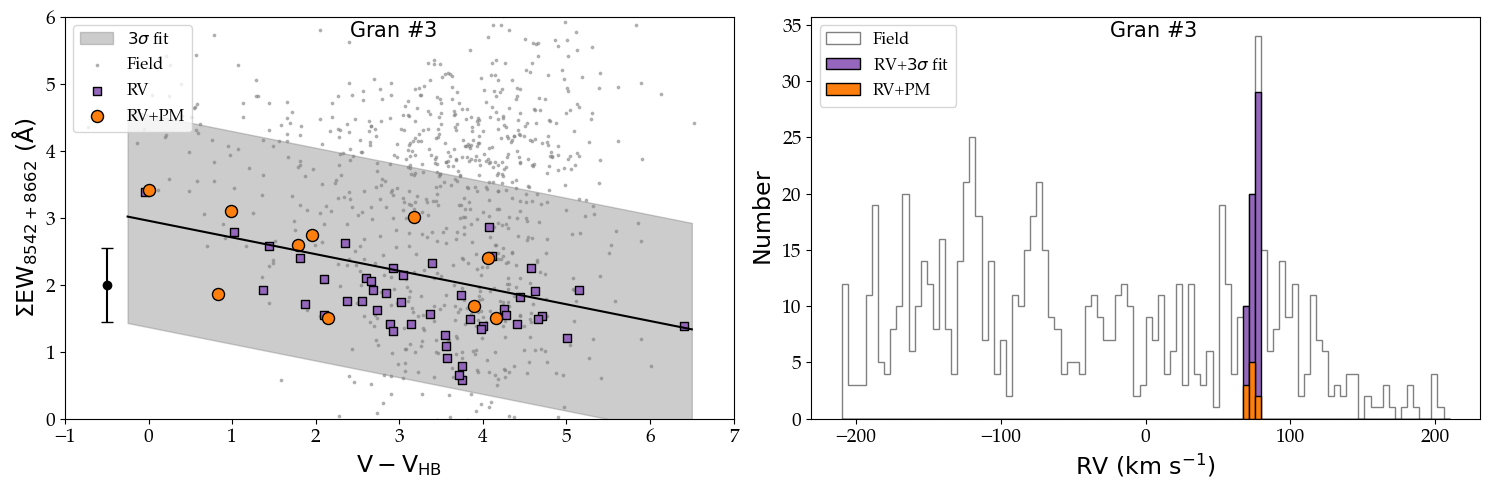}
    \includegraphics[scale=0.35]{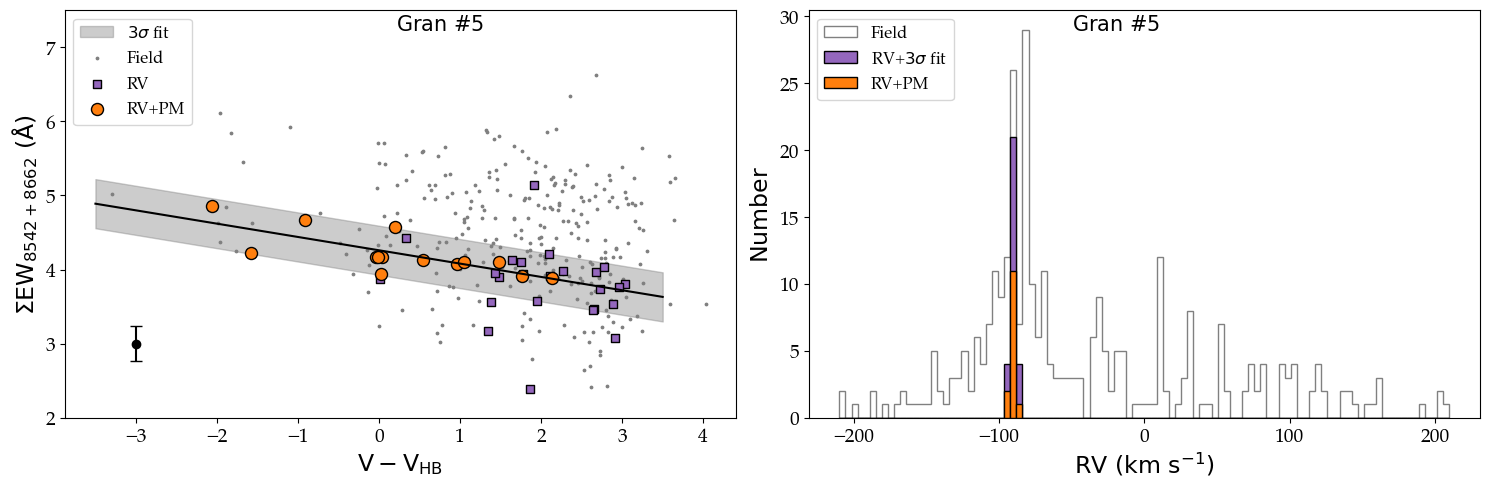}\\
    \caption{({\bf Upper left panel}): EW of the two most prominent lines in the CaT versus the V magnitude normalised to the HB level for the GC {\tt Gran~1}.
    Circular orange symbols represent the PM-selected stars matched with the MUSE extracted spectra, purple squares show the stars selected by RV within 5 \kms\ from the cluster value, 
    and grey points represent the field stars present in the cube. 
    The error bars of the black point represent the mean EW error for the stars in the MUSE field.
    The line fitted to the PM-selected stars is also shown with a shaded area and is equivalent to 3 times the intercept uncertainty.
    ({\bf Upper right panel}): RV histogram of all the sources in the MUSE cube colour-coded by a selection procedure: orange for the PM, purple of the stars with RV that lie within the shaded area,
    and grey for the field stars in the FoV.
    ({\bf Upper middle left and upper middle right panels}): Same as in the upper panels but for {\tt Gran~2}.
    ({\bf Lower middle left and lower middle right panels}): Same as in the upper panels but for {\tt Gran~3}.
    ({\bf Lower left and lower right panels}): Same as in the upper panels but for {\tt Gran~5}.}
    \label{fig:MUSEcube_spectraGran1-2-3-5}
\end{figure*}

At this point, we crossmatched the cluster members selected with Gaia with the stars for which we could extract a spectrum from MUSE. 
Figure~\ref{fig:MUSEcube_spectraGran1-2-3-5} shows the derived EWs and RVs for {\tt Gran~1, 2, 3} and {\tt 5}.
Cluster members selected having PMs and consistent with a single RV peak are shown as orange circles.
A few more stars having RV consistent with the cluster mean value (within $\pm 5$ \kms) are shown with purple squares. 
All the other field stars are shown as small grey points. 
In the left panels, for {\tt Gran~1, 2, 3} and {\tt Gran~5}, we notice that PM-selected stars follow the expected trend for a cluster in the 
$\Sigma {\rm EW_{8542+8662}}$-(V-V$_{\rm HB}$) plane \citep[c.f.,][]{husser20}.
To illustrate the errors on the EW determination, we include, as a black point with error bars, the mean uncertainty for all the stars detected in the MUSE cube.
Using this trend, we can confidently extend the selection of members for those stars having RVs and EWs consistent with the cluster.
The selection was performed extending three times the standard deviation of the intercept of the fitted line.  
The new members included allowed us to refine the mean cluster RV, adding significance to the derived value.
Notice that the RV peak is narrow, also confirmed with a low standard deviation as listed in Table~\ref{tbl:spectra}, as expected for a real GC.
Particular attention is needed on the {\tt Gran~3} case, given the slightly broader distribution of the RV peak, also noticeable in Table~\ref{tbl:spectra} 
with a higher velocity dispersion by $\sim 1$ \kms\ more than the other two observed clusters.
Moreover, in all the cluster RV histograms, a 4.2 \kms\ binning was used.
The same method applied to the candidate cluster {\tt C1}, did not yield a clear RV peak. 
Therefore, considering that the CMD of this candidate is also very sparse, we conclude that this overdensity is not a real cluster.

Mean RVs and metallicities of each cluster, calculated with the information from Figure~\ref{fig:MUSEcube_spectraGran1-2-3-5}, are reported in Table~\ref{tbl:spectra}. 
We found that both {\tt Gran~2} and {\tt 3} clusters are metal-poor, consistent with the extended HBs seen in their CMDs.
Using the relation by \cite{husser20}, we derive a metallicity for {\tt Gran~1, 2, 3} and {\tt 5} of 
${\rm [Fe/H] = -1.19 \pm 0.19}$, ${\rm -2.07 \pm 0.17}$, ${\rm -2.37 \pm 0.18}$ and ${\rm -1.56 \pm 0.17}$ dex, respectively.
Note that uncertainties in \feh\ are in agreement with broad error estimates for a mono-metallic population of stars, as found by \cite{husser20} in other MW GCs.

\begin{figure*}
    \centering
    \includegraphics[scale=0.32]{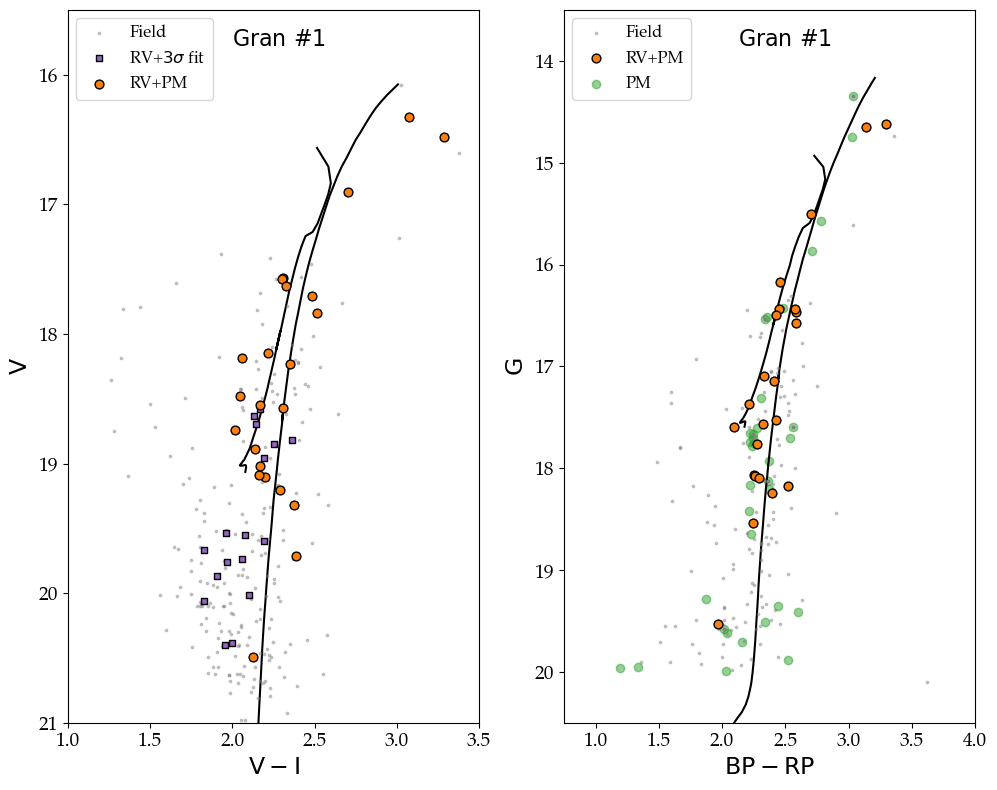}
    \includegraphics[scale=0.32]{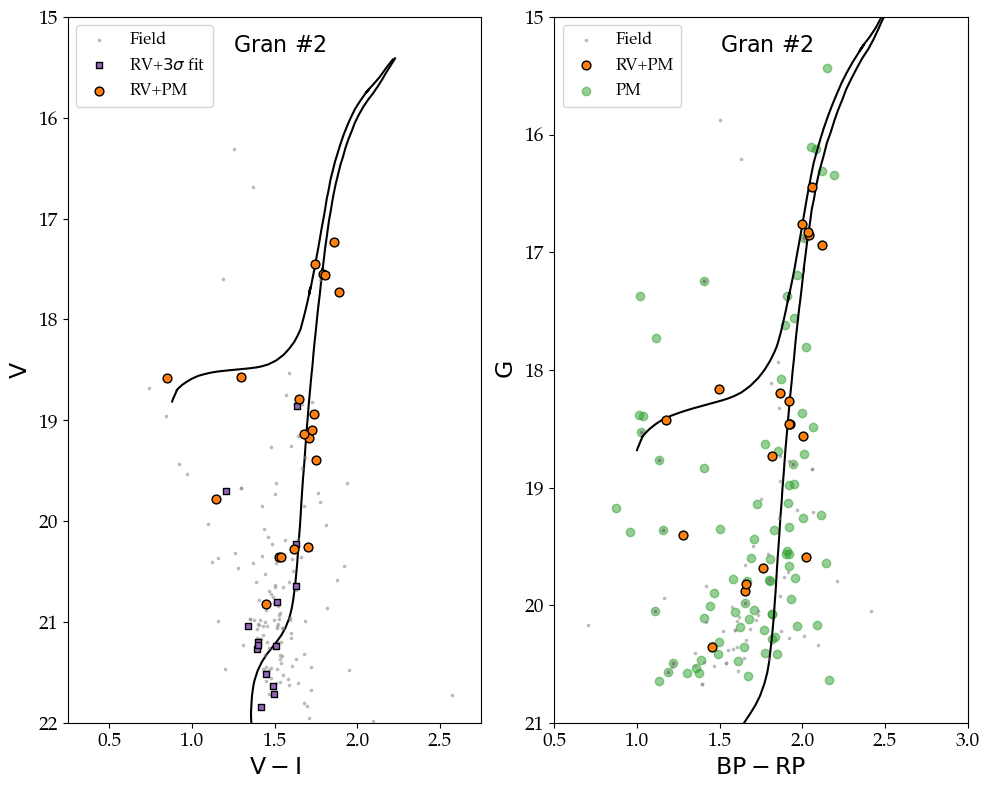}\\
    \includegraphics[scale=0.32]{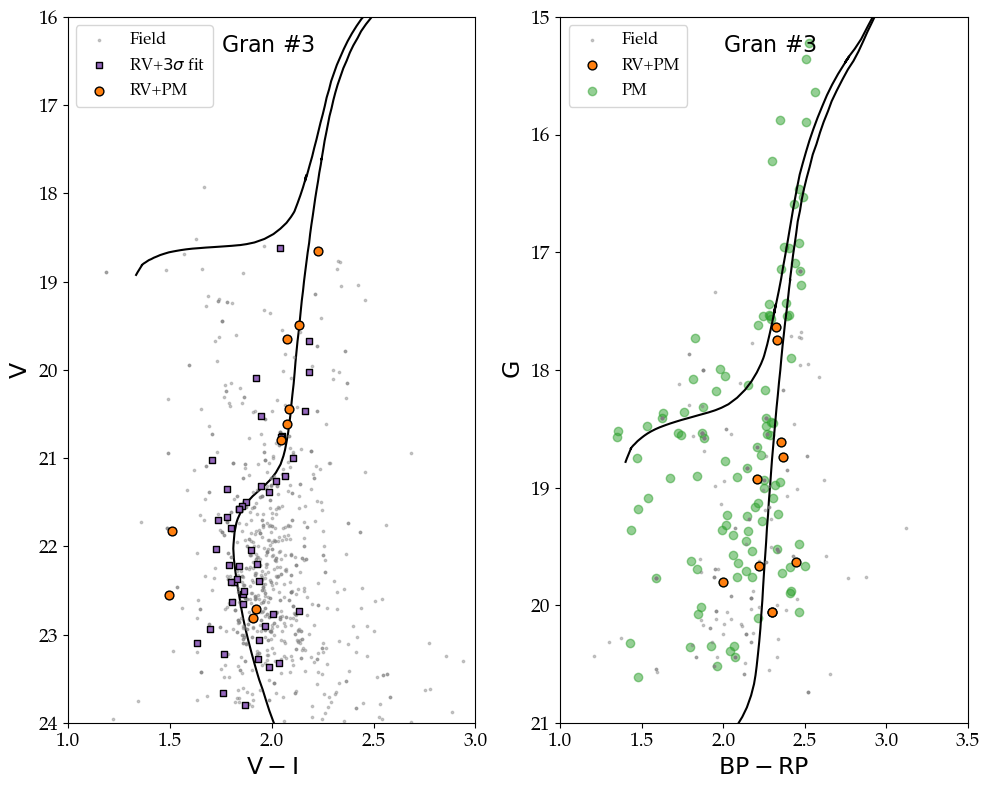}
    \includegraphics[scale=0.32]{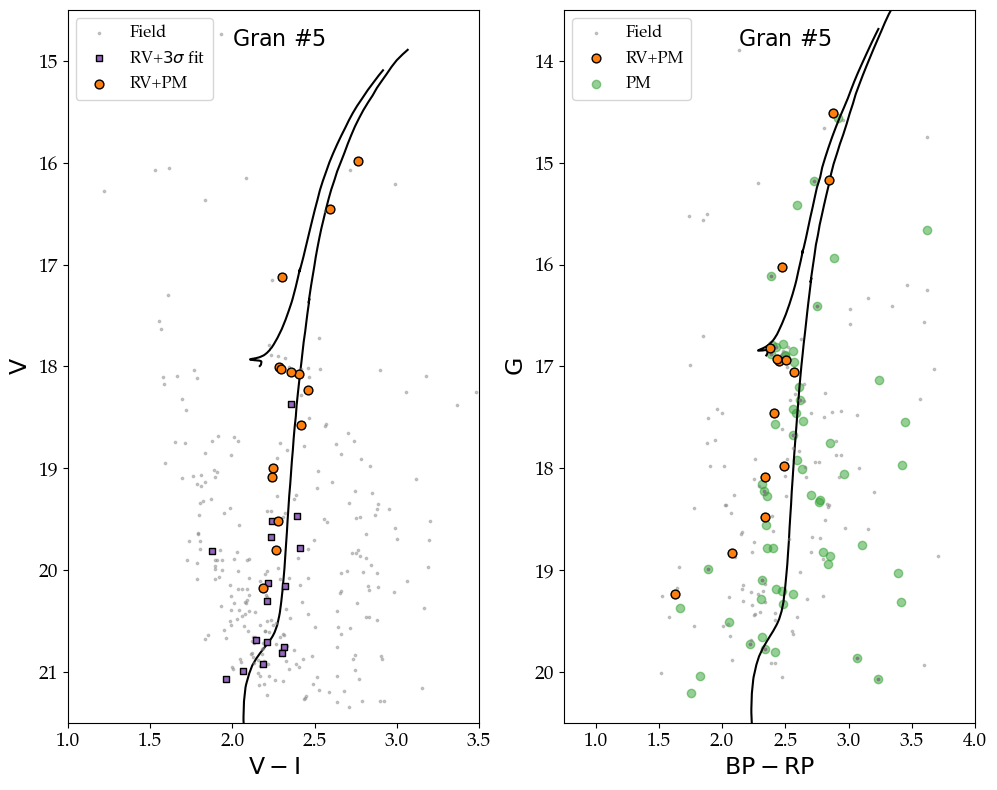}\\
    \caption{({\bf Upper left and upper middle left}): CMDs of {\tt Gran~1} with the MUSE and Gaia colours. 
    Orange circles, purple squares and grey points stand for the PM-selected stars with available RVs, the MUSE stars with RVs similar to the cluster that also are located within the shaded area of 
    Fig~\ref{fig:MUSEcube_spectraGran1-2-3-5}, and the field stars that do not correspond to the other two groups.
    The Gaia CMD includes the same stars with MUSE counterparts as orange circles, the Gaia-only PM selected stars as shaded green circles, and the field stars with MUSE detections. 
    The selected PARSEC isochrones are identical for all the diagrams, with an age of 12 Gyr, and the determined cluster metallicity, shifted to the RC/HB.
    ({\bf Upper middle right and upper right}): Same as in the left panels but for {\tt Gran~2}.
    As can be expected for the V-band, the location of the stars at V$\sim 18.5$ mags reveals the HB for this cluster. There are also hints of two BS stars off the isochrone.
    ({\bf Lower left and lower middle left}): Same as in the left panels but for {\tt Gran~3}.
    The cluster SGB is clearly visible by its characteristic shape at (V-I, V) $\sim$ (2.0, 21.5) mag.
    ({\bf Lower middle right and lower right}): Same as in the left panels but for {\tt Gran~5}.
    The cluster RC is visible at V$\sim$18 mag or G$\sim$17 mag.}
    \label{fig:MUSE_CMD_CaT}
\end{figure*}

Figure~\ref{fig:MUSE_CMD_CaT} shows an optical CMD for bonafide cluster members, where V and I images were obtained from the MUSE pipeline, by convolving the datacubes with the filter passbands.
The optical VI magnitudes shown here are calibrated, and give the same distance modulus derived above from the Gaia (G, BP-RP) CMD.
As the figure shows, a RC can be identified for {\tt Gran~1} and {\tt 5}, a feature that is seen in other GCs at this metallicity; 
the HB for {\tt Gran~2} is entirely horizontal, as expected for the V filter; 
and the SGB of {\tt Gran~3} is correctly defined and follows with great precision the selected isochrone.
We include in the Gaia CMD the PM selected stars not present within the MUSE FoV to compare the real members with our initial selection.

\begin{figure*}
    \centering
    \includegraphics[scale=0.35]{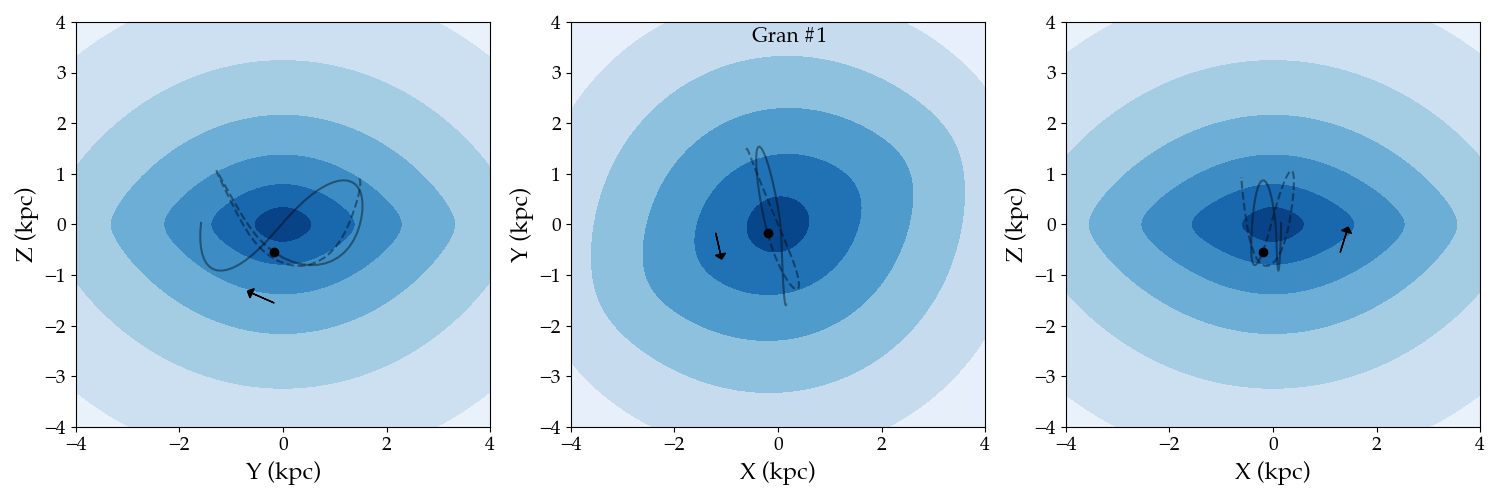}
    \includegraphics[scale=0.35]{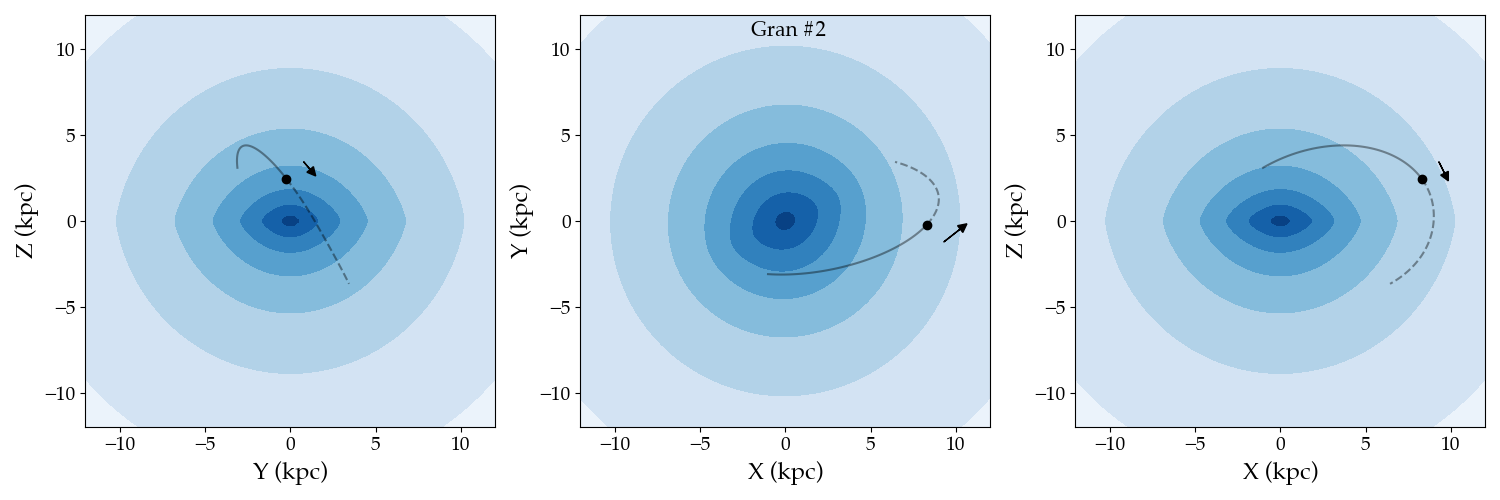}
    \includegraphics[scale=0.35]{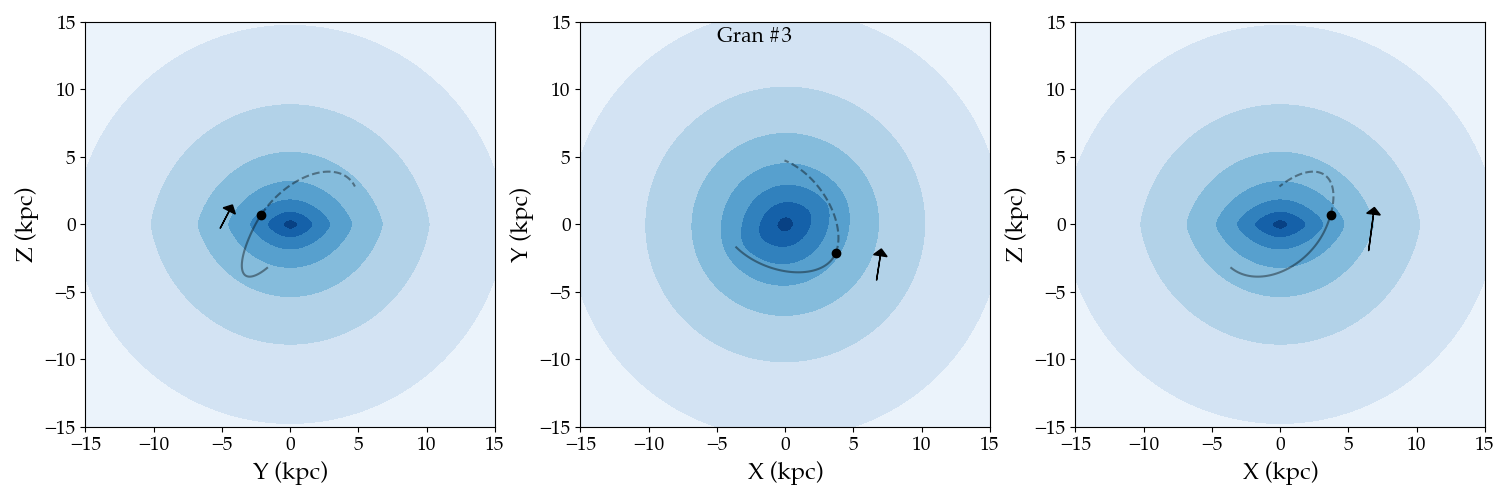}
    \includegraphics[scale=0.35]{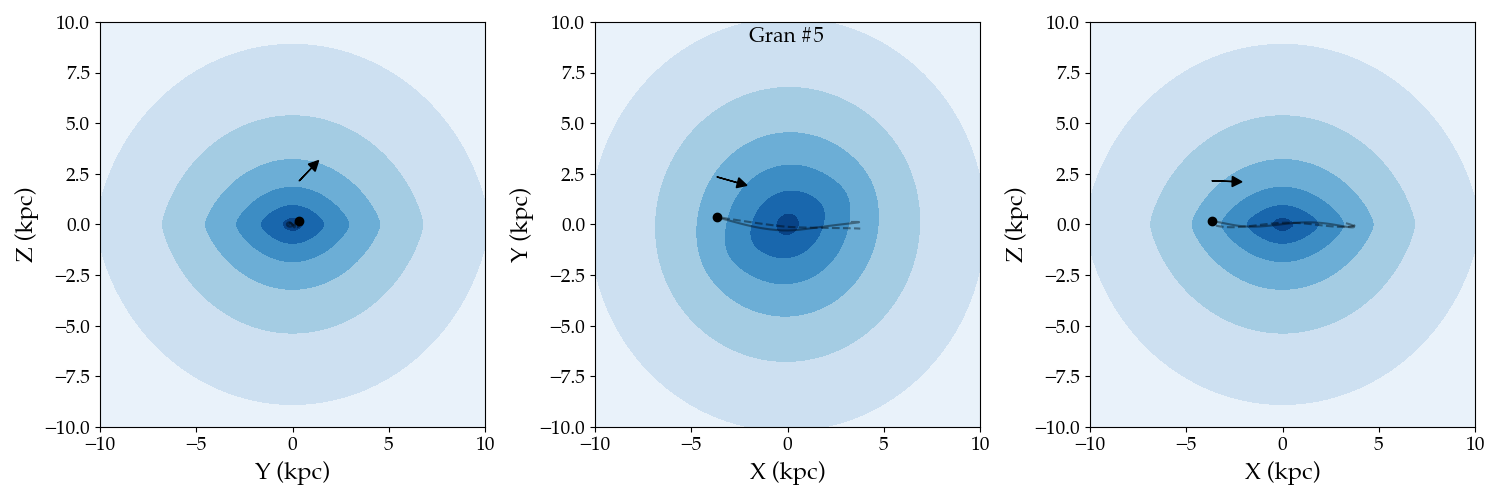}
    \caption{({\bf Upper panels}): Galactocentric cartesian projection (YZ, XY and XZ planes from left to right) of the orbit derived for {\tt Gran~1}. 
    An arrow represents the actual direction of the cluster in time. Backwards and forward orbit integrations through $\sim$100 Myr are presented in solid and dashed lines, respectively.
    The background model contours comprise the equipotential levels of the synthetic built galaxy. The bar is clearly marked in the central parts of the projections.
    ({\bf Upper middle panels}): The same as in the upper panels but for the orbit that describes {\tt Gran~2} in the Galactocentric cartesian projection. 
    ({\bf Lower middle panels}): The same as in the upper panels but for the orbit that describes {\tt Gran~3} in the Galactocentric cartesian projection. 
    ({\bf Lower panels}): The same as in the upper panels but for the orbit that describes {\tt Gran~5} in the Galactocentric cartesian projection. 
    Note that the cluster orbit is completely constrained up to a maximum Z excursion of $\sim$0.13 kpc.
    As it is evident in this case from both YZ and XZ projections, the orbit is entirely confined within the Galactic plane.}
    \label{fig:orbits}
\end{figure*}

Because we have both their tangential and radial velocities, we are able to determine for the first time estimated orbital parameters for {\tt Gran~1, 2, 3} and {\tt 5}.
We used the galactic dynamics python package {\tt gala} \citep{gala1, gala2} allowing us to simulate the MW potential in a realistic manner, 
and therefore integrate the orbit of these four clusters within it. 
The MW potential includes a halo, a disk and a bulge using the recommended {\tt MWPotential2014} parameters described in \cite{galpy}.
Additionally, we included a rotating bar, with the parameters given in \cite{shen20}.
We then integrated test particles with the {\tt Gran~1, 2, 3} and {\tt 5} space velocities to trace their orbits through the Galaxy.
Integration times were short ($\sim$ 100 Myr) to visualise the clusters immediate response to the Galactic potential, as shown in Figure~\ref{fig:orbits}.
A more extended integration time was also performed to calculate orbital properties such as the maximum $z$ excursion length, 
the amplitude of the orbits, and the total energy and angular momentum (see Table~\ref{tbl:spectra}).
Note that integrated orbital properties were derived using mean cluster coordinates, PMs, and RVs. 
More precise values can be achieved considering its uncertainties and different Galaxy models, however, the presented values serve as a first estimate of the cluster properties.

Recently, major discoveries have been made describing some of the processes involved in the formation of the MW. 
Starting with the formal discovery and analysis of the Gaia-Enceladus/Sausage merger event \citep{belokurov18, helmi18b, myeong19}, 
more evidence has been collected to describe the early merging history of the MW using its GCs \citep{myeong18, kruijssen19b, woody21}.
Following the \cite{massari19} definition of the different groups in the orbital energy space ($E_{\rm tot}$-$L_z$), we 
assigned the most probable classification to our MUSE observed clusters.
Take in consideration that orbital motions could be slightly different using another Galaxy model.
According to the $z$-axis angular momentum and total energy of the derived orbit, we estimated that {\tt Gran~1} belongs to the Main Progenitor or the so-called ``low-energy'' group,
while {\tt Gran~2, 3} and {\tt 5} appears to be related to the Gaia-Enceladus/Sausage structure. {\tt Gran~5}, could also be related to the ``low-energy'' group, since its orbital parameters
are located in the limits of both classifications.

\begin{table*}
\caption{Chemodynamical parameters derived from the MUSE data. The level of the HB in each cluster is represented by $V_{\rm HB}$, while 
the orbit eccentricity, maximum ${\rm Z}$ excursion, pericenter and apocenter radius, angular momentum in the $z$ direction 
and total orbital energy, are presented by $e$, $z_{\rm max}$, $r_{\rm peri}$, $r_{\rm apo}$, $L_z$, $E_{\rm tot}$, respectively.}
\begin{tabular}{c|c|c|c|c|c|c|c|c|c}
\label{tbl:spectra}
GC     &   ${\rm RV}$     &       \feh        & $V_{\rm HB}$  &  $e$   &  $z_{\rm max}$  & $r_{\rm peri}$ & $r_{\rm apo}$ &          $L_z$         &     $E_{\rm tot}$    \\
       &     (\kms)       &       (dex)       &    (mag)      &        &      (kpc)      &    (kpc)       &     (kpc)     & (kpc$^{2}$ Myr$^{-1}$) & (kpc$^2$ Myr$^{-2}$) \\
\hline\hline                     
Gran 1 & 32.30 $\pm$ 1.87 &  -1.19 $\pm$ 0.19 &     19.08     &  0.76  &       0.38      &     0.31       &     2.22     &          0.03          &         -0.21        \\
Gran 2 & 53.22 $\pm$ 1.67 &  -2.07 $\pm$ 0.17 &     18.59     &  0.34  &       5.44      &     4.59       &     9.24     &          0.79          &         -0.16        \\
\hline
Gran 3 & 74.32 $\pm$ 2.70 &  -2.37 $\pm$ 0.18 &     18.65     &  0.08  &       3.88      &     4.66       &     5.47     &          0.69          &         -0.17        \\
Gran 5 & -90.40 $\pm$ 1.93 & -1.56 $\pm$ 0.17 &     18.04     &  0.90  &       0.13      &     0.20       &     3.75     &         -0.04          &         -0.19        \\
\hline\hline
\end{tabular}
\end{table*}

\section{Summary}
\label{sec:summary}

We have detected and analysed seven new cluster candidates located towards the Galactic bulge, resulting in the discovery of five new GCs, dynamically confirmed through PMs or RVs.

All five of them exhibit narrow sequences in the CMD, coherent motion in PM-space and compact on-sky projections.

Based on the computed distances and orbits, when available, we concluded that three of them are halo GCs 
({\tt Gran~2, 3} and {\tt 4}), while {\tt Gran~3} and {\tt 5} present bulge-like properties.
MUSE cubes have been analysed to extract RVs, metallicities and orbital properties of {\tt Gran~1, 2, 3} and {\tt 5}, reconfirming their cluster nature, initially derived with Gaia PMs.
With ${\rm [Fe/H] \lesssim -2}$ dex and distances greater than $\sim 12$ kpc, {\tt Gran~2} and {\tt 3} are clearly part of the fairly unexplored regions of the Galaxy that lie behind the bulge.
Our preliminary analysis indicates that {\tt Gran~1} is located in between the Main Progenitor and the ``low-energy'' group, 
{\tt Gran~2} and {\tt 3} present dynamical signatures similar to the ones exhibited by the Gaia-Enceladus/Sausage structure, 
while {\tt Gran~5} could be part of the Main Progenitor or the ``low-energy'' group.

This work proves that new GCs can be discovered in the far side of the MW, even with optical surveys and behind high extinction regions produced by the disk and bulge.

The new clusters, especially {\tt Gran~1}, lie in the lower part of the MW GC luminosity function, an under-represented regime, as discussed in \cite{baumgardt18}.
Their velocity dispersions are comparable with typical known GCs, representing a low-luminosity population of objects that had escaped detection until now.

\section*{Acknowledgements}

We thank the referee for the valuable feedback to improve the quality of the article and H. Baumgardt, who realised that there was an error within orbit determination of the clusters.

This work is part of the Ph.D. thesis of F. G., funded by grant CONICYT-PCHA Doctorado Nacional 2017-21171485. 
F. G. also acknowledges CONICYT-Pasant\'ia Doctoral en el Extranjero 2019-75190166 and ESO SSDF 19/20 (ST) GAR funding. 
M. Z. acknowledge support from FONDECYT Regular grant No. 1191505.
E. V. acknowledges the Excellence Cluster ORIGINS Funded by the Deutsche Forschungsgemeinschaft (DFG, German Research Foundation) under Germany's Excellence Strategy – EXC-2094 – 390783311.
A. R. A. acknowledges support from FONDECYT through grant 3180203.
Support for a few authors is provided by the BASAL Center for Astrophysics and Associated Technologies (CATA) through grant PFB-06, and the Ministry for the Economy, Development, and Tourism, Programa Iniciativa Cient\'ifica Milenio through grant IC120009, awarded to the Millennium Institute of Astrophysics (M.A.S.). 

This work has made use of data from the European Space
Agency (ESA) mission Gaia (https://www.cosmos.esa.int/gaia), processed by the Gaia Data Processing and Analysis Consortium (DPAC, https:
//www.cosmos.esa.int/web/gaia/dpac/consortium). Funding for the
DPAC has been provided by national institutions, in particular the institutions
participating in the Gaia Multilateral Agreement.

We gratefully acknowledge the use of data from the VVV ESO Public Survey
program ID 179.B-2002 taken with the VISTA telescope, and data products
from the Cambridge Astronomical Survey Unit (CASU). The VVV Survey data
are made public at the ESO Archive.
Based on observations taken within the ESO VISTA Public Survey VVV, Program ID 179.B-2002.

This research has made use of the VizieR catalogue access tool, CDS, 
Strasbourg, France \citep{simbad}. The original description 
of the VizieR service was published in \citep{vizier}.
 
This research made use of: 
TOPCAT \citep{topcat}, IPython \citep{ipython}, numpy \citep{numpy}, matplotlib \citep{matplotlib}, Astropy, a community developed core Python package for Astronomy \citep{astropy1,astropy2},
galpy: A Python Library for Galactic Dynamics \citep{galpy}, scanninglaw \citep{Gaiaverse1, Gaiaverse2, Gaiaverse3}, dustmaps \citep{green2018}, and gala \citep{gala1, gala2}. 
This research has made use of NASA’s Astrophysics Data System.\\

\section*{Data Availability}

This project used data obtained with the Multi Unit Spectroscopic Explorer (MUSE; proposal 0103.D-0368 and 105.20MY.001, P.I.: F. Gran), from the European Space Agency (ESA) mission Gaia 
(\url{https://www. cosmos.esa.int/gaia}), processed by the Gaia Data Processing and Analysis Consortium (DPAC, \url{https://www. cosmos.esa.int/web/gaia/dpac/consortium})
and from the MW-BULGE-PSFPHOT compilation \citep{surot19, surot20}. All the data used in this study is publicly available.



\bibliographystyle{mnras}
\bibliography{NewGCs} 

\begin{thebibliography}{}
\makeatletter
\relax
\def\mn@urlcharsother{\let\do\@makeother \do\$\do\&\do\#\do\^\do\_\do\%\do\~}
\def\mn@doi{\begingroup\mn@urlcharsother \@ifnextchar [ {\mn@doi@}
  {\mn@doi@[]}}
\def\mn@doi@[#1]#2{\def\@tempa{#1}\ifx\@tempa\@empty \href
  {http://dx.doi.org/#2} {doi:#2}\else \href {http://dx.doi.org/#2} {#1}\fi
  \endgroup}
\def\mn@eprint#1#2{\mn@eprint@#1:#2::\@nil}
\def\mn@eprint@arXiv#1{\href {http://arxiv.org/abs/#1} {{\tt arXiv:#1}}}
\def\mn@eprint@dblp#1{\href {http://dblp.uni-trier.de/rec/bibtex/#1.xml}
  {dblp:#1}}
\def\mn@eprint@#1:#2:#3:#4\@nil{\def\@tempa {#1}\def\@tempb {#2}\def\@tempc
  {#3}\ifx \@tempc \@empty \let \@tempc \@tempb \let \@tempb \@tempa \fi \ifx
  \@tempb \@empty \def\@tempb {arXiv}\fi \@ifundefined
  {mn@eprint@\@tempb}{\@tempb:\@tempc}{\expandafter \expandafter \csname
  mn@eprint@\@tempb\endcsname \expandafter{\@tempc}}}

\bibitem[\protect\citeauthoryear{{Alonso-Garc{\'\i}a}, {D{\'e}k{\'a}ny},
  {Catelan}, {Contreras Ramos}, {Gran}, {Amigo}, {Leyton}  \&
  {Minniti}}{{Alonso-Garc{\'\i}a} et~al.}{2015}]{alonsogarcia15}
{Alonso-Garc{\'\i}a} J.,  {D{\'e}k{\'a}ny} I.,  {Catelan} M.,  {Contreras
  Ramos} R.,  {Gran} F.,  {Amigo} P.,  {Leyton} P.,   {Minniti} D.,  2015,
  \mn@doi [\aj] {10.1088/0004-6256/149/3/99}, \href
  {https://ui.adsabs.harvard.edu/abs/2015AJ....149...99A} {149, 99}

\bibitem[\protect\citeauthoryear{{Anders} et~al.,}{{Anders}
  et~al.}{2019}]{anders19}
{Anders} F.,  et~al., 2019, \mn@doi [\aap] {10.1051/0004-6361/201935765}, \href
  {https://ui.adsabs.harvard.edu/abs/2019A&A...628A..94A} {628, A94}

\bibitem[\protect\citeauthoryear{{Andrae} et~al.,}{{Andrae}
  et~al.}{2018}]{andrae18}
{Andrae} R.,  et~al., 2018, \mn@doi [\aap] {10.1051/0004-6361/201732516}, \href
  {https://ui.adsabs.harvard.edu/abs/2018A&A...616A...8A} {616, A8}

\bibitem[\protect\citeauthoryear{{Andrews} \& {Lindsay}}{{Andrews} \&
  {Lindsay}}{1967}]{andrews67}
{Andrews} A.~D.,  {Lindsay} E.~M.,  1967, Irish Astronomical Journal, \href
  {https://ui.adsabs.harvard.edu/abs/1967IrAJ....8..126A} {8, 126}

\bibitem[\protect\citeauthoryear{{Armandroff} \& {Da Costa}}{{Armandroff} \&
  {Da Costa}}{1991}]{armandroff91}
{Armandroff} T.~E.,  {Da Costa} G.~S.,  1991, \mn@doi [\aj] {10.1086/115769},
  \href {https://ui.adsabs.harvard.edu/abs/1991AJ....101.1329A} {101, 1329}

\bibitem[\protect\citeauthoryear{{Arsenault} et~al.,}{{Arsenault}
  et~al.}{2008}]{arsenault08}
{Arsenault} R.,  et~al., 2008, in {Hubin} N.,  {Max} C.~E.,   {Wizinowich}
  P.~L.,  eds,  Society of Photo-Optical Instrumentation Engineers (SPIE)
  Conference Series Vol. 7015, Adaptive Optics Systems. p. 701524,
  \mn@doi{10.1117/12.790359}

\bibitem[\protect\citeauthoryear{{Astropy Collaboration} et~al.,}{{Astropy
  Collaboration} et~al.}{2013}]{astropy1}
{Astropy Collaboration} et~al., 2013, \mn@doi [\aap]
  {10.1051/0004-6361/201322068}, \href
  {https://ui.adsabs.harvard.edu/abs/2013A&A...558A..33A} {558, A33}

\bibitem[\protect\citeauthoryear{{Astropy Collaboration} et~al.,}{{Astropy
  Collaboration} et~al.}{2018}]{astropy2}
{Astropy Collaboration} et~al., 2018, \mn@doi [\aj] {10.3847/1538-3881/aabc4f},
  \href {https://ui.adsabs.harvard.edu/abs/2018AJ....156..123A} {156, 123}

\bibitem[\protect\citeauthoryear{{Bacon} et~al.,}{{Bacon}
  et~al.}{2010}]{bacon10}
{Bacon} R.,  et~al., 2010, in {McLean} I.~S.,  {Ramsay} S.~K.,   {Takami} H.,
  eds,  Society of Photo-Optical Instrumentation Engineers (SPIE) Conference
  Series Vol. 7735, Ground-based and Airborne Instrumentation for Astronomy
  III. p. 773508, \mn@doi{10.1117/12.856027}

\bibitem[\protect\citeauthoryear{{Barbuy}, {Chiappini}  \& {Gerhard}}{{Barbuy}
  et~al.}{2018}]{barbuy18}
{Barbuy} B.,  {Chiappini} C.,   {Gerhard} O.,  2018, \mn@doi [\araa]
  {10.1146/annurev-astro-081817-051826}, \href
  {https://ui.adsabs.harvard.edu/abs/2018ARA&A..56..223B} {56, 223}

\bibitem[\protect\citeauthoryear{{Barbuy} et~al.,}{{Barbuy}
  et~al.}{2021}]{barbuy21}
{Barbuy} B.,  et~al., 2021, \mn@doi [\aap] {10.1051/0004-6361/202039761}, \href
  {https://ui.adsabs.harvard.edu/abs/2021A&A...648A..16B} {648, A16}

\bibitem[\protect\citeauthoryear{{Bastian} \& {Lardo}}{{Bastian} \&
  {Lardo}}{2018}]{bastian18}
{Bastian} N.,  {Lardo} C.,  2018, \mn@doi [\araa]
  {10.1146/annurev-astro-081817-051839}, \href
  {https://ui.adsabs.harvard.edu/abs/2018ARA&A..56...83B} {56, 83}

\bibitem[\protect\citeauthoryear{{Baumgardt} \& {Hilker}}{{Baumgardt} \&
  {Hilker}}{2018}]{baumgardt18}
{Baumgardt} H.,  {Hilker} M.,  2018, \mn@doi [\mnras] {10.1093/mnras/sty1057},
  \href {https://ui.adsabs.harvard.edu/abs/2018MNRAS.478.1520B} {478, 1520}

\bibitem[\protect\citeauthoryear{{Baumgardt} \& {Vasiliev}}{{Baumgardt} \&
  {Vasiliev}}{2021}]{baumgardt21}
{Baumgardt} H.,  {Vasiliev} E.,  2021, \mn@doi [\mnras]
  {10.1093/mnras/stab1474}, \href
  {https://ui.adsabs.harvard.edu/abs/2021MNRAS.505.5957B} {505, 5957}

\bibitem[\protect\citeauthoryear{{Baumgardt}, {Hilker}, {Sollima}  \&
  {Bellini}}{{Baumgardt} et~al.}{2019}]{baumgardt19}
{Baumgardt} H.,  {Hilker} M.,  {Sollima} A.,   {Bellini} A.,  2019, \mn@doi
  [\mnras] {10.1093/mnras/sty2997}, \href
  {https://ui.adsabs.harvard.edu/abs/2019MNRAS.482.5138B} {482, 5138}

\bibitem[\protect\citeauthoryear{{Baumgardt}, {Sollima}  \&
  {Hilker}}{{Baumgardt} et~al.}{2020}]{baumgardt20}
{Baumgardt} H.,  {Sollima} A.,   {Hilker} M.,  2020, \mn@doi [\pasa]
  {10.1017/pasa.2020.38}, \href
  {https://ui.adsabs.harvard.edu/abs/2020PASA...37...46B} {37, e046}

\bibitem[\protect\citeauthoryear{{Belokurov}, {Erkal}, {Evans}, {Koposov}  \&
  {Deason}}{{Belokurov} et~al.}{2018}]{belokurov18}
{Belokurov} V.,  {Erkal} D.,  {Evans} N.~W.,  {Koposov} S.~E.,   {Deason}
  A.~J.,  2018, \mn@doi [\mnras] {10.1093/mnras/sty982}, \href
  {https://ui.adsabs.harvard.edu/abs/2018MNRAS.478..611B} {478, 611}

\bibitem[\protect\citeauthoryear{{Bianchini}, {Sills}, {van de Ven}  \&
  {Sippel}}{{Bianchini} et~al.}{2017}]{bianchini17}
{Bianchini} P.,  {Sills} A.,  {van de Ven} G.,   {Sippel} A.~C.,  2017, \mn@doi
  [\mnras] {10.1093/mnras/stx1114}, \href
  {https://ui.adsabs.harvard.edu/abs/2017MNRAS.469.4359B} {469, 4359}

\bibitem[\protect\citeauthoryear{{Bica}, {Pavani}, {Bonatto}  \& {Lima}}{{Bica}
  et~al.}{2019}]{bica19}
{Bica} E.,  {Pavani} D.~B.,  {Bonatto} C.~J.,   {Lima} E.~F.,  2019, \mn@doi
  [\aj] {10.3847/1538-3881/aaef8d}, \href
  {https://ui.adsabs.harvard.edu/abs/2019AJ....157...12B} {157, 12}

\bibitem[\protect\citeauthoryear{{Bland-Hawthorn} \&
  {Gerhard}}{{Bland-Hawthorn} \& {Gerhard}}{2016}]{MWreview16}
{Bland-Hawthorn} J.,  {Gerhard} O.,  2016, \mn@doi [\araa]
  {10.1146/annurev-astro-081915-023441}, \href
  {https://ui.adsabs.harvard.edu/abs/2016ARA&A..54..529B} {54, 529}

\bibitem[\protect\citeauthoryear{{Bonatto} \& {Bica}}{{Bonatto} \&
  {Bica}}{2008}]{bonatto08}
{Bonatto} C.,  {Bica} E.,  2008, \mn@doi [\aap] {10.1051/0004-6361:20078687},
  \href {https://ui.adsabs.harvard.edu/abs/2008A&A...479..741B} {479, 741}

\bibitem[\protect\citeauthoryear{{Borissova} et~al.,}{{Borissova}
  et~al.}{2014}]{borissova14}
{Borissova} J.,  et~al., 2014, \mn@doi [\aap] {10.1051/0004-6361/201322483},
  \href {https://ui.adsabs.harvard.edu/abs/2014A&A...569A..24B} {569, A24}

\bibitem[\protect\citeauthoryear{{Boubert} \& {Everall}}{{Boubert} \&
  {Everall}}{2020}]{Gaiaverse2}
{Boubert} D.,  {Everall} A.,  2020, \mn@doi [\mnras] {10.1093/mnras/staa2305},
  \href {https://ui.adsabs.harvard.edu/abs/2020MNRAS.497.4246B} {497, 4246}

\bibitem[\protect\citeauthoryear{{Boubert}, {Everall}  \& {Holl}}{{Boubert}
  et~al.}{2020}]{Gaiaverse1}
{Boubert} D.,  {Everall} A.,   {Holl} B.,  2020, \mn@doi [\mnras]
  {10.1093/mnras/staa2050}, \href
  {https://ui.adsabs.harvard.edu/abs/2020MNRAS.497.1826B} {497, 1826}

\bibitem[\protect\citeauthoryear{{Boubert}, {Everall}, {Fraser}, {Gration}  \&
  {Holl}}{{Boubert} et~al.}{2021}]{Gaiaverse3}
{Boubert} D.,  {Everall} A.,  {Fraser} J.,  {Gration} A.,   {Holl} B.,  2021,
  \mn@doi [\mnras] {10.1093/mnras/staa3791}, \href
  {https://ui.adsabs.harvard.edu/abs/2021MNRAS.501.2954B} {501, 2954}

\bibitem[\protect\citeauthoryear{{Bovy}}{{Bovy}}{2015}]{galpy}
{Bovy} J.,  2015, \mn@doi [\apjs] {10.1088/0067-0049/216/2/29}, \href
  {https://ui.adsabs.harvard.edu/abs/2015ApJS..216...29B} {216, 29}

\bibitem[\protect\citeauthoryear{Bradley et~al.,}{Bradley
  et~al.}{2020}]{photutils}
Bradley L.,  et~al., 2020, astropy/photutils: 1.0.0,
  \mn@doi{10.5281/zenodo.4044744}, \url
  {https://doi.org/10.5281/zenodo.4044744}

\bibitem[\protect\citeauthoryear{{Bressan}, {Marigo}, {Girardi}, {Salasnich},
  {Dal Cero}, {Rubele}  \& {Nanni}}{{Bressan} et~al.}{2012}]{PARSEC1}
{Bressan} A.,  {Marigo} P.,  {Girardi} L.,  {Salasnich} B.,  {Dal Cero} C.,
  {Rubele} S.,   {Nanni} A.,  2012, \mn@doi [\mnras]
  {10.1111/j.1365-2966.2012.21948.x}, \href
  {https://ui.adsabs.harvard.edu/abs/2012MNRAS.427..127B} {427, 127}

\bibitem[\protect\citeauthoryear{{Brown}}{{Brown}}{2021}]{brown21}
{Brown} A. G.~A.,  2021, arXiv e-prints, \href
  {https://ui.adsabs.harvard.edu/abs/2021arXiv210211712B} {p. arXiv:2102.11712}

\bibitem[\protect\citeauthoryear{{Cantat-Gaudin} \& {Anders}}{{Cantat-Gaudin}
  \& {Anders}}{2020}]{cantatgaudin20a}
{Cantat-Gaudin} T.,  {Anders} F.,  2020, \mn@doi [\aap]
  {10.1051/0004-6361/201936691}, \href
  {https://ui.adsabs.harvard.edu/abs/2020A&A...633A..99C} {633, A99}

\bibitem[\protect\citeauthoryear{{Cantat-Gaudin} et~al.,}{{Cantat-Gaudin}
  et~al.}{2019}]{cantatgaudin19}
{Cantat-Gaudin} T.,  et~al., 2019, \mn@doi [\aap]
  {10.1051/0004-6361/201834453}, \href
  {https://ui.adsabs.harvard.edu/abs/2019A&A...624A.126C} {624, A126}

\bibitem[\protect\citeauthoryear{{Cantat-Gaudin} et~al.,}{{Cantat-Gaudin}
  et~al.}{2020}]{cantatgaudin20b}
{Cantat-Gaudin} T.,  et~al., 2020, \mn@doi [\aap]
  {10.1051/0004-6361/202038192}, \href
  {https://ui.adsabs.harvard.edu/abs/2020A&A...640A...1C} {640, A1}

\bibitem[\protect\citeauthoryear{{Cardelli}, {Clayton}  \& {Mathis}}{{Cardelli}
  et~al.}{1989}]{cardelli89}
{Cardelli} J.~A.,  {Clayton} G.~C.,   {Mathis} J.~S.,  1989, \mn@doi [\apj]
  {10.1086/167900}, \href
  {https://ui.adsabs.harvard.edu/abs/1989ApJ...345..245C} {345, 245}

\bibitem[\protect\citeauthoryear{{Carlberg}}{{Carlberg}}{2020}]{carlberg20}
{Carlberg} R.~G.,  2020, \mn@doi [\apj] {10.3847/1538-4357/ab80bf}, \href
  {https://ui.adsabs.harvard.edu/abs/2020ApJ...893..116C} {893, 116}

\bibitem[\protect\citeauthoryear{{Castro-Ginard}, {Jordi}, {Luri},
  {Cantat-Gaudin}  \& {Balaguer-N{\'u}{\~n}ez}}{{Castro-Ginard}
  et~al.}{2019}]{castroginard19}
{Castro-Ginard} A.,  {Jordi} C.,  {Luri} X.,  {Cantat-Gaudin} T.,
  {Balaguer-N{\'u}{\~n}ez} L.,  2019, \mn@doi [\aap]
  {10.1051/0004-6361/201935531}, \href
  {https://ui.adsabs.harvard.edu/abs/2019A&A...627A..35C} {627, A35}

\bibitem[\protect\citeauthoryear{{Castro-Ginard} et~al.,}{{Castro-Ginard}
  et~al.}{2020}]{castroginard20}
{Castro-Ginard} A.,  et~al., 2020, \mn@doi [\aap]
  {10.1051/0004-6361/201937386}, \href
  {https://ui.adsabs.harvard.edu/abs/2020A&A...635A..45C} {635, A45}

\bibitem[\protect\citeauthoryear{{Catelan}, {Pritzl}  \& {Smith}}{{Catelan}
  et~al.}{2004}]{catelan04}
{Catelan} M.,  {Pritzl} B.~J.,   {Smith} H.~A.,  2004, \mn@doi [\apjs]
  {10.1086/422916}, \href
  {https://ui.adsabs.harvard.edu/abs/2004ApJS..154..633C} {154, 633}

\bibitem[\protect\citeauthoryear{{Cenarro}, {Cardiel}, {Gorgas}, {Peletier},
  {Vazdekis}  \& {Prada}}{{Cenarro} et~al.}{2001}]{cenarro01}
{Cenarro} A.~J.,  {Cardiel} N.,  {Gorgas} J.,  {Peletier} R.~F.,  {Vazdekis}
  A.,   {Prada} F.,  2001, \mn@doi [\mnras] {10.1046/j.1365-8711.2001.04688.x},
  \href {https://ui.adsabs.harvard.edu/abs/2001MNRAS.326..959C} {326, 959}

\bibitem[\protect\citeauthoryear{{Chen}, {Girardi}, {Bressan}, {Marigo},
  {Barbieri}  \& {Kong}}{{Chen} et~al.}{2014}]{PARSEC3}
{Chen} Y.,  {Girardi} L.,  {Bressan} A.,  {Marigo} P.,  {Barbieri} M.,   {Kong}
  X.,  2014, \mn@doi [\mnras] {10.1093/mnras/stu1605}, \href
  {https://ui.adsabs.harvard.edu/abs/2014MNRAS.444.2525C} {444, 2525}

\bibitem[\protect\citeauthoryear{{Chen}, {Bressan}, {Girardi}, {Marigo}, {Kong}
   \& {Lanza}}{{Chen} et~al.}{2015}]{PARSEC2}
{Chen} Y.,  {Bressan} A.,  {Girardi} L.,  {Marigo} P.,  {Kong} X.,   {Lanza}
  A.,  2015, \mn@doi [\mnras] {10.1093/mnras/stv1281}, \href
  {https://ui.adsabs.harvard.edu/abs/2015MNRAS.452.1068C} {452, 1068}

\bibitem[\protect\citeauthoryear{{Contreras Ramos} et~al.,}{{Contreras Ramos}
  et~al.}{2017}]{contrerasramos17}
{Contreras Ramos} R.,  et~al., 2017, \mn@doi [\aap]
  {10.1051/0004-6361/201731462}, \href
  {https://ui.adsabs.harvard.edu/abs/2017A&A...608A.140C} {608, A140}

\bibitem[\protect\citeauthoryear{{Czesla}, {Schr{\"o}ter}, {Schneider},
  {Huber}, {Pfeifer}, {Andreasen}  \& {Zechmeister}}{{Czesla}
  et~al.}{2019}]{pyastronomy}
{Czesla} S.,  {Schr{\"o}ter} S.,  {Schneider} C.~P.,  {Huber} K.~F.,  {Pfeifer}
  F.,  {Andreasen} D.~T.,   {Zechmeister} M.,  2019, {PyA: Python
  astronomy-related packages} (\mn@eprint {ascl} {1906.010})

\bibitem[\protect\citeauthoryear{{Davidge}}{{Davidge}}{2000}]{davidge00}
{Davidge} T.~J.,  2000, \mn@doi [\apjs] {10.1086/313292}, \href
  {https://ui.adsabs.harvard.edu/abs/2000ApJS..126..105D} {126, 105}

\bibitem[\protect\citeauthoryear{{Dias}, {Alessi}, {Moitinho}  \&
  {L{\'e}pine}}{{Dias} et~al.}{2002}]{dias02}
{Dias} W.~S.,  {Alessi} B.~S.,  {Moitinho} A.,   {L{\'e}pine} J.~R.~D.,  2002,
  \mn@doi [\aap] {10.1051/0004-6361:20020668}, \href
  {https://ui.adsabs.harvard.edu/abs/2002A&A...389..871D} {389, 871}

\bibitem[\protect\citeauthoryear{{Dias}, {Monteiro}, {Caetano}, {L{\'e}pine},
  {Assafin}  \& {Oliveira}}{{Dias} et~al.}{2014}]{dias14}
{Dias} W.~S.,  {Monteiro} H.,  {Caetano} T.~C.,  {L{\'e}pine} J.~R.~D.,
  {Assafin} M.,   {Oliveira} A.~F.,  2014, \mn@doi [\aap]
  {10.1051/0004-6361/201323226}, \href
  {https://ui.adsabs.harvard.edu/abs/2014A&A...564A..79D} {564, A79}

\bibitem[\protect\citeauthoryear{{Djorgovski}}{{Djorgovski}}{1987}]{djorgovski87}
{Djorgovski} S.,  1987, \mn@doi [\apjl] {10.1086/184903}, \href
  {https://ui.adsabs.harvard.edu/abs/1987ApJ...317L..13D} {317, L13}

\bibitem[\protect\citeauthoryear{{Errani}, {Pe{\~n}arrubia}  \&
  {Walker}}{{Errani} et~al.}{2018}]{errani18}
{Errani} R.,  {Pe{\~n}arrubia} J.,   {Walker} M.~G.,  2018, \mn@doi [\mnras]
  {10.1093/mnras/sty2505}, \href
  {https://ui.adsabs.harvard.edu/abs/2018MNRAS.481.5073E} {481, 5073}

\bibitem[\protect\citeauthoryear{Ester, Kriegel, Sander  \& Xu}{Ester
  et~al.}{1996}]{dbscan}
Ester M.,  Kriegel H.-P.,  Sander J.,   Xu X.,  1996, in Proceedings of the
  Second International Conference on Knowledge Discovery and Data Mining.
  KDD’96.
AAAI Press, p. 226–231

\bibitem[\protect\citeauthoryear{{Ferraro} et~al.,}{{Ferraro}
  et~al.}{2021}]{ferraro21}
{Ferraro} F.~R.,  et~al., 2021, \mn@doi [Nature Astronomy]
  {10.1038/s41550-020-01267-y}, \href
  {https://ui.adsabs.harvard.edu/abs/2021NatAs...5..311F} {5, 311}

\bibitem[\protect\citeauthoryear{{Forbes} \& {Bridges}}{{Forbes} \&
  {Bridges}}{2010}]{forbes10}
{Forbes} D.~A.,  {Bridges} T.,  2010, \mn@doi [\mnras]
  {10.1111/j.1365-2966.2010.16373.x}, \href
  {https://ui.adsabs.harvard.edu/abs/2010MNRAS.404.1203F} {404, 1203}

\bibitem[\protect\citeauthoryear{{Gaia Collaboration} et~al.,}{{Gaia
  Collaboration} et~al.}{2016}]{gaia_mission}
{Gaia Collaboration} et~al., 2016, \mn@doi [\aap]
  {10.1051/0004-6361/201629272}, \href
  {https://ui.adsabs.harvard.edu/abs/2016A&A...595A...1G} {595, A1}

\bibitem[\protect\citeauthoryear{{Gaia Collaboration} et~al.,}{{Gaia
  Collaboration} et~al.}{2018}]{brown18}
{Gaia Collaboration} et~al., 2018, \mn@doi [\aap]
  {10.1051/0004-6361/201833051}, \href
  {https://ui.adsabs.harvard.edu/abs/2018A&A...616A...1G} {616, A1}

\bibitem[\protect\citeauthoryear{{Gaia Collaboration} et~al.,}{{Gaia
  Collaboration} et~al.}{2019}]{gaiadr2_vars}
{Gaia Collaboration} et~al., 2019, \mn@doi [\aap]
  {10.1051/0004-6361/201833304}, \href
  {https://ui.adsabs.harvard.edu/abs/2019A&A...623A.110G} {623, A110}

\bibitem[\protect\citeauthoryear{{Gaia Collaboration} et~al.,}{{Gaia
  Collaboration} et~al.}{2021}]{brown20}
{Gaia Collaboration} et~al., 2021, \mn@doi [\aap]
  {10.1051/0004-6361/202039657}, \href
  {https://ui.adsabs.harvard.edu/abs/2021A&A...649A...1G} {649, A1}

\bibitem[\protect\citeauthoryear{{Garro} et~al.,}{{Garro}
  et~al.}{2020}]{garro20}
{Garro} E.~R.,  et~al., 2020, \mn@doi [\aap] {10.1051/0004-6361/202039233},
  \href {https://ui.adsabs.harvard.edu/abs/2020A&A...642L..19G} {642, L19}

\bibitem[\protect\citeauthoryear{{Gehrels}}{{Gehrels}}{1986}]{gehrels86}
{Gehrels} N.,  1986, \mn@doi [\apj] {10.1086/164079}, \href
  {https://ui.adsabs.harvard.edu/abs/1986ApJ...303..336G} {303, 336}

\bibitem[\protect\citeauthoryear{{Gran} et~al.,}{{Gran} et~al.}{2016}]{gran16}
{Gran} F.,  et~al., 2016, \mn@doi [\aap] {10.1051/0004-6361/201527511}, \href
  {https://ui.adsabs.harvard.edu/abs/2016A&A...591A.145G} {591, A145}

\bibitem[\protect\citeauthoryear{{Gran} et~al.,}{{Gran} et~al.}{2019}]{gran19}
{Gran} F.,  et~al., 2019, \mn@doi [\aap] {10.1051/0004-6361/201834986}, \href
  {https://ui.adsabs.harvard.edu/abs/2019A&A...628A..45G} {628, A45}

\bibitem[\protect\citeauthoryear{{Gratton}, {Bragaglia}, {Carretta}, {D'Orazi},
  {Lucatello}  \& {Sollima}}{{Gratton} et~al.}{2019}]{gratton19}
{Gratton} R.,  {Bragaglia} A.,  {Carretta} E.,  {D'Orazi} V.,  {Lucatello} S.,
   {Sollima} A.,  2019, \mn@doi [\aapr] {10.1007/s00159-019-0119-3}, \href
  {https://ui.adsabs.harvard.edu/abs/2019A&ARv..27....8G} {27, 8}

\bibitem[\protect\citeauthoryear{Green}{Green}{2018}]{green2018}
Green G.~M.,  2018, \mn@doi [Journal of Open Source Software]
  {10.21105/joss.00695}, 3, 695

\bibitem[\protect\citeauthoryear{{Harris}}{{Harris}}{1996}]{harris96}
{Harris} W.~E.,  1996, \mn@doi [\aj] {10.1086/118116}, \href
  {https://ui.adsabs.harvard.edu/abs/1996AJ....112.1487H} {112, 1487}

\bibitem[\protect\citeauthoryear{{Harris}}{{Harris}}{2010}]{harris10}
{Harris} W.~E.,  2010, arXiv e-prints, \href
  {https://ui.adsabs.harvard.edu/abs/2010arXiv1012.3224H} {p. arXiv:1012.3224}

\bibitem[\protect\citeauthoryear{Harris et~al.,}{Harris et~al.}{2020}]{numpy}
Harris C.~R.,  et~al., 2020, \mn@doi [Nature] {10.1038/s41586-020-2649-2}, 585,
  357

\bibitem[\protect\citeauthoryear{{Hartke}, {Kakkad}, {Reyes}, {Moya-Sierralta},
  {Reyes}, {Kravtsov}, {Kolb}  \& {Selman}}{{Hartke} et~al.}{2020}]{hartke20}
{Hartke} J.,  {Kakkad} D.,  {Reyes} C.,  {Moya-Sierralta} C.,  {Reyes} A.,
  {Kravtsov} T.,  {Kolb} J.,   {Selman} F.,  2020, in Society of Photo-Optical
  Instrumentation Engineers (SPIE) Conference Series. p. 114480V,
  \mn@doi{10.1117/12.2560793}

\bibitem[\protect\citeauthoryear{{Helmi}, {Babusiaux}, {Koppelman}, {Massari},
  {Veljanoski}  \& {Brown}}{{Helmi} et~al.}{2018}]{helmi18b}
{Helmi} A.,  {Babusiaux} C.,  {Koppelman} H.~H.,  {Massari} D.,  {Veljanoski}
  J.,   {Brown} A. G.~A.,  2018, \mn@doi [\nat] {10.1038/s41586-018-0625-x},
  \href {https://ui.adsabs.harvard.edu/abs/2018Natur.563...85H} {563, 85}

\bibitem[\protect\citeauthoryear{{Holmberg}, {Lauberts}, {Schuster}  \&
  {West}}{{Holmberg} et~al.}{1974}]{holmberg74}
{Holmberg} E.~B.,  {Lauberts} A.,  {Schuster} H.~E.,   {West} R.~M.,  1974,
  \aaps, \href {https://ui.adsabs.harvard.edu/abs/1974A&AS...18..463H} {18,
  463}

\bibitem[\protect\citeauthoryear{{Holmberg}, {Lauberts}, {Schuster}  \&
  {West}}{{Holmberg} et~al.}{1978}]{holmberg78}
{Holmberg} E.~B.,  {Lauberts} A.,  {Schuster} H.~E.,   {West} R.~M.,  1978,
  \aaps, \href {https://ui.adsabs.harvard.edu/abs/1978A&AS...34..285H} {34,
  285}

\bibitem[\protect\citeauthoryear{{Horta} et~al.,}{{Horta}
  et~al.}{2021}]{horta21}
{Horta} D.,  et~al., 2021, \mn@doi [\mnras] {10.1093/mnras/staa2987}, \href
  {https://ui.adsabs.harvard.edu/abs/2021MNRAS.500.1385H} {500, 1385}

\bibitem[\protect\citeauthoryear{{Hughes}, {Pfeffer}, {Martig}, {Reina-Campos},
  {Bastian}, {Crain}  \& {Kruijssen}}{{Hughes} et~al.}{2020}]{hughes20}
{Hughes} M.~E.,  {Pfeffer} J.~L.,  {Martig} M.,  {Reina-Campos} M.,  {Bastian}
  N.,  {Crain} R.~A.,   {Kruijssen} J.~M.~D.,  2020, \mn@doi [\mnras]
  {10.1093/mnras/stz3341}, \href
  {https://ui.adsabs.harvard.edu/abs/2020MNRAS.491.4012H} {491, 4012}

\bibitem[\protect\citeauthoryear{{Hunt} \& {Reffert}}{{Hunt} \&
  {Reffert}}{2021}]{hunt21}
{Hunt} E.~L.,  {Reffert} S.,  2021, \mn@doi [\aap]
  {10.1051/0004-6361/202039341}, \href
  {https://ui.adsabs.harvard.edu/abs/2021A&A...646A.104H} {646, A104}

\bibitem[\protect\citeauthoryear{Hunter}{Hunter}{2007}]{matplotlib}
Hunter J.~D.,  2007, \mn@doi [Computing in Science \& Engineering]
  {10.1109/MCSE.2007.55}, 9, 90

\bibitem[\protect\citeauthoryear{{Husser} et~al.,}{{Husser}
  et~al.}{2020}]{husser20}
{Husser} T.-O.,  et~al., 2020, \mn@doi [\aap] {10.1051/0004-6361/201936508},
  \href {https://ui.adsabs.harvard.edu/abs/2020A&A...635A.114H} {635, A114}

\bibitem[\protect\citeauthoryear{{Ibata}, {Irwin}, {Lewis}  \&
  {Stolte}}{{Ibata} et~al.}{2001}]{ibata01}
{Ibata} R.,  {Irwin} M.,  {Lewis} G.~F.,   {Stolte} A.,  2001, \mn@doi [\apjl]
  {10.1086/318894}, \href
  {https://ui.adsabs.harvard.edu/abs/2001ApJ...547L.133I} {547, L133}

\bibitem[\protect\citeauthoryear{{Ibata}, {Malhan}, {Martin}  \&
  {Starkenburg}}{{Ibata} et~al.}{2018}]{ibata18}
{Ibata} R.~A.,  {Malhan} K.,  {Martin} N.~F.,   {Starkenburg} E.,  2018,
  \mn@doi [\apj] {10.3847/1538-4357/aadba3}, \href
  {https://ui.adsabs.harvard.edu/abs/2018ApJ...865...85I} {865, 85}

\bibitem[\protect\citeauthoryear{{Kamann}, {Wisotzki}  \& {Roth}}{{Kamann}
  et~al.}{2013}]{kamman13}
{Kamann} S.,  {Wisotzki} L.,   {Roth} M.~M.,  2013, \mn@doi [\aap]
  {10.1051/0004-6361/201220476}, \href
  {https://ui.adsabs.harvard.edu/abs/2013A&A...549A..71K} {549, A71}

\bibitem[\protect\citeauthoryear{{Kharchenko}, {Piskunov}, {Schilbach},
  {R{\"o}ser}  \& {Scholz}}{{Kharchenko} et~al.}{2013}]{kharchenko13}
{Kharchenko} N.~V.,  {Piskunov} A.~E.,  {Schilbach} E.,  {R{\"o}ser} S.,
  {Scholz} R.~D.,  2013, \mn@doi [\aap] {10.1051/0004-6361/201322302}, \href
  {https://ui.adsabs.harvard.edu/abs/2013A&A...558A..53K} {558, A53}

\bibitem[\protect\citeauthoryear{{Kharchenko}, {Piskunov}, {Schilbach},
  {R{\"o}ser}  \& {Scholz}}{{Kharchenko} et~al.}{2016}]{kharchenko16}
{Kharchenko} N.~V.,  {Piskunov} A.~E.,  {Schilbach} E.,  {R{\"o}ser} S.,
  {Scholz} R.~D.,  2016, \mn@doi [\aap] {10.1051/0004-6361/201527292}, \href
  {https://ui.adsabs.harvard.edu/abs/2016A&A...585A.101K} {585, A101}

\bibitem[\protect\citeauthoryear{{King}}{{King}}{1962}]{king62}
{King} I.,  1962, \mn@doi [\aj] {10.1086/108756}, \href
  {https://ui.adsabs.harvard.edu/abs/1962AJ.....67..471K} {67, 471}

\bibitem[\protect\citeauthoryear{{Kisku} et~al.,}{{Kisku}
  et~al.}{2021}]{kisku21}
{Kisku} S.,  et~al., 2021, \mn@doi [\mnras] {10.1093/mnras/stab525}, \href
  {https://ui.adsabs.harvard.edu/abs/2021MNRAS.tmp..569K} {}

\bibitem[\protect\citeauthoryear{{Kruijssen}}{{Kruijssen}}{2019}]{kruijssen19a}
{Kruijssen} J.~M.~D.,  2019, \mn@doi [\mnras] {10.1093/mnrasl/slz052}, \href
  {https://ui.adsabs.harvard.edu/abs/2019MNRAS.486L..20K} {486, L20}

\bibitem[\protect\citeauthoryear{{Kruijssen}, {Pfeffer}, {Reina-Campos},
  {Crain}  \& {Bastian}}{{Kruijssen} et~al.}{2019}]{kruijssen19b}
{Kruijssen} J.~M.~D.,  {Pfeffer} J.~L.,  {Reina-Campos} M.,  {Crain} R.~A.,
  {Bastian} N.,  2019, \mn@doi [\mnras] {10.1093/mnras/sty1609}, \href
  {https://ui.adsabs.harvard.edu/abs/2019MNRAS.486.3180K} {486, 3180}

\bibitem[\protect\citeauthoryear{{Lauberts}}{{Lauberts}}{1982}]{lauberts82}
{Lauberts} A.,  1982, {ESO/Uppsala survey of the ESO(B) atlas}

\bibitem[\protect\citeauthoryear{{Leaman}, {VandenBerg}  \& {Mendel}}{{Leaman}
  et~al.}{2013}]{leaman13}
{Leaman} R.,  {VandenBerg} D.~A.,   {Mendel} J.~T.,  2013, \mn@doi [\mnras]
  {10.1093/mnras/stt1540}, \href
  {https://ui.adsabs.harvard.edu/abs/2013MNRAS.436..122L} {436, 122}

\bibitem[\protect\citeauthoryear{{Lindegren} et~al.,}{{Lindegren}
  et~al.}{2018}]{lindegren18}
{Lindegren} L.,  et~al., 2018, \mn@doi [\aap] {10.1051/0004-6361/201832727},
  \href {https://ui.adsabs.harvard.edu/abs/2018A&A...616A...2L} {616, A2}

\bibitem[\protect\citeauthoryear{{Mackey} \& {Gilmore}}{{Mackey} \&
  {Gilmore}}{2004}]{mackey04}
{Mackey} A.~D.,  {Gilmore} G.~F.,  2004, \mn@doi [\mnras]
  {10.1111/j.1365-2966.2004.08343.x}, \href
  {https://ui.adsabs.harvard.edu/abs/2004MNRAS.355..504M} {355, 504}

\bibitem[\protect\citeauthoryear{{Marigo} et~al.,}{{Marigo}
  et~al.}{2017}]{PARSEC5}
{Marigo} P.,  et~al., 2017, \mn@doi [\apj] {10.3847/1538-4357/835/1/77}, \href
  {https://ui.adsabs.harvard.edu/abs/2017ApJ...835...77M} {835, 77}

\bibitem[\protect\citeauthoryear{{Massari}, {Koppelman}  \& {Helmi}}{{Massari}
  et~al.}{2019}]{massari19}
{Massari} D.,  {Koppelman} H.~H.,   {Helmi} A.,  2019, \mn@doi [\aap]
  {10.1051/0004-6361/201936135}, \href
  {https://ui.adsabs.harvard.edu/abs/2019A&A...630L...4M} {630, L4}

\bibitem[\protect\citeauthoryear{{Minniti} et~al.,}{{Minniti}
  et~al.}{2010}]{vvv}
{Minniti} D.,  et~al., 2010, \mn@doi [\na] {10.1016/j.newast.2009.12.002},
  \href {https://ui.adsabs.harvard.edu/abs/2010NewA...15..433M} {15, 433}

\bibitem[\protect\citeauthoryear{{Minniti} et~al.,}{{Minniti}
  et~al.}{2011a}]{minniti11a}
{Minniti} D.,  et~al., 2011a, \mn@doi [\aap] {10.1051/0004-6361/201015795},
  \href {https://ui.adsabs.harvard.edu/abs/2011A&A...527A..81M} {527, A81}

\bibitem[\protect\citeauthoryear{{Minniti}, {Saito}, {Alonso-Garc{\'\i}a},
  {Lucas}  \& {Hempel}}{{Minniti} et~al.}{2011b}]{minniti11b}
{Minniti} D.,  {Saito} R.~K.,  {Alonso-Garc{\'\i}a} J.,  {Lucas} P.~W.,
  {Hempel} M.,  2011b, \mn@doi [\apjl] {10.1088/2041-8205/733/2/L43}, \href
  {https://ui.adsabs.harvard.edu/abs/2011ApJ...733L..43M} {733, L43}

\bibitem[\protect\citeauthoryear{{Minniti}, {Alonso-Garc{\'\i}a}, {Braga},
  {Contreras Ramos}, {Hempel}, {Palma}, {Pullen}  \& {Saito}}{{Minniti}
  et~al.}{2017a}]{minniti17b}
{Minniti} D.,  {Alonso-Garc{\'\i}a} J.,  {Braga} V.,  {Contreras Ramos} R.,
  {Hempel} M.,  {Palma} T.,  {Pullen} J.,   {Saito} R.~K.,  2017a, \mn@doi
  [Research Notes of the American Astronomical Society]
  {10.3847/2515-5172/aa9ab7}, \href
  {https://ui.adsabs.harvard.edu/abs/2017RNAAS...1...16M} {1, 16}

\bibitem[\protect\citeauthoryear{{Minniti}, {Alonso-Garc{\'\i}a}  \&
  {Pullen}}{{Minniti} et~al.}{2017b}]{minniti17c}
{Minniti} D.,  {Alonso-Garc{\'\i}a} J.,   {Pullen} J.,  2017b, \mn@doi
  [Research Notes of the American Astronomical Society]
  {10.3847/2515-5172/aaa3ed}, \href
  {https://ui.adsabs.harvard.edu/abs/2017RNAAS...1...54M} {1, 54}

\bibitem[\protect\citeauthoryear{{Minniti} et~al.,}{{Minniti}
  et~al.}{2017c}]{minniti17a}
{Minniti} D.,  et~al., 2017c, \mn@doi [\apjl] {10.3847/2041-8213/aa95b8}, \href
  {https://ui.adsabs.harvard.edu/abs/2017ApJ...849L..24M} {849, L24}

\bibitem[\protect\citeauthoryear{{Moni Bidin} et~al.,}{{Moni Bidin}
  et~al.}{2011}]{monibidin11}
{Moni Bidin} C.,  et~al., 2011, \mn@doi [\aap] {10.1051/0004-6361/201117488},
  \href {https://ui.adsabs.harvard.edu/abs/2011A&A...535A..33M} {535, A33}

\bibitem[\protect\citeauthoryear{{Myeong}, {Evans}, {Belokurov}, {Sand ers}  \&
  {Koposov}}{{Myeong} et~al.}{2018}]{myeong18}
{Myeong} G.~C.,  {Evans} N.~W.,  {Belokurov} V.,  {Sand ers} J.~L.,   {Koposov}
  S.~E.,  2018, \mn@doi [\apjl] {10.3847/2041-8213/aad7f7}, \href
  {https://ui.adsabs.harvard.edu/abs/2018ApJ...863L..28M} {863, L28}

\bibitem[\protect\citeauthoryear{{Myeong}, {Vasiliev}, {Iorio}, {Evans}  \&
  {Belokurov}}{{Myeong} et~al.}{2019}]{myeong19}
{Myeong} G.~C.,  {Vasiliev} E.,  {Iorio} G.,  {Evans} N.~W.,   {Belokurov} V.,
  2019, \mn@doi [\mnras] {10.1093/mnras/stz1770}, \href
  {https://ui.adsabs.harvard.edu/abs/2019MNRAS.488.1235M} {488, 1235}

\bibitem[\protect\citeauthoryear{{Ochsenbein}, {Bauer}  \&
  {Marcout}}{{Ochsenbein} et~al.}{2000}]{vizier}
{Ochsenbein} F.,  {Bauer} P.,   {Marcout} J.,  2000, \mn@doi [\aaps]
  {10.1051/aas:2000169}, \href
  {https://ui.adsabs.harvard.edu/abs/2000A&AS..143...23O} {143, 23}

\bibitem[\protect\citeauthoryear{{Ortolani}, {Bica}  \& {Barbuy}}{{Ortolani}
  et~al.}{1995}]{ortolani95}
{Ortolani} S.,  {Bica} E.,   {Barbuy} B.,  1995, \aap, \href
  {https://ui.adsabs.harvard.edu/abs/1995A&A...296..680O} {296, 680}

\bibitem[\protect\citeauthoryear{{Ortolani}, {Bica}  \& {Barbuy}}{{Ortolani}
  et~al.}{2006}]{ortolani06}
{Ortolani} S.,  {Bica} E.,   {Barbuy} B.,  2006, \mn@doi [\apjl]
  {10.1086/507108}, \href
  {https://ui.adsabs.harvard.edu/abs/2006ApJ...646L.115O} {646, L115}

\bibitem[\protect\citeauthoryear{{Ortolani}, {Nardiello}, {P{\'e}rez-Villegas},
  {Bica}  \& {Barbuy}}{{Ortolani} et~al.}{2019}]{ortolani19}
{Ortolani} S.,  {Nardiello} D.,  {P{\'e}rez-Villegas} A.,  {Bica} E.,
  {Barbuy} B.,  2019, \mn@doi [\aap] {10.1051/0004-6361/201834477}, \href
  {https://ui.adsabs.harvard.edu/abs/2019A&A...622A..94O} {622, A94}

\bibitem[\protect\citeauthoryear{{Palma} et~al.,}{{Palma}
  et~al.}{2019}]{palma19}
{Palma} T.,  et~al., 2019, \mn@doi [\mnras] {10.1093/mnras/stz1489}, \href
  {https://ui.adsabs.harvard.edu/abs/2019MNRAS.487.3140P} {487, 3140}

\bibitem[\protect\citeauthoryear{{Pastorelli} et~al.,}{{Pastorelli}
  et~al.}{2019}]{PARSEC6}
{Pastorelli} G.,  et~al., 2019, \mn@doi [\mnras] {10.1093/mnras/stz725}, \href
  {https://ui.adsabs.harvard.edu/abs/2019MNRAS.485.5666P} {485, 5666}

\bibitem[\protect\citeauthoryear{Pedregosa et~al.,}{Pedregosa
  et~al.}{2011}]{scikit-learn}
Pedregosa F.,  et~al., 2011, Journal of Machine Learning Research, 12, 2825

\bibitem[\protect\citeauthoryear{P\'erez \& Granger}{P\'erez \&
  Granger}{2007}]{ipython}
P\'erez F.,  Granger B.~E.,  2007, \mn@doi [Computing in Science and
  Engineering] {10.1109/MCSE.2007.53}, 9, 21

\bibitem[\protect\citeauthoryear{{Price-Jones} et~al.,}{{Price-Jones}
  et~al.}{2020}]{pricejones20}
{Price-Jones} N.,  et~al., 2020, \mn@doi [\mnras] {10.1093/mnras/staa1905},
  \href {https://ui.adsabs.harvard.edu/abs/2020MNRAS.496.5101P} {496, 5101}

\bibitem[\protect\citeauthoryear{Price-Whelan}{Price-Whelan}{2017}]{gala1}
Price-Whelan A.~M.,  2017, \mn@doi [Journal of Open Source Software]
  {10.21105/joss.00388}, 2, 388

\bibitem[\protect\citeauthoryear{{Price-Whelan}, {Sesar}, {Johnston}  \&
  {Rix}}{{Price-Whelan} et~al.}{2016}]{pricewhelan16}
{Price-Whelan} A.~M.,  {Sesar} B.,  {Johnston} K.~V.,   {Rix} H.-W.,  2016,
  \mn@doi [\apj] {10.3847/0004-637X/824/2/104}, \href
  {https://ui.adsabs.harvard.edu/abs/2016ApJ...824..104P} {824, 104}

\bibitem[\protect\citeauthoryear{Price-Whelan et~al.,}{Price-Whelan
  et~al.}{2020}]{gala2}
Price-Whelan A.,  et~al., 2020, adrn/gala: v1.3,
  \mn@doi{10.5281/zenodo.4159870}, \url
  {https://doi.org/10.5281/zenodo.4159870}

\bibitem[\protect\citeauthoryear{{Queiroz} et~al.,}{{Queiroz}
  et~al.}{2018}]{queiroz18}
{Queiroz} A.~B.~A.,  et~al., 2018, \mn@doi [\mnras] {10.1093/mnras/sty330},
  \href {https://ui.adsabs.harvard.edu/abs/2018MNRAS.476.2556Q} {476, 2556}

\bibitem[\protect\citeauthoryear{{Renzini}}{{Renzini}}{2017}]{renzini17}
{Renzini} A.,  2017, \mn@doi [\mnras] {10.1093/mnrasl/slx057}, \href
  {https://ui.adsabs.harvard.edu/abs/2017MNRAS.469L..63R} {469, L63}

\bibitem[\protect\citeauthoryear{{Rossi}, {Ortolani}, {Barbuy}, {Bica}  \&
  {Bonfanti}}{{Rossi} et~al.}{2015}]{rossi15}
{Rossi} L.~J.,  {Ortolani} S.,  {Barbuy} B.,  {Bica} E.,   {Bonfanti} A.,
  2015, \mn@doi [\mnras] {10.1093/mnras/stv748}, \href
  {https://ui.adsabs.harvard.edu/abs/2015MNRAS.450.3270R} {450, 3270}

\bibitem[\protect\citeauthoryear{{Rutledge}, {Hesser}, {Stetson}, {Mateo},
  {Simard}, {Bolte}, {Friel}  \& {Copin}}{{Rutledge} et~al.}{1997}]{rutledge97}
{Rutledge} G.~A.,  {Hesser} J.~E.,  {Stetson} P.~B.,  {Mateo} M.,  {Simard} L.,
   {Bolte} M.,  {Friel} E.~D.,   {Copin} Y.,  1997, \mn@doi [\pasp]
  {10.1086/133958}, \href
  {https://ui.adsabs.harvard.edu/abs/1997PASP..109..883R} {109, 883}

\bibitem[\protect\citeauthoryear{{Saracino} et~al.,}{{Saracino}
  et~al.}{2015}]{saracino15}
{Saracino} S.,  et~al., 2015, \mn@doi [\apj] {10.1088/0004-637X/806/2/152},
  \href {https://ui.adsabs.harvard.edu/abs/2015ApJ...806..152S} {806, 152}

\bibitem[\protect\citeauthoryear{{Shen} \& {Zheng}}{{Shen} \&
  {Zheng}}{2020}]{shen20}
{Shen} J.,  {Zheng} X.-W.,  2020, \mn@doi [Research in Astronomy and
  Astrophysics] {10.1088/1674-4527/20/10/159}, \href
  {https://ui.adsabs.harvard.edu/abs/2020RAA....20..159S} {20, 159}

\bibitem[\protect\citeauthoryear{{Skrutskie} et~al.,}{{Skrutskie}
  et~al.}{2006}]{2mass}
{Skrutskie} M.~F.,  et~al., 2006, \mn@doi [\aj] {10.1086/498708}, \href
  {https://ui.adsabs.harvard.edu/abs/2006AJ....131.1163S} {131, 1163}

\bibitem[\protect\citeauthoryear{{Soszy{\'n}ski} et~al.,}{{Soszy{\'n}ski}
  et~al.}{2019}]{soszynski19}
{Soszy{\'n}ski} I.,  et~al., 2019, \mn@doi [\actaa]
  {10.32023/0001-5237/69.4.2}, \href
  {https://ui.adsabs.harvard.edu/abs/2019AcA....69..321S} {69, 321}

\bibitem[\protect\citeauthoryear{{Str{\"o}bele} et~al.,}{{Str{\"o}bele}
  et~al.}{2012}]{strobele12}
{Str{\"o}bele} S.,  et~al., 2012, in {Ellerbroek} B.~L.,  {Marchetti} E.,
  {V{\'e}ran} J.-P.,  eds,  Society of Photo-Optical Instrumentation Engineers
  (SPIE) Conference Series Vol. 8447, Adaptive Optics Systems III. p. 844737,
  \mn@doi{10.1117/12.926110}

\bibitem[\protect\citeauthoryear{{Str{\"o}bele}, {Kasper}  \&
  {Madec}}{{Str{\"o}bele} et~al.}{2020}]{strobele20}
{Str{\"o}bele} S.,  {Kasper} M.,   {Madec} P.~Y.,  2020, in Society of
  Photo-Optical Instrumentation Engineers (SPIE) Conference Series. p. 114481B,
  \mn@doi{10.1117/12.2561224}

\bibitem[\protect\citeauthoryear{{Stuik}, {Bacon}, {Conzelmann}, {Delabre},
  {Fedrigo}, {Hubin}, {Le Louarn}  \& {Str{\"o}bele}}{{Stuik}
  et~al.}{2006}]{stuik06}
{Stuik} R.,  {Bacon} R.,  {Conzelmann} R.,  {Delabre} B.,  {Fedrigo} E.,
  {Hubin} N.,  {Le Louarn} M.,   {Str{\"o}bele} S.,  2006, \mn@doi [\nar]
  {10.1016/j.newar.2005.10.015}, \href
  {https://ui.adsabs.harvard.edu/abs/2006NewAR..49..618S} {49, 618}

\bibitem[\protect\citeauthoryear{{Surot} et~al.,}{{Surot}
  et~al.}{2019}]{surot19}
{Surot} F.,  et~al., 2019, \mn@doi [\aap] {10.1051/0004-6361/201833550}, \href
  {https://ui.adsabs.harvard.edu/abs/2019A&A...623A.168S} {623, A168}

\bibitem[\protect\citeauthoryear{{Surot}, {Valenti}, {Gonzalez}, {Zoccali},
  {S{\"o}kmen}, {Hidalgo}  \& {Minniti}}{{Surot} et~al.}{2020}]{surot20}
{Surot} F.,  {Valenti} E.,  {Gonzalez} O.~A.,  {Zoccali} M.,  {S{\"o}kmen} E.,
  {Hidalgo} S.~L.,   {Minniti} D.,  2020, \mn@doi [\aap]
  {10.1051/0004-6361/202038346}, \href
  {https://ui.adsabs.harvard.edu/abs/2020A&A...644A.140S} {644, A140}

\bibitem[\protect\citeauthoryear{{Tang}, {Bressan}, {Rosenfield}, {Slemer},
  {Marigo}, {Girardi}  \& {Bianchi}}{{Tang} et~al.}{2014}]{PARSEC4}
{Tang} J.,  {Bressan} A.,  {Rosenfield} P.,  {Slemer} A.,  {Marigo} P.,
  {Girardi} L.,   {Bianchi} L.,  2014, \mn@doi [\mnras]
  {10.1093/mnras/stu2029}, \href
  {https://ui.adsabs.harvard.edu/abs/2014MNRAS.445.4287T} {445, 4287}

\bibitem[\protect\citeauthoryear{{Taylor}}{{Taylor}}{2005}]{topcat}
{Taylor} M.~B.,  2005, in {Shopbell} P.,  {Britton} M.,   {Ebert} R.,  eds,
  Astronomical Society of the Pacific Conference Series Vol. 347, Astronomical
  Data Analysis Software and Systems XIV. p.~29

\bibitem[\protect\citeauthoryear{{Valcarce}, {Catelan}  \&
  {Sweigart}}{{Valcarce} et~al.}{2012}]{PGPUC}
{Valcarce} A.~A.~R.,  {Catelan} M.,   {Sweigart} A.~V.,  2012, \mn@doi [\aap]
  {10.1051/0004-6361/201219510}, \href
  {https://ui.adsabs.harvard.edu/abs/2012A&A...547A...5V} {547, A5}

\bibitem[\protect\citeauthoryear{{Valenti}, {Ferraro}  \& {Origlia}}{{Valenti}
  et~al.}{2010}]{valenti10}
{Valenti} E.,  {Ferraro} F.~R.,   {Origlia} L.,  2010, \mn@doi [\mnras]
  {10.1111/j.1365-2966.2009.15991.x}, \href
  {https://ui.adsabs.harvard.edu/abs/2010MNRAS.402.1729V} {402, 1729}

\bibitem[\protect\citeauthoryear{{Vasiliev}}{{Vasiliev}}{2019}]{vasilev19}
{Vasiliev} E.,  2019, \mn@doi [\mnras] {10.1093/mnras/stz171}, \href
  {https://ui.adsabs.harvard.edu/abs/2019MNRAS.484.2832V} {484, 2832}

\bibitem[\protect\citeauthoryear{{V{\'a}squez}, {Zoccali}, {Hill}, {Gonzalez},
  {Saviane}, {Rejkuba}  \& {Battaglia}}{{V{\'a}squez} et~al.}{2015}]{vasquez15}
{V{\'a}squez} S.,  {Zoccali} M.,  {Hill} V.,  {Gonzalez} O.~A.,  {Saviane} I.,
  {Rejkuba} M.,   {Battaglia} G.,  2015, \mn@doi [\aap]
  {10.1051/0004-6361/201526534}, \href
  {https://ui.adsabs.harvard.edu/abs/2015A&A...580A.121V} {580, A121}

\bibitem[\protect\citeauthoryear{{V{\'a}squez} et~al.,}{{V{\'a}squez}
  et~al.}{2018}]{vasquez18}
{V{\'a}squez} S.,  et~al., 2018, \mn@doi [\aap] {10.1051/0004-6361/201833525},
  \href {https://ui.adsabs.harvard.edu/abs/2018A&A...619A..13V} {619, A13}

\bibitem[\protect\citeauthoryear{Virtanen et~al.,}{Virtanen
  et~al.}{2020}]{scipy}
Virtanen P.,  et~al., 2020, \mn@doi [Nature Methods]
  {10.1038/s41592-019-0686-2}, \href {https://rdcu.be/b08Wh} {17, 261}

\bibitem[\protect\citeauthoryear{{Wang} \& {Chen}}{{Wang} \&
  {Chen}}{2019}]{wang19}
{Wang} S.,  {Chen} X.,  2019, \mn@doi [\apj] {10.3847/1538-4357/ab1c61}, \href
  {https://ui.adsabs.harvard.edu/abs/2019ApJ...877..116W} {877, 116}

\bibitem[\protect\citeauthoryear{{Webb} \& {Carlberg}}{{Webb} \&
  {Carlberg}}{2021}]{webb21}
{Webb} J.~J.,  {Carlberg} R.~G.,  2021, \mn@doi [\mnras]
  {10.1093/mnras/stab353}, \href
  {https://ui.adsabs.harvard.edu/abs/2021MNRAS.502.4547W} {502, 4547}

\bibitem[\protect\citeauthoryear{{Wenger} et~al.,}{{Wenger}
  et~al.}{2000}]{simbad}
{Wenger} M.,  et~al., 2000, \mn@doi [\aaps] {10.1051/aas:2000332}, \href
  {https://ui.adsabs.harvard.edu/abs/2000A&AS..143....9W} {143, 9}

\bibitem[\protect\citeauthoryear{{Woody} \& {Schlaufman}}{{Woody} \&
  {Schlaufman}}{2021}]{woody21}
{Woody} T.,  {Schlaufman} K.~C.,  2021, \mn@doi [\aj]
  {10.3847/1538-3881/abff5f}, \href
  {https://ui.adsabs.harvard.edu/abs/2021AJ....162...42W} {162, 42}

\bibitem[\protect\citeauthoryear{{Zoccali}}{{Zoccali}}{2019}]{zoccali19}
{Zoccali} M.,  2019, Boletin de la Asociacion Argentina de Astronomia La Plata
  Argentina, \href {https://ui.adsabs.harvard.edu/abs/2019BAAA...61..137Z} {61,
  137}

\bibitem[\protect\citeauthoryear{{van den Bergh} \& {Hagen}}{{van den Bergh} \&
  {Hagen}}{1975}]{BH}
{van den Bergh} S.,  {Hagen} G.~L.,  1975, \mn@doi [\aj] {10.1086/111707},
  \href {https://ui.adsabs.harvard.edu/abs/1975AJ.....80...11V} {80, 11}

\makeatother
\end{thebibliography}



\appendix

\section{Gaia (BP-RP, G) recovered CMDs of the discovered GCs.}
\label{sec:AP_Gaia_CMDs}
\begin{figure*}
    \centering
    \includegraphics[scale=0.35]{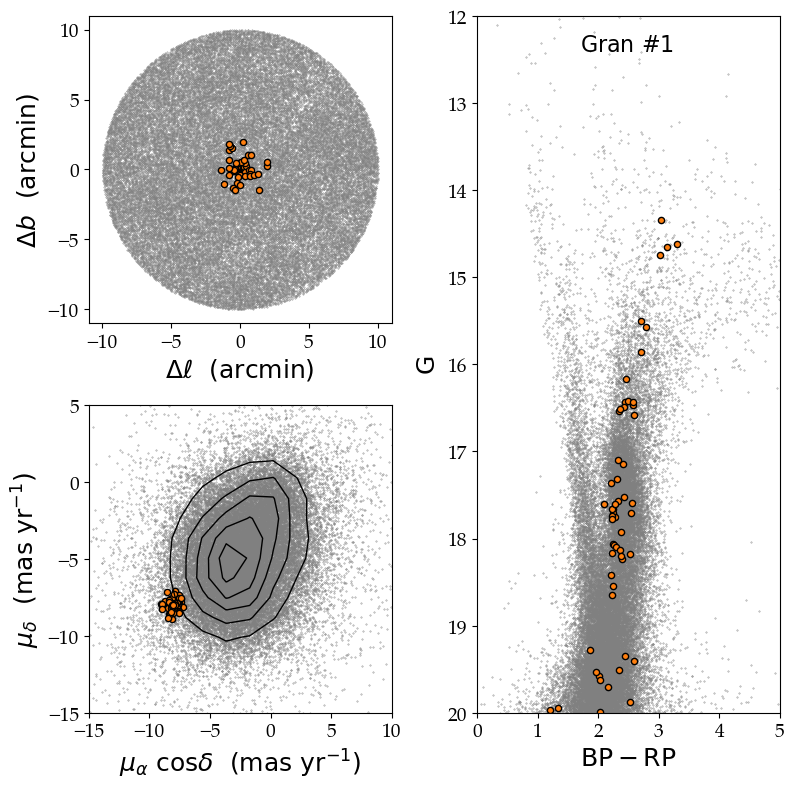}
    \includegraphics[scale=0.35]{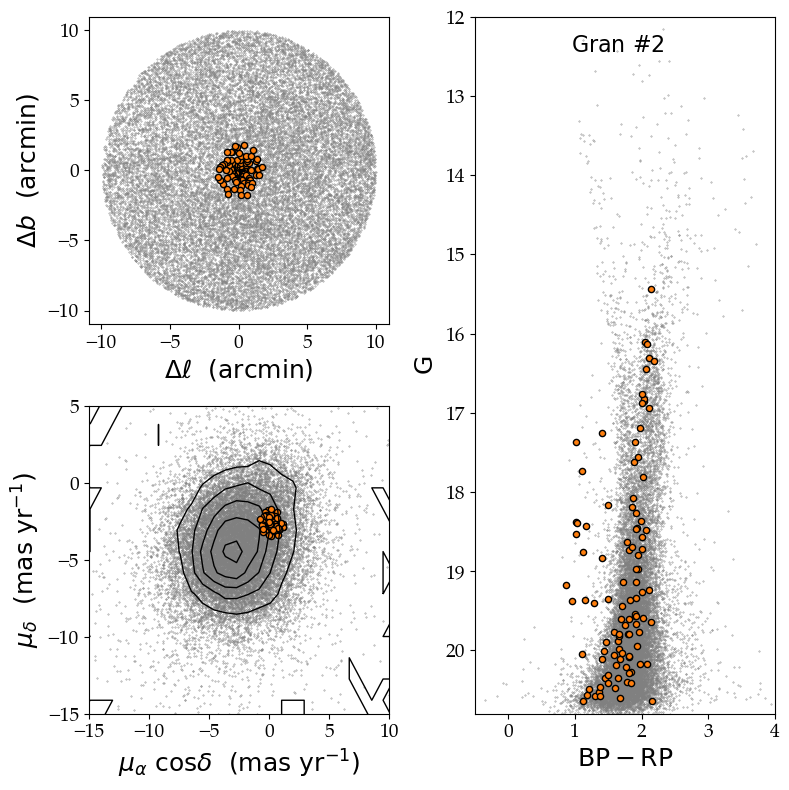}\\
    \includegraphics[scale=0.35]{plots/Gaia_CMD/Gran3_final.png}
    \includegraphics[scale=0.35]{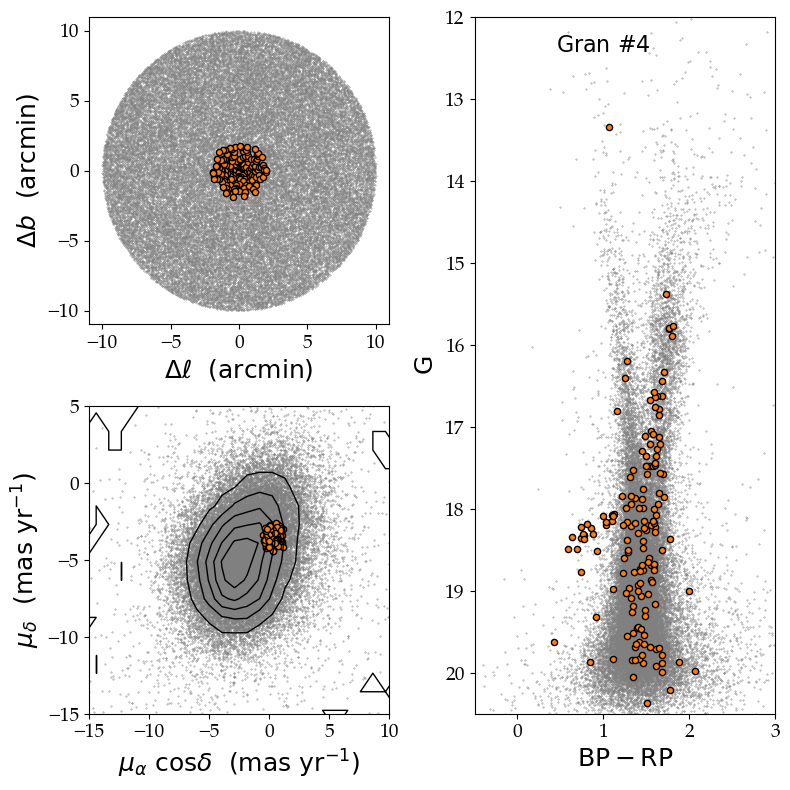}\\
    \includegraphics[scale=0.35]{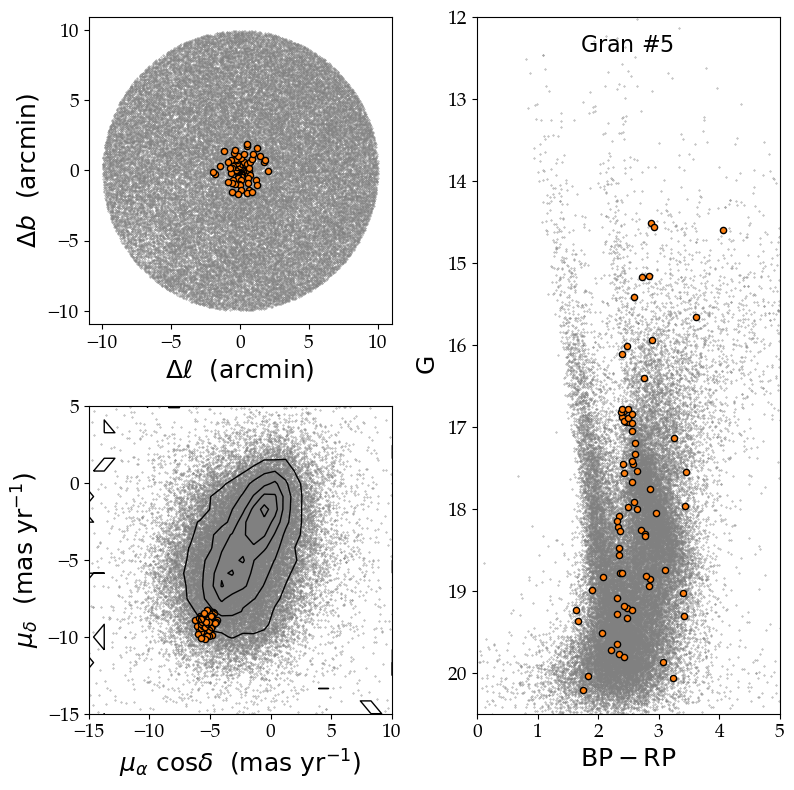}
    \caption{Gaia CMDs selected using only PM information. The panels follow the same structure and content as in Fig.~\ref{fig:Gaia_CMD}.}
    \label{fig:AP_Gaia_CMDs}
\end{figure*}

\section{Gaia-VVV (G-K$_{\rm s}$, K$_{\rm s}$) CMDs of the analysed clusters in the VVV footprint}
\label{sec:AP_GaiaVVV_CMDs}
\begin{figure*}
    \centering
    \includegraphics[scale=0.35]{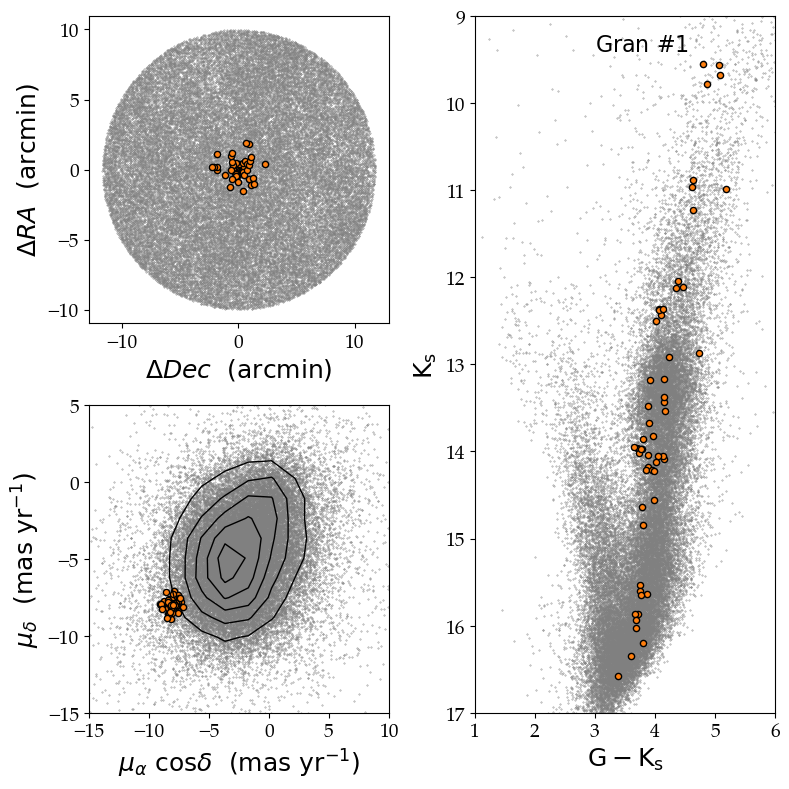}
    \includegraphics[scale=0.35]{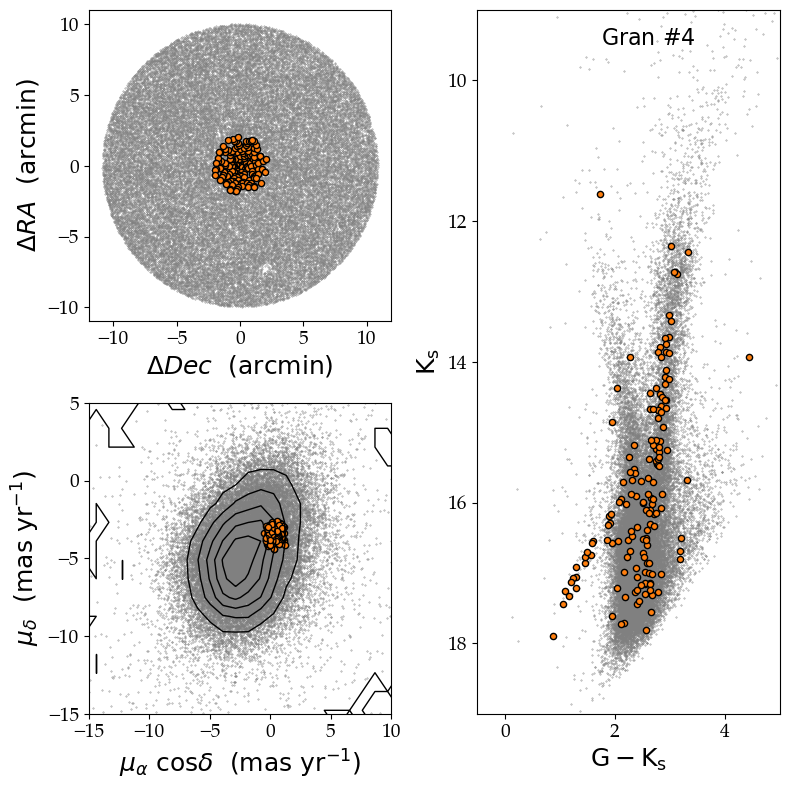}\\
    \includegraphics[scale=0.35]{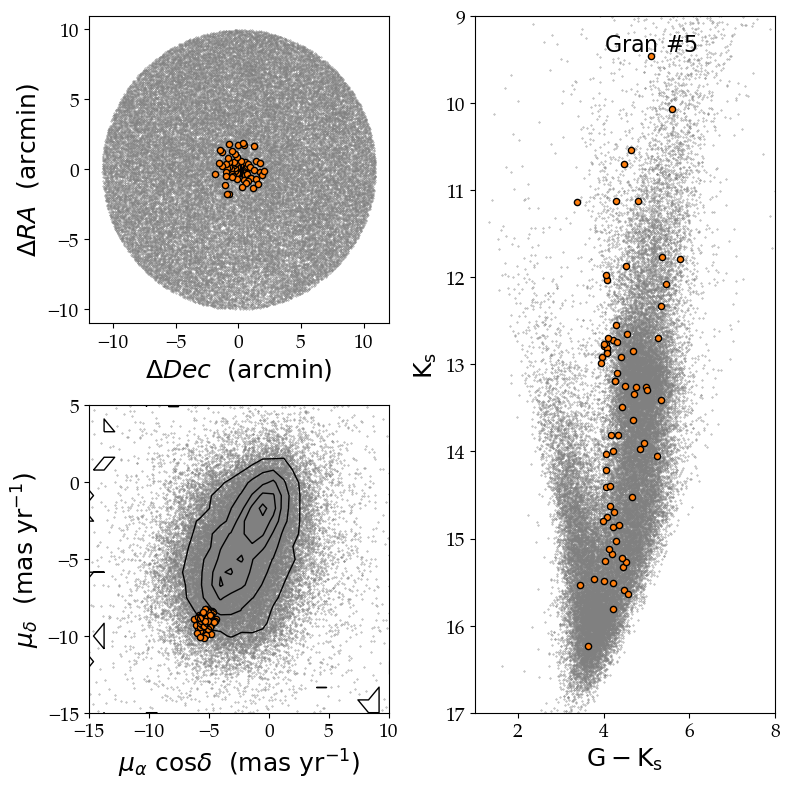}
    \caption{Optical-near-IR CMDs for the clusters within the VVV area. The panels follow the same structure and content as in Fig.~\ref{fig:GaiaVVV}.}
    \label{fig:AP_GaiaVVV_CMDs}
\end{figure*}

\section{Radial profiles of the clusters}
\label{sec:AP_radial_profiles}
\begin{figure*}
    \centering
    \includegraphics[scale=0.35]{plots/Radial_profile/Gran1_radial_profile.png}
    \includegraphics[scale=0.35]{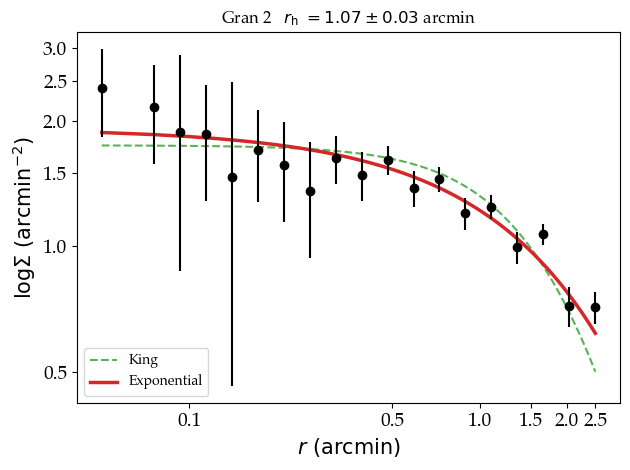}\\
    \includegraphics[scale=0.35]{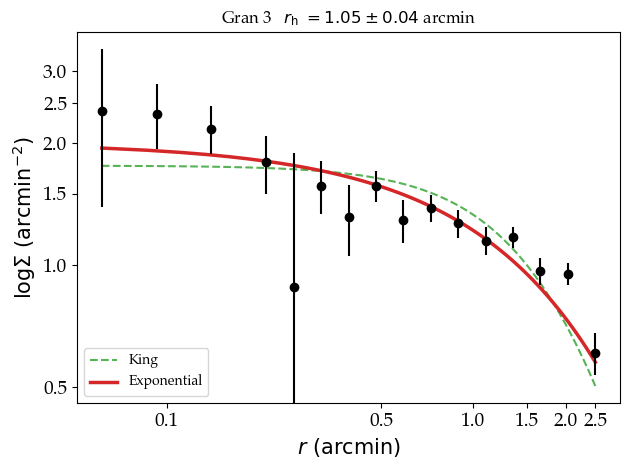}
    \includegraphics[scale=0.35]{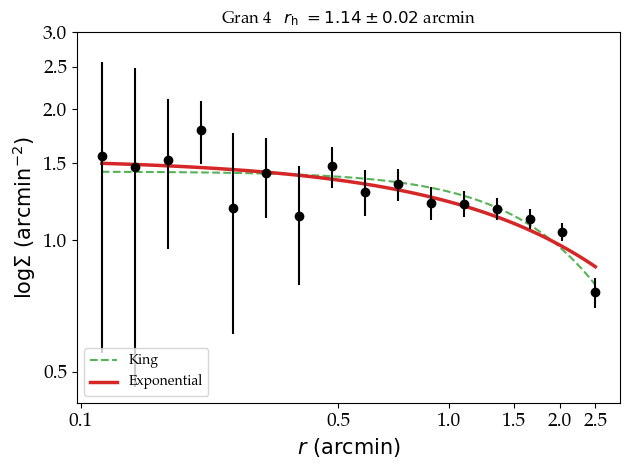}\\
    \includegraphics[scale=0.35]{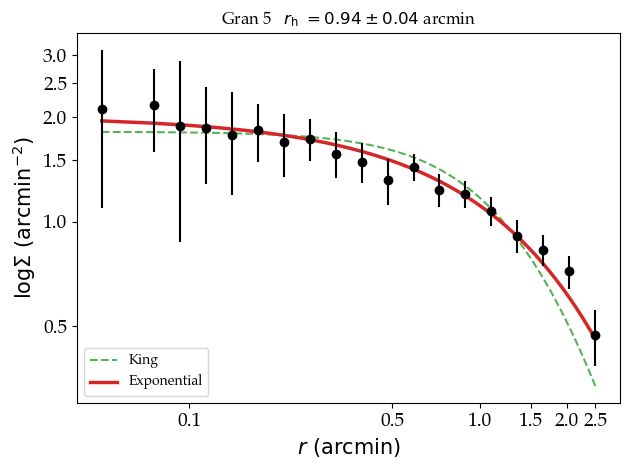}
    \caption{Radial luminosity profiles of the discovered clusters {\tt Gran 1-2-3-4-5}. 
    As explained in Sec.~\ref{sec:photometric} and Fig.~\ref{fig:gran1_profile}, the exponential profile is preferred in all the cases.}
    \label{fig:AP_rprofiles}
\end{figure*}

\section{Known GCs: BH~261 and Djorg~1}
\label{sec:AP_BH261Djorg1}
\begin{figure*}
    \centering
    \includegraphics[scale=0.35]{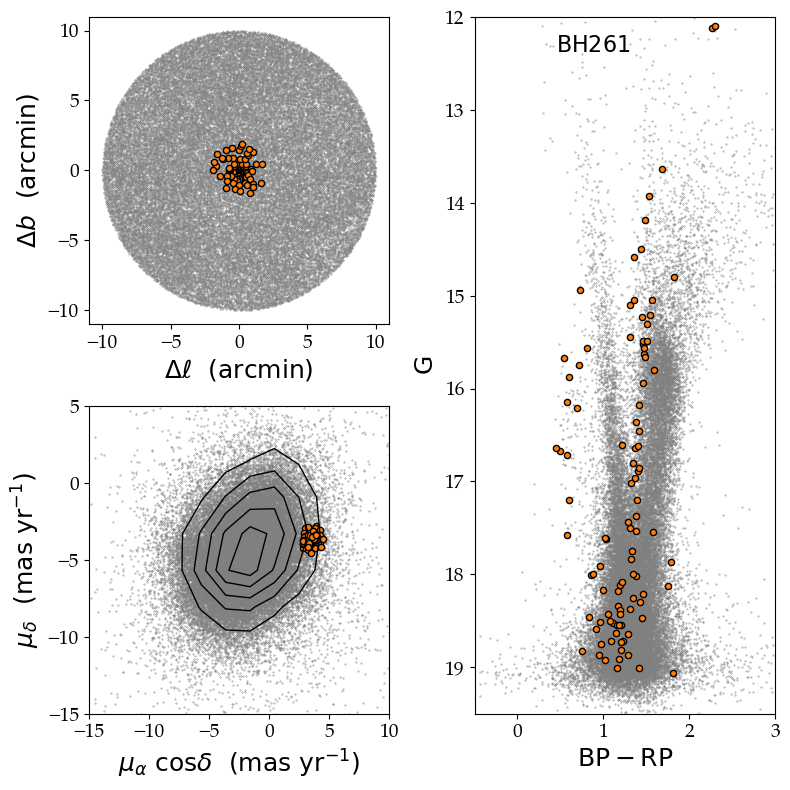}
    \includegraphics[scale=0.35]{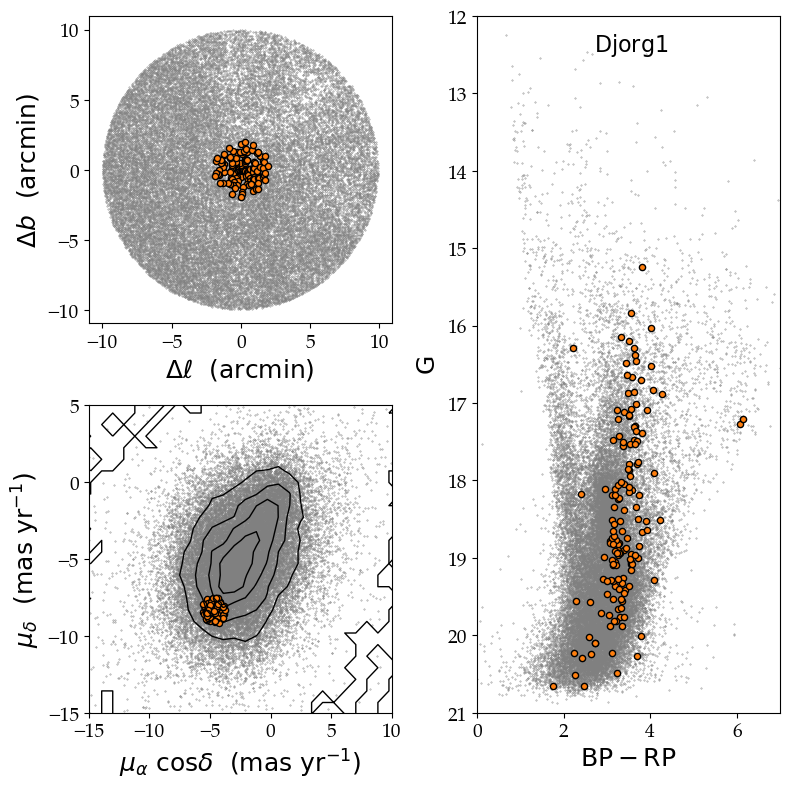}\\
    \includegraphics[scale=0.35]{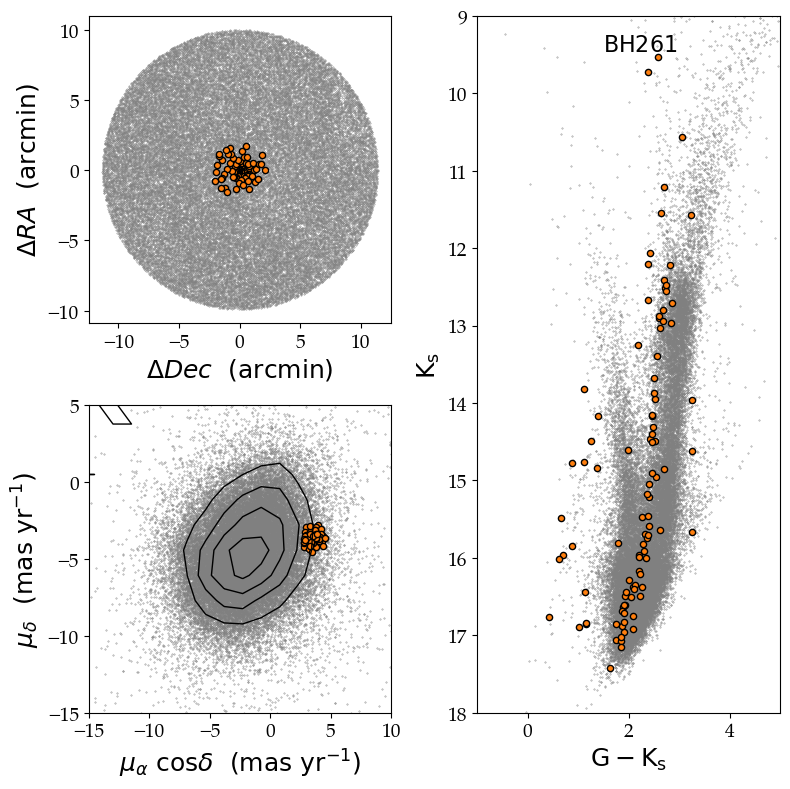}
    \includegraphics[scale=0.35]{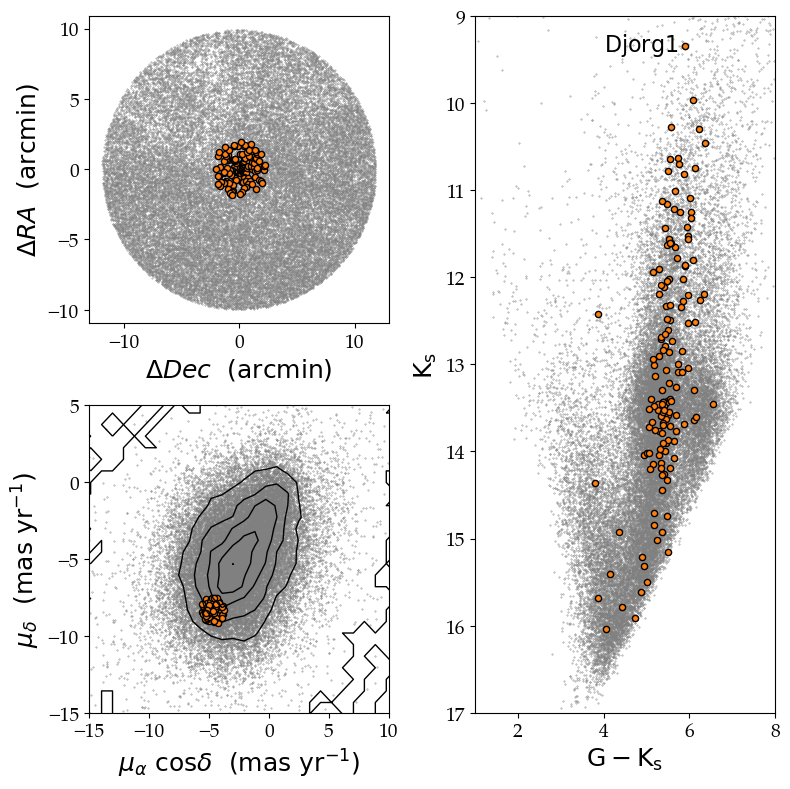}\\
    \includegraphics[scale=0.35]{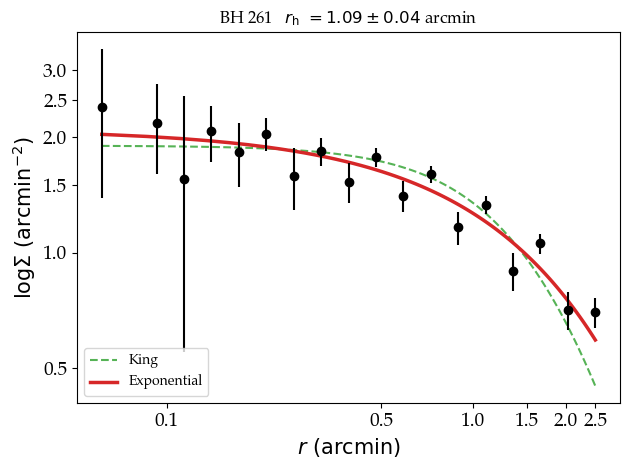}
    \includegraphics[scale=0.35]{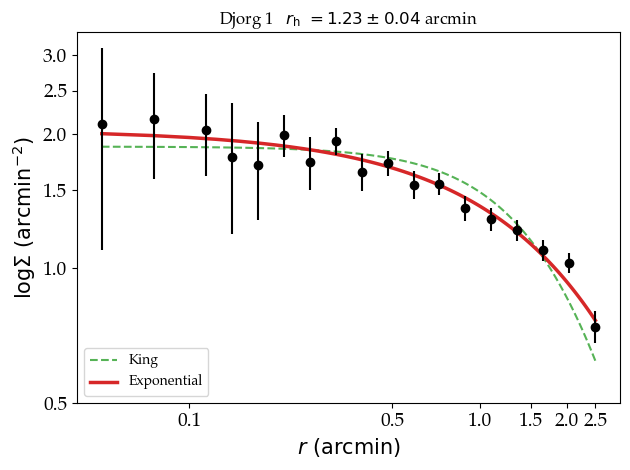}\\
    \caption{Gaia recovered CMDs for the GC BH~261 and Djorg~1 using only PM information. 
    The panels follow the same structure and content of the Fig.~\ref{fig:Gaia_CMD}, Fig.~\ref{fig:GaiaVVV}, 
    and Fig.~\ref{fig:gran1_profile} for the upper, middle and lower pair of panels, respectively.}
    \label{fig:AP_BH261Djorg1_plots}
\end{figure*}

\begin{figure*}
    \centering
    \includegraphics[scale=0.35]{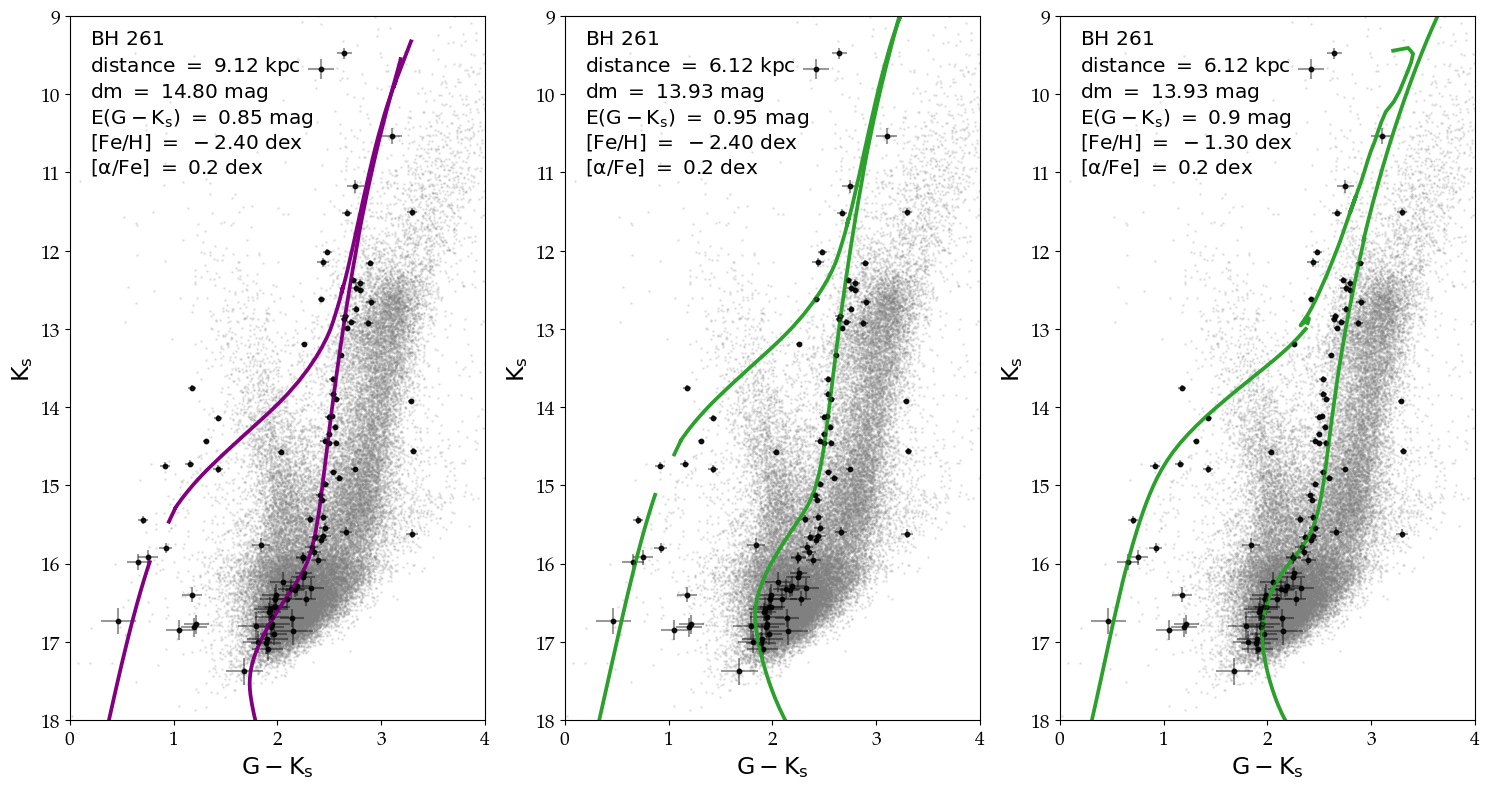}
    \caption{Isochrone comparison for the GC BH~261. 
    ({\bf Left panel}): Derived distance and metallicity of this work for the cluster ($9.12$ kpc and $-2.4$ dex) 
    traced by a combination of a PGPUC (SGB, RGB and HB stages) and PARSEC (AGB phase) isochrones.
    Detailed information regarding the distance modulus, extinction value, metallicity and $\alpha$-enhancement values 
    can be found in the text within each panel.
    Black points with error bars and grey background represent the Gaia PM-selected cluster members 
    and the field stars within 10 arcmin of the cluster centre, respectively.
    ({\bf Middle panel}): Same as the left panel, but using the literature distance to fit the ${\rm [Fe/H] = -2.4}$ dex isochrone.
    The extinction value was shifted to fit the RGB stars present in the cluster.
    ({\bf Right panel}): Same as the left panel, but using the \protect\cite{ortolani06} derived metallicity ($-1.3$ dex) and distance ($6.12$ kpc).}
    \label{fig:AP_BH261_iso}
\end{figure*}

\section{GC candidate {\tt C1}}
\label{sec:AP_C1}

\begin{figure*}
    \centering
    \includegraphics[scale=0.30]{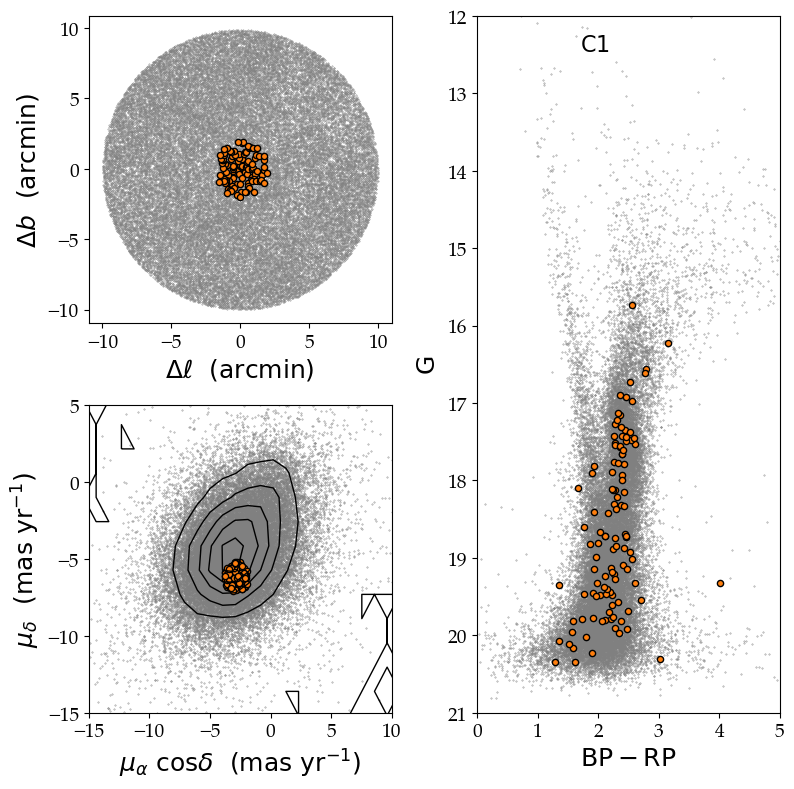}
    \includegraphics[scale=0.30]{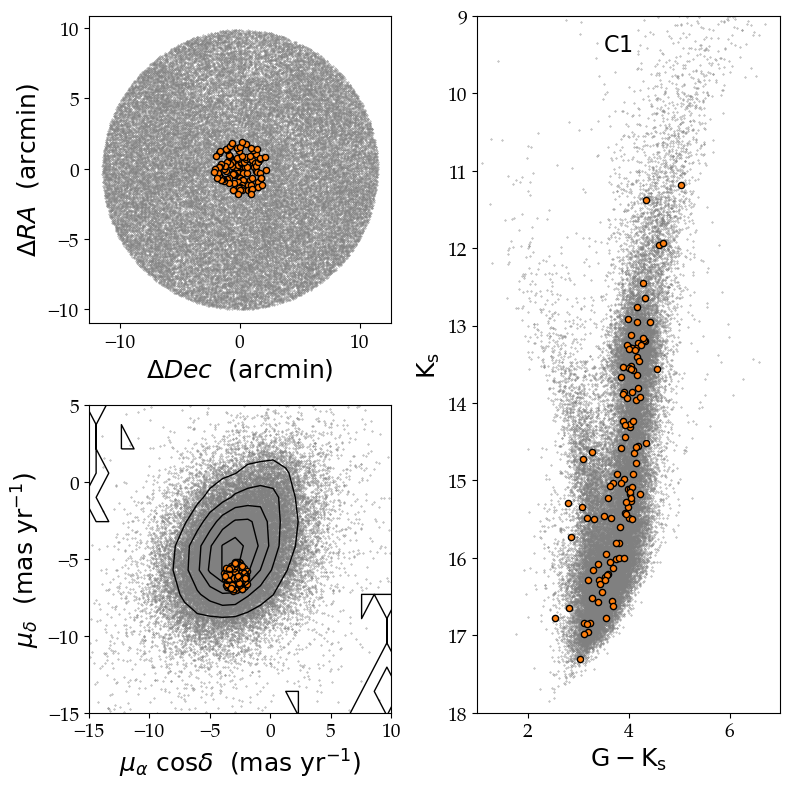}\\
    \includegraphics[scale=0.30]{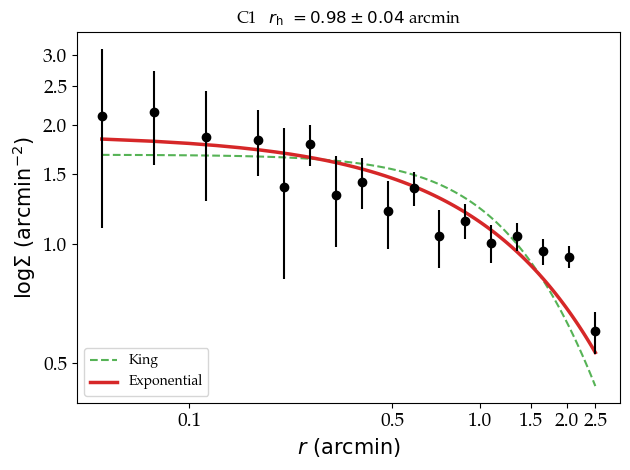}\\
    \caption{Gaia, Gaia-VVV CMDs, and radial profile of the discarded GC candidate {\tt C1}. 
    The panels follow the same structure and content of the Fig.~\ref{fig:Gaia_CMD}, Fig.~\ref{fig:GaiaVVV}, 
    and Fig.~\ref{fig:gran1_profile} for the upper, middle and lower pair of panels, respectively.}
    \label{fig:AP_C1}
\end{figure*}

\begin{figure*}
    \centering
    \includegraphics[scale=0.30]{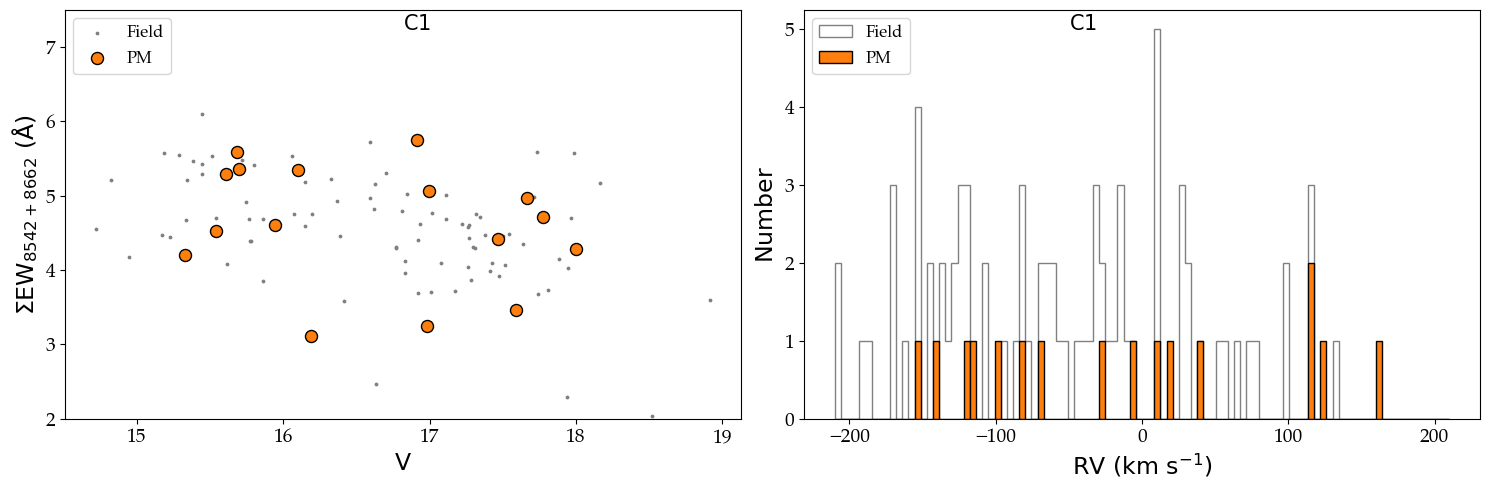}
    \caption{({\bf Left panel}): EW of the two most prominent lines in the CaT versus the V magnitude for {\tt C1}.
    Circular orange symbols represent the PM-selected stars matched with the MUSE extracted spectra and grey points represent the field stars present in the cube. 
    Note that no clear relation between the EW and V-magnitude is observed, contrary to the expected GC behaviour.
    ({\bf Right panel}): RV histogram of all the sources in the MUSE cube colour-coded by a selection procedure: orange for the PM and grey for the field stars in the FoV.
    No RV peak is seen within the 16 matched Gaia PM-selected stars. }
    \label{fig:AP_C1_MUSE}
\end{figure*}




\bsp	
\label{lastpage}
\end{document}